\documentclass[11pt,a4paper]{article}
\usepackage{jheppub}
\pdfoutput=1

\usepackage{amsmath,amssymb,amsfonts,graphics,graphicx,amscd,amsfonts,epsf,epsfig,color}
\graphicspath{{./figs/}}
\usepackage{subcaption}

\usepackage[textsize=tiny]{todonotes}

\allowdisplaybreaks[4]

\subheader{DMUS-MP-20/03, LLNL-JRNL-809361, RIKEN-iTHEMS-Report-20, UTHEP-749}

\title{Partial Deconfinement at Strong Coupling on the Lattice}

\author[a]{Hiromasa Watanabe,}
\author[b]{Georg Bergner,}
\author[c]{Norbert Bodendorfer,}
\author[d]{Shotaro Shiba Funai,}
\author[e]{Masanori Hanada,}
\author[f,g]{Enrico Rinaldi,}
\author[c]{Andreas Sch\"{a}fer,}
\author[h,i]{and Pavlos Vranas}

\affiliation[a]{
Graduate School of Pure and Applied Sciences, University of Tsukuba,\\
Tsukuba, Ibaraki 305-8571, Japan}
\emailAdd{watanabe@het.ph.tsukuba.ac.jp}
\affiliation[b]{
University of Jena, Institute for Theoretical Physics,\\
Max-Wien-Platz 1, D-07743 Jena, Germany}
\affiliation[c]{
University of Regensburg, Institute of Theoretical Physics,\\
Universit\"{a}tsstrasse 31, D-93053, Germany}
\affiliation[d]{
Physics and Biology Unit, Okinawa Institute of Science and Technology (OIST),\\
1919-1 Tancha, Onna-son, Kunigami-gun, Okinawa 904-0495,Japan}
\affiliation[e]{
Department of Mathematics, University of Surrey, Guildford, Surrey, GU2 7XH, UK}
\affiliation[f]{
Arithmer Inc., R\&D Headquarters, Minato, Tokyo 106-6040, Japan}
\affiliation[g]{
RIKEN iTHEMS Program, Wako, Saitama 351-0198, Japan}
\affiliation[h]{
Nuclear and Chemical Sciences Division, Lawrence Livermore National Laboratory,\\
Livermore CA 94550, USA}
\affiliation[i]{
Nuclear Science Division, Lawrence Berkeley National Laboratory,\\
Berkeley, CA 94720, USA}

\abstract{
We provide evidence for partial deconfinement --- the deconfinement of a SU($M$) subgroup of the SU($N$) gauge group ---
by using lattice Monte Carlo simulations.
We take matrix models as concrete examples.
By appropriately fixing the gauge, we observe that the $M\times M$ submatrices deconfine.
This gives direct evidence for partial deconfinement at strong coupling.
We discuss the applications to QCD and holography.
}
\begin{document}
\maketitle

\section{Introduction}\label{sec:introduction}
\hspace{0.51cm}

\begin{figure}[htbp]
\begin{center}
\scalebox{0.4}{
\includegraphics{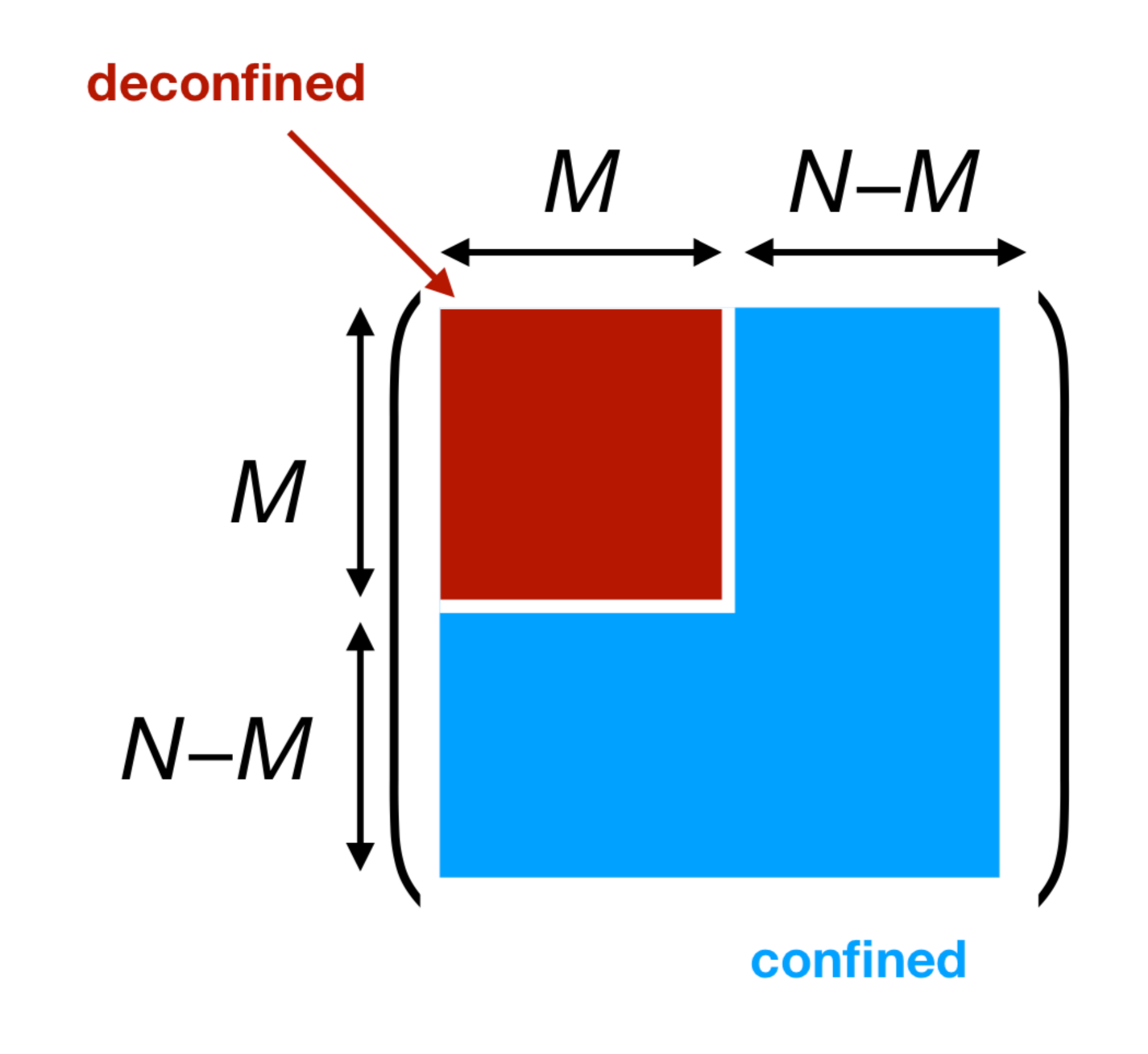}}
\end{center}
\caption{Pictorial representation of partial deconfinement for gauge and adjoint matter degrees of freedom.
Only the $M\times M$-block shown in red is excited.
This picture is taken from Ref.~\cite{Hanada:2019czd}.
In this figure, one specific embedding of SU($M$) into SU($N$) is shown.}
\label{fig:matrix-partial-deconfinement}
\end{figure}

Partial deconfinement~\cite{Hanada:2016pwv,Berenstein:2018lrm,Hanada:2018zxn,Hanada:2019czd,Hanada:2019kue,Hanada:2020uvt} is the coexistence of confined and deconfined sectors in the space of colors (see Fig.~\ref{fig:matrix-partial-deconfinement}: the partially deconfined phase lies between the completely confined phase ($M=0$ in Fig.~\ref{fig:matrix-partial-deconfinement}) and the completely deconfined phase ($M=N$ in Fig.~\ref{fig:matrix-partial-deconfinement}).
A precise definition is given in Sec.~\ref{sec:review}).
Intuitively, the only difference from the common two-phase coexistence, say liquid and solid water at zero degree Celsius and one atmosphere, is that the separation is taking place in the internal space (the space of color degrees of freedom), rather than the usual space.
As we will review in Sec.~\ref{sec:review}, this difference leads to various interesting consequences, some of which are unexpected at first glance.
In Ref.~\cite{Hanada:2020uvt} it was shown that partial deconfinement is analogous to superfluidity, in the sense that the confined and deconfined sectors resemble the superfluid and normal fluid.
This rather unexpected analogy is explained in Sec.~\ref{sec:BEC}, which makes it easier to grasp the physical mechanism of partial deconfinement.

So far, most of the studies of partial deconfinement have been limited to weakly coupled theories: there was no firm evidence beyond heuristic arguments that justified the existence of partial deconfinement at strong coupling.
As a consequence, a sharp definition of partial deconfinement at strong coupling was missing.
In this paper, we focus on a strongly-coupled theory and use numerical methods based on Lattice Monte Carlo simulations to investigate signals of partial deconfinement.
Our results show that, at least for the model we study, the definition of partial deconfinement used at weak coupling can be adopted without change at strong coupling.
In particular, the simplest order parameter is $M$ itself, which is hidden in the theory in a nontrivial manner, as we will see in Sec.~\ref{sec:BEC}.
We show that the value of $M$ determined in a few different manners coincide, which provides us with a strong consistency check.
We will also discuss an explicit gauge fixing which makes the two-phase coexistence manifest.

As a concrete setup, we consider the gauged bosonic matrix model with SU($N$) gauge group.
This theory's action is the dimensional reduction of the $(d+1)$-dimensional Yang-Mills action to $(0+1)$-dimensions.
With the Euclidean signature that will be used in Lattice Monte Carlo simulations, the action is given by
\begin{eqnarray}
S
=
N\int_0^\beta dt\ {\rm Tr}\left\{
\frac{1}{2}(D_t X_I)^2
-
\frac{1}{4}[X_I,X_J]^2
\right\}.
\label{eq:action-bos-BFSS}
\end{eqnarray}
Here $I,J$ run through $1,2,\cdots, d$ (in this paper we focus on $d=9$), $\beta$ is the circumference of the temporal circle which is related to temperature $T$ by $\beta=T^{-1}$, and $X_I$'s are $N\times N$ hermitian matrices.
The covariant derivative $D_t$ is defined by $D_tX_I=\partial_tX_I-i[A_t,X_I]$, where $A_t$ is the gauge field.
This model is often simply called ``the Yang-Mills matrix model'', or \emph{bosonic BFSS}, because it is the bosonic part of the Banks-Fischler-Shenker-Susskind matrix model~\cite{Banks:1996vh,deWit:1988wri} when $d=9$.
The reasons to consider this model are its simplicity and its close relation to the gauge/gravity duality.
Moreover, this model has been studied in the past, both in the weak coupling and the strong coupling regime, and its phase structure at finite temperature is also known~\cite{Bergner:2019rca}.
Building upon the current literature on this model, we can focus on decisive signs of partial deconfinement.

In the large-$N$ limit, this model exhibits a confinement/deconfinement transition characterized by the increase of the entropy from $O(N^0)$ to $O(N^2)$~\cite{Witten:1998zw}.
Concerning this finite temperature transition, partial deconfinement is the phenomenon where only an SU($M$) subgroup of the SU($N$) gauge group deconfines, as pictorially shown in Fig.~\ref{fig:matrix-partial-deconfinement}.
To characterize partial deconfinement, it is convenient to define a continuous parameter identified by $\frac{M}{N}$ whose value can change from 0 to 1.
Partial deconfinement happens between the completely confined phase, with $\frac{M}{N}=0$, and the completely deconfined phase, with $\frac{M}{N}=1$.
These two phases have entropy and energy of order $N^0$ and $N^2$ (up to the zero-point energy), respectively.

Now, suppose the energy is of order $\epsilon N^2$, where $\epsilon$ is an order $N^0$ number and much smaller than 1.
What kind of quantum states are dominant in the microcanonical ensemble at such intermediate energy range?
The system cannot be in the confined phase because the energy is much larger than $N^0$, but it cannot be in the deconfined phase either because the energy is much smaller than $N^2$.
In partial deconfinement, where an SU($M$) subgroup deconfines, the energy and entropy are of order $M^2$, and hence by taking $M\sim\sqrt{\epsilon}N$ such intermediate values of the energy and entropy can be explained.
Other explanations that may seem natural are discussed in Sec.~\ref{sec:review}, together with the precise characterization of partial deconfinement.

The numerical study in Ref.~\cite{Bergner:2019rca} appears to be consistent with the existence of this partially-deconfined phase. (See also Refs.~\cite{Azuma:2014cfa,Morita:2020liy} regarding the phase diagram,
and Refs.~\cite{Aharony:2004ig,Kawahara:2007fn} for pioneering numerical studies with limited numerical resources
which clarified the qualitative nature of the transition.)
In the $d=9$ Yang-Mills matrix model, the partially-deconfined phase has negative specific heat.
Hence, in the canonical ensemble, it is the maximum of the free energy and not preferred thermodynamically.
\footnote{In the microcanonical ensemble, the entropy is maximized at each fixed energy.
In the canonical ensemble, the free energy is minimized at each fixed temperature.}
Still, this phase is stable in the microcanonical ensemble.
From the point of view of gauge/gravity duality, this phase is interpreted~\cite{Hanada:2016pwv,Hanada:2018zxn} as the dual of a small black hole~\cite{Aharony:1999ti,Aharony:2003sx,Dias:2016eto}.

In order to investigate partial deconfinement directly in the nonperturbative regime, we rely on Lattice Monte Carlo simulations.
These simulations utilize the path integral formalism of quantum mechanics, where the sum over paths is replaced with a sum over `important' field configurations.
On the other hand, partial deconfinement is simpler to analyze in the Hamiltonian formalism, with access to the characteristics of individual states in the Hilbert space.
Since the field configurations in the path integral are different from the wave functions describing the states, the signals of partial deconfinement become more intricate.
Therefore, we devise the following strategy.
In Sec.~\ref{sec:Gaussian-matrix-model}, we consider the analytically solvable case of the gauged Gaussian Matrix Model
\begin{eqnarray}
S
=
N\int_0^\beta dt\ {\rm Tr}\left\{
\frac{1}{2}(D_t X_I)^2
+
\frac{1}{2}X_I^2
\right\},
\label{eq:action_Gaussian}
\end{eqnarray}
where $I=1,2,\cdots,d$.
With guidance from analytical results, we derive a few nontrivial features of partial deconfinement in terms of the {\it master field}.
We show that the features of the master field can be seen in the lattice configurations and exemplify them with numerical evidence.
Then, in Sec.~\ref{sec:Yang-Mills-MM} we move on to investigate the nonperturbative Yang-Mills matrix model, which is the original target of our study.
The goal is to establish the coexistence of the confined and deconfined sectors in the strongly-interacting theory. For this purpose we will determine whether the features of the master field we discovered in the Gaussian matrix model can be applied to the field configurations in our target theory.
If that is the case, we can demonstrate that the partially-deconfined phase is an intermediate phase in the confinement/deconfinement transition even at strong coupling.
In Sec.~\ref{sec:conclusion} we conclude the paper with some discussions regarding the future applications.

\section{Review of partial deconfinement}\label{sec:review}
\hspace{0.51cm}

We already mentioned in the introduction that partial deconfinement is related to phase coexistence in color space.
As such, it can be seen as a phenomenon characterized by deconfinement of an SU($M$) subgroup within the SU($N$) gauge group of a large-$N$ gauge theory.
In this section, we start with a formal definition, which can be applied to any large-$N$ gauge theory, including strongly-coupled theories, without introducing any new concept which could be confusing to the readers.
Moreover, we provide explicit examples to clarify the physical meaning behind the formal definition: first, the motivation from the gravity side in Sec.~\ref{sec:gravity-dual}, then an intuitive picture in the gauge theory side in Sec.~\ref{sec:intuitive-picture}, and lastly, the BEC-confinement correspondence which makes the meaning of the formal definition clear in Sec.~\ref{sec:BEC}.

\begin{figure}
\begin{center}
\scalebox{0.2}{
\includegraphics{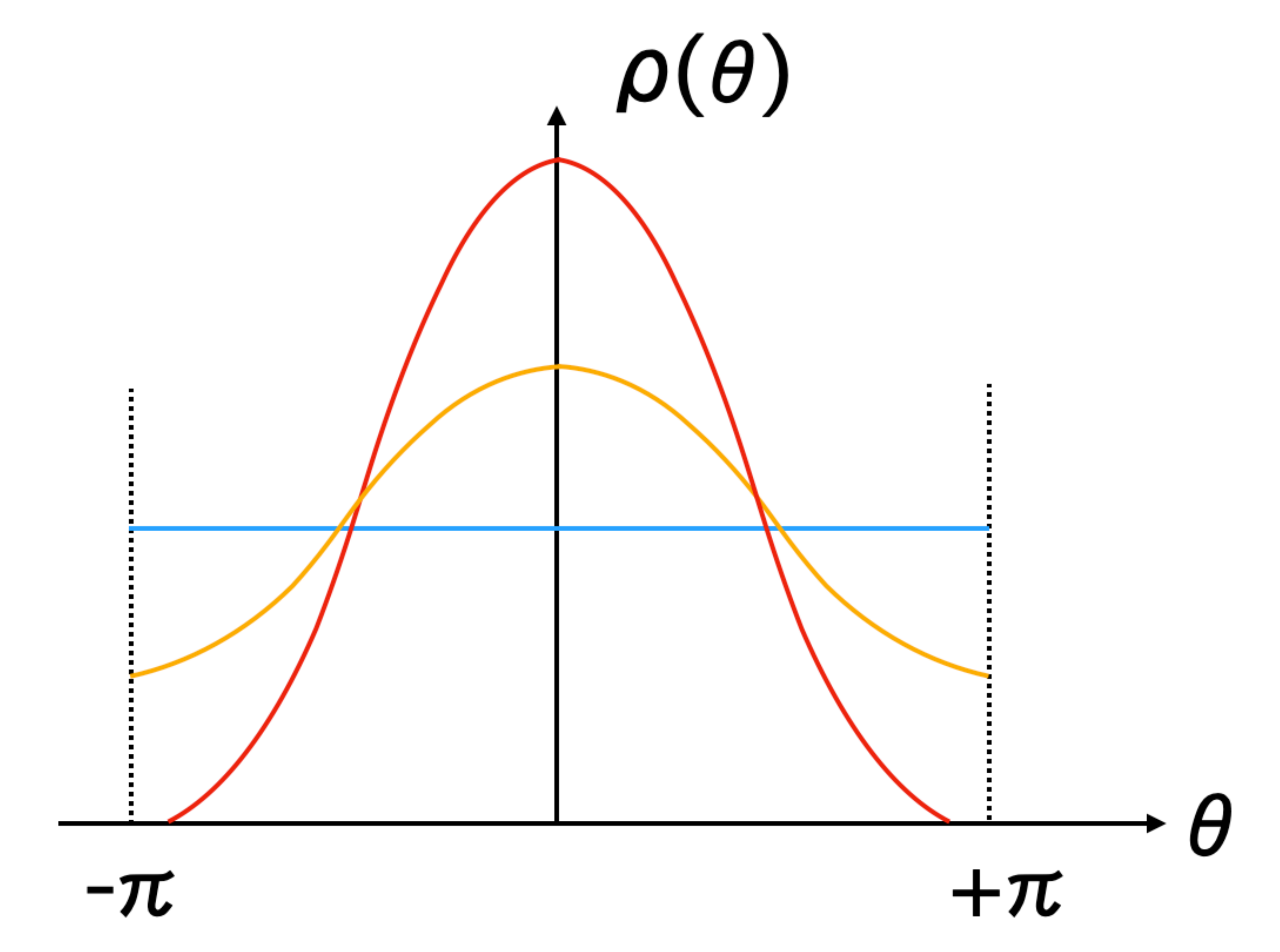}}
\end{center}
\caption{
The distributions of the Polyakov line phases in each phase.
The blue, orange and red lines are the completely-confined, partially-deconfined and completely-deconfined phases, respectively.
The specific form of the distribution function $\rho(\theta)$ for the partially and completely deconfined phases depends on the theory under consideration.
}
\label{fig:phase-distribution}
\end{figure}

For the formal definition we use the distribution of the phases of the Polyakov line.
In gauge theories, the Polyakov line is defined by ${\cal P}e^{i\int_0^\beta dt A_t}$, where ${\cal P}$ stands for path ordering.
This is a $N\times N$ unitary matrix, and the eigenvalues are written as $e^{i\theta_1},\cdots,e^{i\theta_N}$, where the phases $\theta_1,\cdots,\theta_N$ lie between $-\pi$ and $+\pi$.
At large $N$, the phases are described by the distribution function $\rho(\theta)$ which is normalized as $\int_{-\pi}^\pi d\theta\rho(\theta)=1$.
By using $\rho(\theta)$, the completely-confined phase, partially-deconfined phase (equivalently, partially-confined phase) and completely-deconfined phase are defined as follows~\cite{Hanada:2020uvt}:\footnote{
The definition provided here requires to strictly take the large-$N$ limit.
At the qualitative level, the same picture can be valid at finite $N$.
Imagine that we put finite amount of water in our freezer. We do not have to take the infinite-volume limit in order to see the coexistence of liquid phase and solid phase. Of course, it is not clear whether atoms at the interface of two phases belong to solid or liquid, but it is a minor correction to the phase-coexistence picture. Exactly the same situation can be realized at finite $N$, namely most colors are in the confined sector or deconfined sector, but some colors may be somewhere in between.
In any event, the large-$N$ limit is needed for a formal definition, so we will focus on the large-$N$ limit in this paper. 
}
\begin{itemize}
\item
The completely-confined phase refers to an equilibrium state with the uniform phase distribution, $\rho(\theta)=\frac{1}{2\pi}$. See blue line in Fig.~\ref{fig:phase-distribution}.
Note that this is the state which is traditionally called simply the `confined phase'.
\item
The partially-deconfined phase refers to an equilibrium state with the nonuniform phase distribution which is strictly positive, $\rho(\theta)>0$. See orange line in Fig.~\ref{fig:phase-distribution}.
\item
The completely-deconfined phase refers to an equilibrium state with the nonuniform phase distribution which is zero in a finite range, in other words it is the `gapped' eigenvalue distribution.
See red line in Fig.~\ref{fig:phase-distribution}; there is a gap in the distribution around $\theta=\pm\pi$.
\end{itemize}

Note that, because this definition does not refer to the center symmetry, it can also be applied to theories without center symmetry, such as the $N_f$ -flavor large-$N_c$ QCD with $N_f/N_c$ finite.
(In fact, we believe that partial-deconfinement applies to a rather broad range of theories.)
The actual physical meaning of partial deconfinement is not immediately clear from this formal definition.
To illustrate it let us focus on the behavior of color degrees of freedom. As depicted in Fig. 1, two phases -- confined and deconfined -- coexist in the space of colors~\cite{Hanada:2016pwv,Berenstein:2018lrm,Hanada:2018zxn,Hanada:2019czd,Hanada:2019kue,Hanada:2020uvt}.
The readers may wonder why we consider the possibility of such a seemingly-exotic phase, and how it is related to the formal definition given above. Therefore, we will provide an extensive review below.

\subsection{Motivation from gravity}
\label{sec:gravity-dual}
\hspace{0.51cm}

Partial deconfinement has been originally proposed in order to solve a few puzzles associated with the confinement/deconfinement phase transition of gauge theories in light of the gauge/gravity duality.
In particular, the original motivation~\cite{Hanada:2016pwv} was to explain how the gauge/gravity duality relates the thermodynamics of gauge theories to the physics of black holes: this is a very nontrivial problem studied by several papers in the literature, as we discuss below.

The clearest example of the connection between thermal phase transitions in gauge theories and black holes can be seen in the duality between the thermodynamics of the 4d ${\cal N}=4$ super Yang-Mills theory (SYM) and the type IIB superstring theory on AdS$_5\times$S$^5$~\cite{Witten:1998zw}.
In this duality, the confined and deconfined phases on the gauge theory side are dual to the thermal AdS geometry and the `large' black hole in AdS space, respectively.
In the canonical ensemble, these two phases are separated by a first order phase transition.
This transition is called the Hawking-Page transition.

It was immediately realized that between these two phases there must be an intermediate region.
From the gravity side of the duality, this is simply a very small black hole, which is approximately the ten-dimensional Schwarzschild black hole~\cite{Aharony:1999ti,Aharony:2003sx}.
The energy of this small black hole scales as $E\sim N^2T^{-7}$, where $N^2$ corresponds to the inverse of the Newton constant.
This is a stable physical state in the microcanonical ensemble.\footnote{
Precisely speaking, the phase diagram is a little bit more complicated.
Below a certain energy, the large BH (wrapping on S$^5$) has a negative specific heat.
In a certain energy range, both the large BH with negative specific heat and
the small BH can be entropy maxima in the microcanonical ensemble.
At some energy the large BH solution becomes unstable against the localization along S$^5$, analogous to the Gregory-Laflamme transition between black hole and black string.
The details of this phase transition have not been fully understood yet.
There are attempts to construct gravity solutions connecting the two phases
(e.g.~Ref.~\cite{Dias:2015pda}),
and a proposal regarding the gauge theory description~\cite{Yaffe:2017axl}.
Our numerical analyses in this paper are not precise enough to see such fine structure, even if it exists.
}
Schwarzschild black holes with negative specific heat are more realistic than charged black holes such as the `large' black holes in AdS, which have positive specific heat.
Therefore, from the physics point of view, it is important to understand this intermediate phase.

\begin{figure}
\begin{center}
\scalebox{0.2}{
\includegraphics{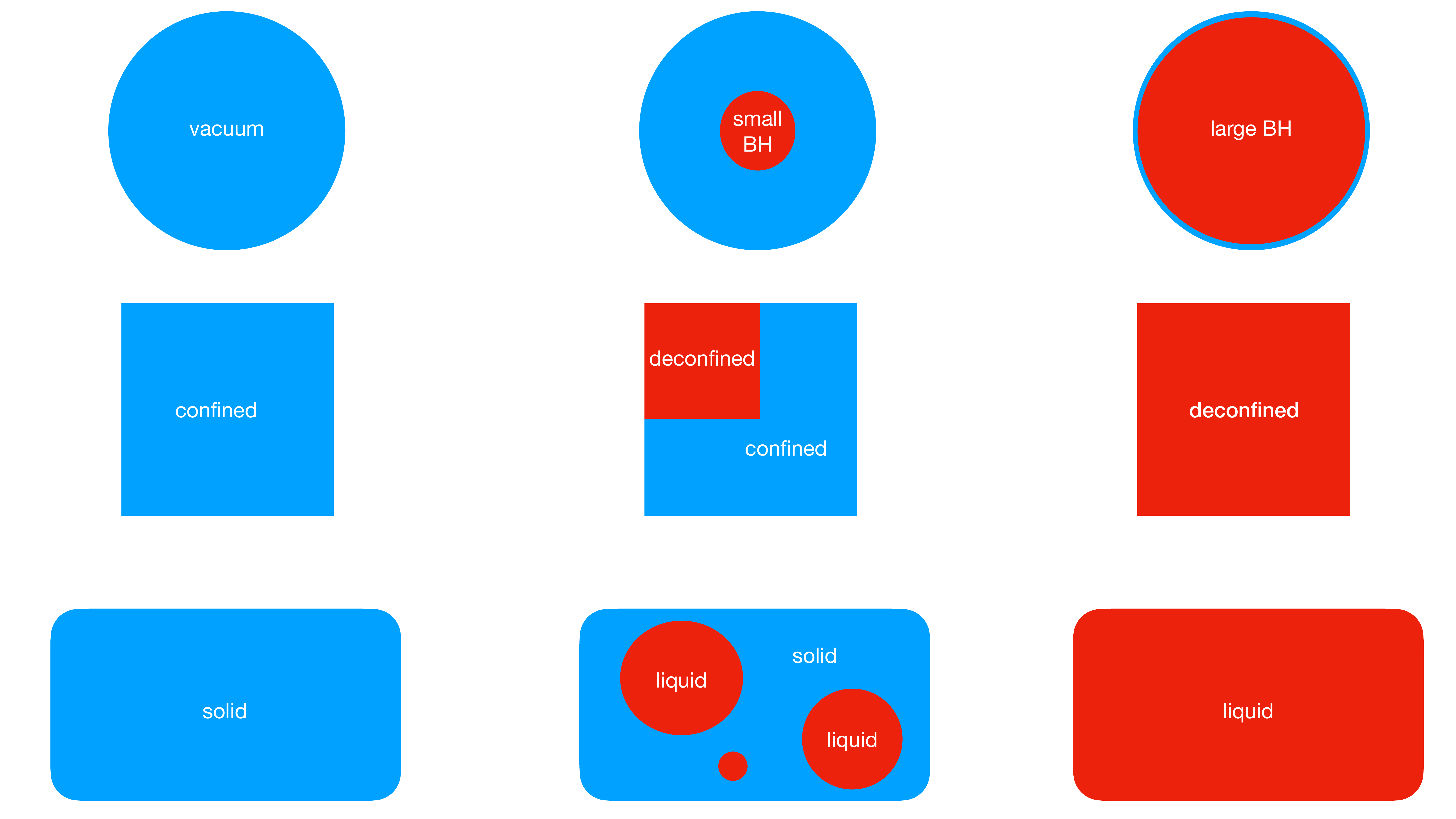}}
\end{center}
\caption{
Analogy between three phases in string theory in AdS space (top), Yang-Mills theory (middle) and water (bottom).
Each column represent a different phase. The two-phase coexistence is shown in the center column. The two-phase coexistence in color space can take place even when the order of phase transition is not one. As explicit examples, see Ref.~\cite{Hanada:2019kue}.
}
\label{fig:correspondence}
\end{figure}

One of the ``puzzles'' is how to interpret this small black hole state on the gauge side of the duality.
In fact, how can a healthy quantum field theory lead to a stable state with a negative specific heat?
Partial deconfinement gives a natural answer to this problem~\cite{Hanada:2016pwv}, introducing a new phase with negative specific heat in thermal phase transitions.
The small deconfined sector is regarded as a small black hole, just in the same way as a large deconfined sector --- completely deconfined phase --- is regarded as a `large' black hole.
(cfr. the top and middle rows in Fig.~\ref{fig:correspondence}.)
Similar phases with negative specific heat are predicted for other theories too.
For example, the D0-brane quantum mechanics~\cite{Banks:1996vh,deWit:1988wri} is expected to describe the Schwarzschild black hole in eleven dimensions at very low temperature~\cite{Itzhaki:1998dd}: such phase would also be understood as a partially-deconfined phase.

In the context of holography with weakly-curved gravity description, the partially-deconfined phase
has negative specific heat. However in more generic gauge theories it is not necessarily the case~\cite{Hanada:2018zxn,Hanada:2019czd,Hanada:2019kue},
as we will see in Sec.~\ref{sec:intuitive-picture}.

\subsection{Intuitive picture}\label{sec:intuitive-picture}
\hspace{0.51cm}
While partial deconfinement seems like an original and natural explanation for the physics of the small black hole, it was conjectured to be a general mechanism at work in many theories~\cite{Berenstein:2018lrm,Hanada:2018zxn}.
For example, it has been analytically proven in several weakly coupled theories~\cite{Hanada:2019czd,Hanada:2019kue,Hanada:2020uvt},
starting from seminal papers~\cite{Aharony:2003sx,Sundborg:1999ue}.
These pioneering papers pointed out that the confinement/deconfinement transition characterized by a jump of the energy and entropy can exist even in the weak-coupling limit --- it can be kinematical, rather than dynamical --- and the intermediate phase resembling the small black hole can exist  in general.
This intermediate phase lies between the Hagedorn transition and the Gross-Witten-Wadia (GWW) transition~\cite{Aharony:2003sx}.
Those transitions are characterized by the distribution of the phases of the Polyakov line, $\rho(\theta)$.
%
At the Hagedorn transition, $\rho(\theta)$ changes from uniform ($\rho(\theta)=\frac{1}{2\pi}$) to non-uniform.
At the GWW transition, a gap is formed, namely $\rho(\theta)$ becomes zero in a finite range of $\theta$ above the GWW point.
Even large-$N$ QCD could have such an intermediate phase~\cite{Schnitzer:2004qt} and it appears to be similar to the cross-over region of real QCD with $N=3$.
Namely, the GWW transition has to exist, and as we will explain further in Sec.~\ref{sec:BEC}, the GWW transition is naturally identified with the transition from the partially-deconfined phase
to the completely-deconfined phase.
At weak coupling, explicit analytic calculations~\cite{Schnitzer:2004qt,Hanada:2019kue}
show a cross-over-like behavior with a third order GWW phase transition (and hence, it is not really cross-over).
Partial deconfinement gives a precise physical interpretation to this phase~\cite{Hanada:2019kue}.

One way to approach partial deconfinement is to analyze the thermodynamics of large-$N$ gauge theories from the point of view of the microcanonical ensemble~\cite{Hanada:2018zxn,Hanada:2019czd,Hanada:2019kue}.
In the microcanonical ensemble, the energy $E$ is varied as a parameter, and the entropy $S$ is maximized at each fixed energy.
In the confining phase, $E$ and $S$ are of order $N^0$, up to the zero-point energy, while in the deconfining phase they are of order $N^2$.
This is simply due to the counting of degrees of freedom.
In QCD language, we would refer to them as \emph{hadrons/glueballs} and \emph{quarks/gluons}.
Now we consider a specific value of the energy $E=\epsilon N^2$, where $\epsilon$ is a small but order $N^0$ number, such as $0.1$ or $10^{-100}$.
This is a perfectly reasonable choice because the energy can be varied continuously.
On the other hand, a question arises: what kind of phase is realizing this specific energy?
It cannot be the confined phase, because the energy is too large and it cannot be the deconfined phase, because the energy is too small.
The answer is the partially-deconfined phase with $M\sim\sqrt{\epsilon}N$.

In many theories, including QCD, the canonical and microcanonical ensemble give the same result.
The story becomes slightly intricate for systems exhibiting a first-order phase transition in the canonical ensemble, as we will explain in Sec.~\ref{sec:remark-1st-order}.
It is worth noting that unless the volume of ordinary space is sufficiently large, the microcanonical ensemble is a physically more realistic setup than the canonical ensemble.
This is because the canonical ensemble is typically derived from the microcanonical ensemble as follows.
Firstly, let us consider an isolated system in which the energy is conserved, The microcanonical ensemble gives a reasonable statistical description of such a system.
If the space is sufficiently large, the system can be divided into a small sub-system and a large heat bath which are in thermal equilibrium.
Then the small sub-system is described by the canonical ensemble, with the temperature set by the heat bath.
By construction, this derivation of the canonical ensemble assumes sufficiently large spatial volume, and hence, is not applicable to a small volume.

Although partial deconfinement has been discovered only recently, from the discussion above it does not look exotic.
Another way to approach it is to recall the first-order transition in a locally interacting system --- i.e. the transition between the liquid and solid phases of water --- and generalizing it to a system with nonlocal interactions.
In the canonical ensemble, water exhibits a first-order phase transition at the temperature of 0 $^{\circ}$C and pressure of 1 atmosphere.
In the microcanonical ensemble, depending on the energy $E$ the amount of liquid and solid phases change, because of the latent heat at the transition temperature.
When $E$ is small/large the completely solid/liquid phase is observed, while in the intermediate range a mixture of two phases appears.
(cf. the bottom row of Fig.~\ref{fig:correspondence}.)
Equivalently, the mixture of two phases is realized {\it when the energy is not sufficiently small to be in the completely solid phase, and not sufficiently large to be in the completely liquid phase}.
The temperature remains fixed because of the short-range nature of the interactions: as long as the interaction at the interface of two phases can be ignored, the temperature cannot change.

\begin{figure}[htbp]
\centering
  \includegraphics[width=0.97\textwidth]{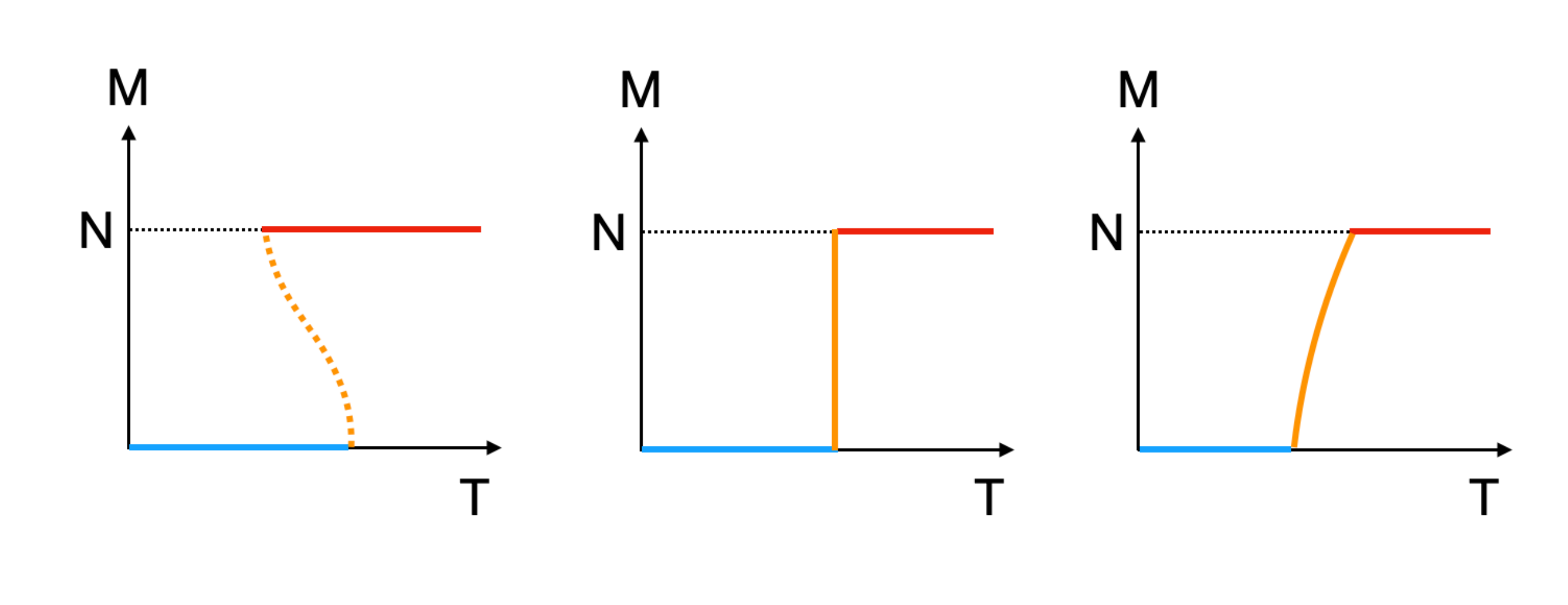}
  \caption{Three basic patterns of $T$-dependence of $M$~\cite{Hanada:2018zxn}.
The blue, orange and red lines are the completely confined, partially-deconfined and completely deconfined phases, respectively.
[Left] First-order transition with hysteresis, e.g.~strongly coupled 4d ${\cal N}=4$ super Yang-Mills theory on S$^3$,
and the Yang-Mills matrix model studied in Sec.~\ref{sec:Yang-Mills-MM}.
[Middle] First-order transition without hysteresis, e.g.~free Yang-Mills and Gaussian matrix model studied in Sec.~\ref{sec:Gaussian-matrix-model}.
[Right] Non-first-order transition, e.g.~
large-$N$ QCD~\cite{Schnitzer:2004qt}.
}\label{fig:M-vs-T-3-types}
\end{figure}

A similar mechanism can exist for the gauge theory phase transition introduced above.
In the space of color degrees of freedom two phases --- confined and deconfined --- can coexist.
However, because the interaction between the color degrees of freedom is all-to-all, the temperature can change in a nontrivial way depending on the details of the theory.
Pictorially we can visualize three possible patterns as shown in Fig.~\ref{fig:M-vs-T-3-types}.
The blue, orange, and red lines represent the completely confined phase, partially-deconfined phase (or equivalently, partially-confined phase) and completely deconfined phase.
These three phases are the counterparts, in color space, of the solid, mixture and liquid phases, respectively.
Let us analyze each pattern individually:
\begin{itemize}
\item The center panel in Fig.~\ref{fig:M-vs-T-3-types} would be the easiest one to understand.
  The Gaussian matrix model studied in Sec.~\ref{sec:Gaussian-matrix-model} belongs to this class.
  The temperature does not change, similarly to the case of the mixture of liquid water and ice.

\item The Yang-Mills matrix model discussed in Sec.~\ref{sec:Yang-Mills-MM} is similar to the left panel in Fig.~\ref{fig:M-vs-T-3-types}.
  In this case, the partially-deconfined phase has a negative specific heat.
  In the canonical ensemble such phase is not favored thermodynamically and to emphasize this feature we used a dotted line.
  Strongly coupled 4d ${\cal N}=4$ Yang-Mills and pure Yang-Mills belong to this class too.
  Depending on the geometry of the ordinary space, instability can set in even in the microcanonical ensemble (see Sec.~\ref{sec:remark-1st-order} for the details).

\end{itemize}

In the above example of liquid and solid water, the local nature of the interaction makes the notion of `separation into two phases' easily understandable, because the interaction between the liquid and solid at the interface of two phases is usually negligible. The same holds in the weak-coupling limit of large-$N$ gauge theories, and the separation to confined and deconfined sectors can be proven precisely~\cite{Hanada:2019czd,Hanada:2019kue}.
See Sec.~\ref{sec:Gaussian-matrix-model} for an example of this kind.
There, a Gaussian matrix model with gauge-singlet constraint is considered.
The dynamical degrees of freedom are harmonic oscillators,
and in the confined sector they remain in the ground state,
while in the deconfined sector they are excited.
Note that, even in the weak-coupling limit, the temperature dependence is not necessarily like in the second panel of Fig.~\ref{fig:M-vs-T-3-types}.
Indeed, large-$N$ QCD with $N_f$ flavor, with $\frac{N_f}{N}$ fixed, shows a nontrivial temperature dependence like the third panel of Fig.~\ref{fig:M-vs-T-3-types},
due to the interplay between the number of deconfined color degrees of freedom
and the number of deconfined flavor degrees of freedom~\cite{Hanada:2019kue}.
With the interaction, the nature of deconfined and confined phases can be modified,
but we expect the coexistence of two phases can persist.
This point is elaborated more in Sec.~\ref{sec:BEC}, and also, numerically studied in this paper.

It can also be instructive to consider possible objections to the picture presented by partial deconfinement.
For example, it can be objected that the phase separation should only take place in ordinary space and not in the internal space of color degrees of freedom.
However, it is true that the confinement/deconfinement transition can take place even in matrix models where the ordinary space does not exist by definition.
In these cases the phase separation can happen only in the internal `space' for these theories.
Sec.~\ref{sec:remark-1st-order} contains more details about this point.

Another objection would be to have all the color degrees of freedom get mildly excited and not separate into distinct groups representing the confined and deconfined phases.
In the case of water this possibility is forbidden because of the finite latent heat.
In the case of the confinement/deconfinement transition we recall that it can take place even in the weak-coupling limit~\cite{Sundborg:1999ue,Aharony:2003sx}.
The simplest example is the gauged Gaussian matrix model where the color degrees of freedom are just quantum harmonic oscillators with quantized excitation levels.
They cannot be `mildly' excited precisely because of quantization: the discreteness of the energy spectrum plays the same role as the latent heat.

A third objection can be raised about the SU($M$) block structure of the deconfined sector of partial deconfinement.
We could say that there might be other arrangements of the excited degrees of freedom.
However, intuitively, since we are maximizing the entropy at fixed energy, it is natural to expect that the solution of such an extremization problem preserves a large symmetry.
A more precise argument can be made by using the equivalence between color confinement at large $N$ and Bose-Einstein condensation~\cite{Hanada:2020uvt},
as we will see in Sec.~\ref{sec:BEC}.
\subsection{BEC-Confinement correspondence and meaning of Polyakov line}\label{sec:BEC}
\hspace{0.51cm}
In order to understand the precise meaning of partial deconfinement,
a tight connection \cite{Hanada:2020uvt} between color confinement in large-$N$ gauge theory
and Bose-Einstein condensation (BEC) \cite{einstein1924quantentheorie} plays a crucial role.

In the analyses in the weak-coupling limit, it was found that the size of the deconfined sector $M$ can be read off from the distribution of the phases of the Polyakov line as~\cite{Hanada:2018zxn,Hanada:2019czd,Hanada:2019kue}
\begin{eqnarray}
{\rm minimum\ of\ }\rho(\theta)
=
\frac{1}{2\pi}\left(1-\frac{M}{N}\right).
\label{eq;ODLRO-vs-Polyakov}
\end{eqnarray}
The mechanism underlying this correspondence was clarified
by identifying a natural --- in retrospect, almost trivial --- connection between BEC
and confinement at large $N$~\cite{Hanada:2020uvt}.
In order to make this paper self-contained, let us repeat the argument presented in
Ref.~\cite{Hanada:2020uvt}.

Let us consider the system of $N$ indistinguishable bosons in $\mathbb{R}^d$,
trapped in the harmonic potential.
There are $N$ harmonic oscillators $\vec{x}_1,\cdots,\vec{x}_N$ consisting of $d$-components, described by the Hamiltonian
\begin{eqnarray}
H
=
\sum_{c=1}^N\left(
\frac{\vec{p}_{c}^{\,\,2}}{2m}
+
\frac{m\omega^2}{2}\vec{x}_{c}^{\,2}
\right).
\end{eqnarray}
The indistinguishability of the particles, which leads to Bose-Einstein statistics,
can be incorporated as the redundancy under the S$_N$-permutation symmetry.
Therefore, this system describes the S$_N$-gauged quantum mechanics of $N$-component vectors.
The extended Hilbert space containing non-gauge-invariant states is spanned by the standard Fock states,
\begin{eqnarray}
|\vec{n}_1,\vec{n}_2,\cdots,\vec{n}_N\rangle
\equiv
\prod_{i=1}^d
\frac{\hat{a}_{i1}^{\dagger n_{i1}}}{\sqrt{n_{i1}!}}
\frac{\hat{a}_{i2}^{\dagger n_{i2}}}{\sqrt{n_{i2}!}}
\cdots
\frac{\hat{a}_{iN}^{\dagger n_{iN}}}{\sqrt{n_{iN}!}}
|0\rangle,
\label{harmonic_oscillator_basis}
\end{eqnarray}
and the physical, gauge-singlet states are defined as the S$_N$-permutation-invariant states.
The partition function is given by \cite{feynman_superfluidity1}
\begin{eqnarray}
Z
=
\sum_{g\in G}{\rm Tr}\left(
\hat{g} e^{-\beta\hat{H}}
\right),
\label{eq:partition-function-identical-bosons}
\end{eqnarray}
where $G={\rm S}_N$ is the gauge group and $\hat{g}$ is a unitary operator acting on the Hilbert space
corresponding to the group element $g\in G$.
The trace is taken over $|\vec{n}_1,\vec{n}_2,\cdots,\vec{n}_N\rangle$ defined by eq.~\eqref{harmonic_oscillator_basis}.
The insertion of $\hat{g}$ leads to the projection to the gauge-invariant Hilbert space, after the sum over the gauge group $G$ is taken.
More explicitly,
\begin{eqnarray}
Z
&=&
\sum_{g\in G}
\sum_{\vec{n}_1,\cdots,\vec{n}_N}
\langle\vec{n}_1,\cdots,\vec{n}_N|
\hat{g} e^{-\beta\hat{H}}
|\vec{n}_1,\cdots,\vec{n}_N\rangle
\nonumber\\
&=&
\sum_{\vec{n}_1,\cdots,\vec{n}_N}
e^{-\beta\left(E_{\vec{n}_1}+\cdots +E_{\vec{n}_N}\right)}
\sum_{g\in {\rm S}_N}
\langle\vec{n}_1,\cdots,\vec{n}_N|
\hat{g}
|\vec{n}_1,\cdots,\vec{n}_N\rangle
\nonumber\\
&=&
\sum_{\vec{n}_1,\cdots,\vec{n}_N}
e^{-\beta\left(E_{\vec{n}_1}+\cdots +E_{\vec{n}_N}\right)}
\sum_{g\in {\rm S}_N}
\langle\vec{n}_1,\cdots,\vec{n}_N
|\vec{n}_{g(1)},\cdots,\vec{n}_{g(N)}\rangle.
\label{eq:partition-function-identical-bosons-2}
\end{eqnarray}
For generic excited states, all $N$ particles are in different states, and only $g=\textbf{1}$ contributes.
On the other hand, for the ground state $|\vec{0},\vec{0},\cdots,\vec{0}\rangle$ all elements $g\in G$ give the same contribution, which leads to an enhancement by a factor $N!$.
Equivalently, generic states are suppressed by a factor $N!$ compared to the ground state.
As identified by Einstein, this is the cause of Bose-Einstein condensation.

The argument above shows how we can define the gauge-singlet condition explicitely. As such, we could extend that reasoning to different gauge groups, for example by replacing $G$ with a more generic gauge group such as SU($N$) and by considering a more generic field content: there is an enhancement factor (the volume of the gauge group)
associated with the ground state (the confined phase).
We have seen that the operator $\hat{g}$ implements the gauge transformation in the Hilbert space
corresponding to a group element $g\in G$. For the SU($N$) group this element $g$ is nothing but the Polyakov line.
Based on this correspondence, the permutation matrix $g$ in \eqref{eq:partition-function-identical-bosons} and \eqref{eq:partition-function-identical-bosons-2} for the system of identical bosons will define the Polyakov line in a natural way~\cite{Hanada:2020uvt}.
In order to determine the phase distribution, we look at permutation matrices which leave a typical state contributing to thermodynamics unchanged.
When the BEC is formed, long cyclic permutations exchanging the particles in BEC become dominant \cite{feynman_superfluidity1}, and then the off-diagonal long-range order (ODLRO) appears \cite{PhysRev.104.576}.
In terms of the Polyakov line, long cyclic permutations contribute to the constant offset in the distribution function $\rho(\theta)$, and the relation \eqref{eq;ODLRO-vs-Polyakov} follows~\cite{Hanada:2020uvt}.
When BEC is formed, the constant offset becomes nonzero; this is the analogue of the Gross-Witten-Wadia phase transition associated with deconfinement~\cite{Aharony:2003sx}
(see also Sec.~\ref{sec:Gaussian-matrix-model}).
The same logic applies to generic gauge theories as well.
This is the reason why the constant distribution of the Polyakov line phases is a good indicator of confinement.
Note that we can go beyond conventional wisdom: even when the phase distribution is not uniform, we can separate the constant offset and non-uniform part. Namely,
\begin{eqnarray}
\rho(\theta)
=
C
+
\tilde{\rho}(\theta),
\end{eqnarray}
where $C\ge 0$ is the minimum of $\rho(\theta)$ and $\tilde{\rho}(\theta)$ is a non-constant distribution
whose minimum is zero.
This $C$ is related to $M$ as
\begin{eqnarray}
C
=
\frac{1}{2\pi}\left(
1-\frac{M}{N}
\right).
\end{eqnarray}
Hence we can fix the ordering of $\theta_i$'s such that $\theta_1,\cdots,\theta_M$ give
the nonuniform part $\tilde{\rho}(\theta)$, while $\theta_{M+1},\cdots,\theta_N$ lead to the constant part $C$.
This separation should correspond to the splitting of the matrix degrees of freedom to the confined and deconfined sectors, and if we see other fields, the phase separation pictorially shown in Fig.~\ref{fig:matrix-partial-deconfinement} has to hold.

As proposed by Fritz London~\cite{1938Natur.141..643L},
superfluid helium is understood as BEC.
Although BEC was first discovered at vanishing coupling,
it can actually take place even at strong coupling.
This historically famous fact, together with intriguing similarities of various quantum field theories at weak and strong coupling (see e.g.~\cite{Aharony:2005bq,Aharony:2004ig}),
motivate our optimism that partial deconfinement can be valid even at strong coupling.
We will give numerical evidence supporting this optimism later in this paper.

\subsection{Remarks on negative specific heat}\label{sec:remark-1st-order}
\hspace{0.51cm}

Here we summarize some remarks about the partially-deconfined phase in the case where it has negative specific heat.
We refer again to the leftmost panel of Fig.~\ref{fig:M-vs-T-3-types}.
If the volume of the ordinary space is large, the phase with a negative specific heat is not stable.
Any small perturbation can trigger a decay to the co-existence of a  completely confined and a completely deconfined phases.
This is not necessarily the case if the volume is small and finite.
In the case of matrix models, such instability cannot exist by definition, because there is no ordinary space by definition.
Moreover, 4d ${\cal N}=4$ super Yang-Mills on S$^3$ does not have such instability.
The partially-deconfined phase in 4d ${\cal N}=4$ super Yang-Mills on S$^3$ is dual to the small black hole phase, which does not have such instability.\footnote{
There is a subtlety regarding this point, near the phase transition associated with the localization on the S$^5$; see \ref{sec:gravity-dual}. }
It can also be understood via a simple dimensional analysis as follows.
For the coexistence of two phases in the ordinary space to appear, the radius of S$^3$, which we call $r$, has to be sufficiently larger than the typical length scale of the system, which is the inverse of the temperature of the partially-deconfined phase, $\beta$.
However $\beta$ and $r$ are of the same order, as explained in Ref.~\cite{Witten:1998zw}.

When the specific heat is negative, the partially-deconfined phase sits at the maximum of the free energy in the canonical ensemble~\cite{Hanada:2018zxn}.
The completely confined and completely deconfined phases are the minima of the free energy.
The difference of free energy between the minima and the maximum between them is of order $N^2$ and the tunneling is parametrically suppressed at large $N$.
In this way, the local minima is completely stabilized at large $N$, even when it is not the global minimum.
This is very different from the metastable states in the locally interacting systems, such as the supercooled water:
even in the thermodynamic limit (large volume) already a small perturbation can destabilize the supercooled water.

\section{Partial deconfinement: the Gaussian matrix model}\label{sec:Gaussian-matrix-model}
\hspace{0.51cm}
Let us consider the gauged Gaussian matrix model. The action is given by Eq.~\eqref{eq:action_Gaussian}.
This model is analytically solvable~\cite{Aharony:2003sx,Sundborg:1999ue} and partial deconfinement has been introduced for the phase transition in Refs.~\cite{Hanada:2019czd,Berenstein:2018lrm}.
We start from this solvable case in order to analyze how partial deconfinement manifests itself in the path integral formalism.

In the previous section we already mentioned that the Gaussian matrix model has a confinement/deconfinement phase transition which is of first order without hysteresis in the canonical ensemble (the center panel of Fig.~\ref{fig:M-vs-T-3-types}).
The critical temperature is $T=T_c=\frac{1}{\log d}$.
In the canonical ensemble, at $T=T_c$, the energy $E$ and entropy $S$ jump from order $N^0$ to order $N^2$, while the Polyakov loop $P=\frac{1}{N}{\rm Tr}{\cal P}e^{i\int_0^\beta dt A_t}$ jumps from 0 to $\frac{1}{2}$ (see Fig.~2 of Ref.~\cite{Hanada:2019czd}.)
We also fix the center symmetry ambiguity in the phase of the Polyakov loop by setting $P=|P|$ for the remainder of this paper.

As functions of the Polyakov loop, the energy and entropy are expressed as\footnote{
The derivation of Eqs.~\eqref{eq:GMM-E}, \eqref{eq:GMM-S} and \eqref{eq:GMM-P-dist}
can be found in Sec.~3 and Appendix~A.1 of Ref.~\cite{Hanada:2019czd}.
}
\begin{eqnarray}
\left.E\right|_{T=T_c}
\equiv
\left.
\frac{N}{\beta}\int dt \sum_I{\rm Tr}X_I^2
\right|_{T=T_c}
=
\frac{d}{2}N^2
+
N^2P^2
\label{eq:GMM-E}
\end{eqnarray}
and
\begin{eqnarray}
\left.S\right|_{T=T_c}
=
\log d \cdot N^2P^2.
\label{eq:GMM-S}
\end{eqnarray}
As mentioned above, at the critical temperature $P$ can take any value from $0$ to $\frac{1}{2}$. Therefore, up to the zero-point energy $\frac{d}{2}N^2$,
the energy $E$ and entropy $S$ changes from $N^0$ to $N^2$, as $P$ changes from $0$ to $\frac{1}{2}$.
The distribution of the phases of the Polyakov line is
\begin{eqnarray}
\left.
\rho^{\rm(P)}(\theta)
\right|_{T=T_c}
=
\frac{1}{2\pi}\left(
1+2P\cos\theta
\right).
\label{eq:GMM-P-dist}
\end{eqnarray}
The free energy $F=E-TS$ does not depend on $P$:
\begin{eqnarray}
\left.F\right|_{T=T_c}
=
\frac{d}{2}N^2.
\end{eqnarray}
Hence all values of $P$ contribute equally to the canonical partition function.
Essentially the same results hold for various weakly-coupled theories~\cite{Aharony:2003sx,Sundborg:1999ue}.

So far we have introduced important relations for the Gaussian matrix model quantities and we want to relate them to partial deconfinement, following the findings in Ref.~\cite{Hanada:2019czd}.
At $T=T_c$, the size of the deconfined sector $M$ jumps from $M=0$ to $M=N$.
This size can be read off from the distribution of the phases of the Polyakov line, as explained in Sec.~\ref{sec:BEC}.
In the case of the Gaussian matrix model, this relation simplifies to \cite{Hanada:2019czd}
\begin{eqnarray}
P
=
\frac{M}{2N}.
\label{eq:P-vs-M-in-GMM}
\end{eqnarray}
This equation connects an observable of the system ($P$) to the \emph{amount} of deconfined degrees of freedom $M/N$.
Below, we will explain how the separation to two phases (confined and deconfined sectors) can be seen by using this $M$.
In Sec.~\ref{sec:GaussianMM-Hamiltonian}, the explicit separation in terms of the quantum states is explained as well.

It follows that the distribution of the phases of the Polyakov loop in Eq.~\eqref{eq:GMM-P-dist} can be written as
\begin{eqnarray}
\left.
\rho^{\rm(P)}(\theta)
\right|_{T=T_c}
=
\frac{1}{2\pi}\left(
1+\frac{M}{N}\cos\theta
\right).
\label{eq:GMM-P-dist-vs-M}
\end{eqnarray}
Similarly, for the energy $E$ and the entropy $S$, we can write
\begin{eqnarray}
\left.E\right|_{T=T_c}
\equiv
\frac{N}{\beta}\int dt \sum_I{\rm Tr}X_I^2
=
\frac{d}{2}N^2
+
\frac{M^2}{4},
\label{eq:GMM-E-vs-M}
\end{eqnarray}
\begin{eqnarray}
\left.S\right|_{T=T_c}
=
\frac{\log d}{4}M^2.
\label{eq:GMM-S-vs-M}
\end{eqnarray}
Note that these relations are valid only at the phase transition temperature $T=T_c$ where partial deconfinement takes place.
Later in this paper, when we perform the numerical analysis at large but finite $N$, we determine $M$ from Eq.~\eqref{eq:P-vs-M-in-GMM} by imposing $2NP \in [M-\frac{1}{2},M+\frac{1}{2}]$.

While the energy and the entropy have a discontinuity at $T=T_c$ in the canonical ensemble, that is not the case for the microcanonical ensemble where the entropy is maximized for each fixed system energy.
Therefore, it follows that $M$ can be defined by the energy itself through Eq.~\eqref{eq:GMM-E-vs-M}.
Moreover, in the microcanonical ensemble there are two phase transitions: one from $M=0$ to $M>0$ and one from $M<N$ to $M=N$.
They are called the Hagedorn transition and Gross-Witten-Wadia (GWW) transition, respectively.

In order to understand the nature of the states in the microcanonical ensemble,
let us rewrite Eqs.~\eqref{eq:GMM-P-dist-vs-M}, \eqref{eq:GMM-E-vs-M} and \eqref{eq:GMM-S-vs-M} as follows:
\begin{eqnarray}
\left.
\rho^{\rm(P)}(\theta)
\right|_{T=T_c}
=
\left(
1-\frac{M}{N}
\right)
\cdot
\frac{1}{2\pi}
+
\frac{M}{N}\cdot\frac{1+\cos\theta}{2\pi},
\label{eq:Pol-Gaussian}
\end{eqnarray}
\begin{eqnarray}
\left.E\right|_{T=T_c}
=
\frac{d}{2}(N^2-M^2)
+
\left(
\frac{d}{2}+\frac{1}{4}
\right)
M^2,
\label{eq:energy-Gaussian}
\end{eqnarray}
\begin{eqnarray}
\left.S\right|_{T=T_c}
=
0\cdot(N^2-M^2)
+
\frac{\log d}{4}M^2.
\label{eq:entropy-Gaussian}
\end{eqnarray}
This way we can separate clearly two different contributions that get summed.
For each equation, the first term of the sum is interpreted as the contribution of the ground state, while the second term is just the value of each observable for an SU($M$) theory at the GWW-transition point (only $M$ degrees of freedom can be excited).
Partial deconfinement shown in Fig.~\ref{fig:matrix-partial-deconfinement} naturally explains this $M$-dependence.

One objection to this idea would be that Fig.~\ref{fig:matrix-partial-deconfinement} does not look gauge invariant.
Hence, in order to prove partial deconfinement in a gauge-invariant manner,
we first show in Sec.~\ref{sec:GaussianMM-Hamiltonian} the explicit construction of the gauge-singlet states in the Hilbert space from the Hamiltonian formalism.
Then, in Sec.~\ref{sec:GaussianMM-PathIntegral}, we consider the path integral formalism, which is used in the lattice Monte Carlo simulation and we discuss the properties of the master field.

\subsection{The Hamiltonian formalism}\label{sec:GaussianMM-Hamiltonian}
\hspace{0.51cm}
Following Ref.~\cite{Hanada:2019czd},
in the large-$N$ limit, 
we explicitly construct the states governing thermodynamics in the gauged Gaussian matrix model.
The Hamiltonian is
\begin{eqnarray}
\hat{H}
=
\frac{1}{2}{\rm Tr}\left(
\hat{P}_I^2+\hat{X}_I^2
\right).
\end{eqnarray}
The creation and annihilation operators are defined as
$\hat{A}_{I}^\dagger=\frac{1}{\sqrt{2}}\left(\hat{X}_I-i\hat{P}_I\right)$
and $\hat{A}_{I}=\frac{1}{\sqrt{2}}\left(\hat{X}_I+i\hat{P}_I\right)$.
The ground state is the Fock vacuum $|0\rangle$ which is annihilated by all annihilation operators:
\begin{eqnarray}
\hat{A}_{I}|0\rangle
=
0.
\end{eqnarray}
The physical states have to be gauge singlets, e.~g.
\begin{eqnarray}
{\rm Tr}\left(
\hat{A}^\dagger_{I}\hat{A}^\dagger_{J}\hat{A}^\dagger_{K}\cdots
\right)|0\rangle
=
\sum_{i,j,k,l\cdots=1}^N\left(
\hat{A}^\dagger_{I,ij}\hat{A}^\dagger_{J,jk}\hat{A}^\dagger_{K,kl}\cdots
\right)|0\rangle
\end{eqnarray}
Let $\hat{A}^{\dagger\prime}_I$ be the truncation of $\hat{A}_I^\dagger$ to the SU($M$)-part.
We can construct the states which are SU($M$)-invariant but not SU($N$)-invariant as
\begin{eqnarray}
{\rm Tr}\left(
\hat{A}^{\dagger\prime}_{I}\hat{A}^{\dagger\prime}_{J}\hat{A}^{\dagger\prime}_{K}\cdots
\right)|0\rangle
=
\sum_{i,j,k,l\cdots=1}^M\left(
\hat{A}^\dagger_{I,ij}\hat{A}^\dagger_{J,jk}\hat{A}^\dagger_{K,kl}\cdots
\right)|0\rangle.
\label{eq:state-SU(M)}
\end{eqnarray}
Note that the indices in the sum run from 1 to $M$ ({\it not} $N$).
Such states are essentially the same as the states in the SU($M$) theory.
By collecting such states with the energy given by Eq.~\eqref{eq:energy-Gaussian}, we can explain the entropy in Eq.~\eqref{eq:entropy-Gaussian} and the distribution of the phases of the Polyakov loop of Eq.~\eqref{eq:Pol-Gaussian}.
Note that the states above are not invariant under the full SU($N$) symmetry.
In order to obtain the SU($N$)-invariant states, we consider all possible embeddings of SU($M$) into SU($N$) and take a linear combination.
Namely, we consider
\begin{eqnarray}
\frac{1}{\sqrt{\cal N}}\int_{{\rm SU}(N)} dU\ {\cal U}\left(|E; {\rm SU}(M)\rangle\right),
\label{eq:SU(N)-symmetrized}
\end{eqnarray}
where $|E; {\rm SU}(M)\rangle$ is SU($M$)-invariant but not SU($N$)-invariant, ${\cal U}$ represents gauge transformations,
${\cal N}$ is the normalization factor, and the integral is taken over all SU($N$) gauge transformations.
Such SU($N$)-symmetrized states dominate the thermodynamics.

At large $N$, the gauge-invariant, SU($N$)-symmetrized state in Eq.~\eqref{eq:SU(N)-symmetrized} is indistinguishable from the state with a particular embedding of SU($M$) such as Eq.~\eqref{eq:state-SU(M)} in the following sense.
Let $|{\rm SU}(M)\rangle_1$ be the state with a particular embedding, and $|{\rm SU}(M)\rangle_2$ be a state obtained by acting with a certain unitary transformation on $|{\rm SU}(M)\rangle_1$.
For example, $|{\rm SU}(M)\rangle_1$ has excitations only in the upper-left $M\times M$ block, while $|{\rm SU}(M)\rangle_2$ has excitations only in the lower-right $M\times M$ block.
Let $\hat{O}$ be a gauge-invariant operator which is a polynomial of $O(N^0)$ matrices.
We consider these `short' operators because they do not change the energy too much and we want to study the properties of the states with energy of order $N^2$.\footnote{
The counterpart of this in the case of a finite-$N$ theory at large volume is to consider only the operators with a compact support,
in order to make sense of the boundary conditions.
}
Then, ${}_2\langle {\rm SU}(M)|\hat{O}|{\rm SU}(M)\rangle_1=0$, because to connect $|{\rm SU}(M)\rangle_1$ to $|{\rm SU}(M)\rangle_2$
it is necessary to act with $O(N^2)$ creation and annihilation operators.
This is essentially a super-selection rule: different embeddings of SU($M$) to SU($N$) belong to different super-selection sectors.
This `superselection' can work even when the embeddings are very close.
Suppose the upper-left $M\times M$ block is deconfined in $|{\rm SU}(M)\rangle_1$,
and $|{\rm SU}(M)\rangle_2$ is obtained by permuting the $M$-th and $M+1$-th rows and columns.
Those two embeddings appear almost identical, but to connect $|{\rm SU}(M)\rangle_1$ and $|{\rm SU}(M)\rangle_2$ the length of the operator $\hat{O}$ has to be of order $N^1$.
More generally, let $V$ be a generator of ${\rm SU}(N)/{\rm SU}(M)$, whose norm $\sqrt{{\rm Tr}(VV^\dagger)}$ is small but of order $N^0$, say $0.1\times N^0$.
Then, if $|{\rm SU}(M)\rangle_2$ is obtained by acting the SU$(N)$ transformation $e^{iV}$,
then ${}_2\langle {\rm SU}(M)|\hat{O}|{\rm SU}(M)\rangle_1=0$ in the large-$N$ limit with fixed $\frac{M}{N}$.
Hence whether we use a particular embedding or a superposition of all embeddings, we get the same expectation value for $\hat{O}$.

\subsection{The path integral formalism and lattice Monte Carlo}\label{sec:GaussianMM-PathIntegral}
\hspace{0.51cm}

So far we only reviewed how partial deconfinement can be seen in terms of the states in the Hamiltonian formalism, following the results in Ref.~\cite{Hanada:2019czd}.
Next, we consider the path integral formalism which is used in the lattice Monte Carlo simulations of this paper.

\subsubsection{Ensemble properties and master field}
\hspace{0.51cm}
An important remark is that the field configurations in the path integral formalism do not have a simple connection to the quantum states in the Hilbert space, other than the fact that the expectation values of gauge-invariant observables agree.
This makes the connection with partial deconfinement a little bit more intricate, in particular when trying to analyze
the configurations obtained in lattice Monte Carlo simulations.
A typical misunderstanding would be ``the lattice configurations are the wave functions describing specific states in the Hilbert space"; the absence of such a simple connection would be illuminated by noting that lattice configurations have to be averaged in order to obtain the expectation values, unlike the wave function.

At large $N$, there is a simplification: statistical fluctuations are suppressed at leading order, and we can expect the master field~\cite{witten19801} to appear and dominate the path integral.~\footnote{For a review of the master field, the readers can refer to Ref.~\cite{Makeenko:1999hq}.}
Note that the master field is {\it not} the wave function representing the state in the Hilbert space.
Still, we can find characteristic features of the master field describing the partially-deconfined phase,
by making a `mapping' between typical states and the master field configuration (typical lattice configurations).
Our strategy is to confirm those features by lattice simulation.

In this paper, we refer to the master field as a lattice configuration in the Euclidean path integral of the theory at large $N$ which gives the correct expectation values for properly normalized quantities such as $E/N^2$ to leading order in the expansion with respect to $1/N$:
\begin{eqnarray}
\left\langle
f(A_t,X_I)
\right\rangle
=
f(A_t^{\rm (master)},X_I^{\rm (master)})
\qquad
({\rm at\ large}\ N).
\end{eqnarray}
Here $f$ can be any properly normalized gauge-invariant quantity following the 't Hooft scaling,
as long as it does not affect the dominant configuration in the path integral,
similarly to the operator $\hat{O}$ considered in Sec.~\ref{sec:GaussianMM-Hamiltonian}.
We need to understand the features of the master field describing the partially deconfined phase in the Gaussian matrix model.
Then we can start looking for the master field in nontrivial theories such as the Yang-Mills matrix model.\footnote{
A few comments regarding the master field in the completely confined and completely deconfined phases in four-dimensional pure Yang-Mills theory at $N=\infty$, in the context of lattice Monte Carlo simulations, can be found in Ref.~\cite{Lucini:2003zr}.}

Because lattice Monte Carlo simulations are based on importance sampling, the sample-by-sample fluctuations of the properly normalized quantities (e.g.~$E/N^2$, which is of order $N^0$ in the large-$N$ limit) are suppressed as $N$ becomes larger.
In the strict large-$N$ limit, configurations appearing in lattice simulations can be identified with the master field.
However, in actual simulations, we can only study large but finite $N$ values.
To learn about the master field in lattice simulations, we simply study the features of configurations sampled by the Markov Chain Monte Carlo (MCMC) algorithm, at sufficiently large $N$, identifying them with master fields.

In the standard lattice Monte Carlo simulations, the canonical ensemble is obtained.
The canonical partition function can be obtained from the microcanonical ensemble as
\begin{eqnarray}
Z(\beta)
=
\int dE \Omega(E)e^{-\beta E}
=
\int dE e^{-\beta (E-TS(E))},
\end{eqnarray}
where $\Omega(E)=e^{S(E)}$ is the density of states at energy $E$.
Near the first-order transition, as a function of $E$, `free energy' $E-TS(E)$ can have multiple saddles.
Those saddles correspond to the maxima of the entropy at the energy $E$, where the microcanonical temperature $\left(\frac{dS}{dE}\right)^{-1}$ equals the canonical temperature $T$.\footnote{
The saddle-point condition $\frac{d}{dE}(E-TS)=0$ leads to $1-T\cdot\frac{dS}{dE}=0$,
which is equivalent to $\left(\frac{dS}{dE}\right)^{-1}=T$.}
We assume that the microcanonical ensemble can be represented by one master field
at each energy $E$, and expect that each saddle point has a corresponding master field.
In the case of the Gaussian matrix model at $T=T_c$, any $M$ between $0$ and $N$ minimizes the free energy, and hence, we need to treat all values of $M$ separately.
We expect there is a master field for each $M$, and we identify them as the dominant configurations for a given $(T,M)$ pair.
This can be found using numerical lattice simulations.

The master field has an ambiguity due to gauge redundancy.
In order to eliminate the redundancy, we perform Monte Carlo simulations in the static diagonal gauge.
In this gauge, the gauge symmetry is fixed up to S$_N$ permutations.
The gauge field takes the form
\begin{eqnarray}
A_t = {\rm diag}\left(\frac{\theta_1}{\beta},\cdots,\frac{\theta_N}{\beta}\right),
\end{eqnarray}
where $\theta_1,\cdots,\theta_N$ are independent of $t$, and $\theta_i \in [-\pi,+\pi)$.
By using them, the Polyakov loop $P$ is expressed as $P=\frac{1}{N}\sum_{j=1}^N e^{i\theta_j}$.

The Polyakov loop phases can be divided into two groups:
\begin{enumerate}
	\item $M$ of them (we can take them to be $\theta_1,\cdots,\theta_M$ without loss of generality) distributed following the density $\frac{1+\cos\theta}{2\pi}$;
	\item $N-M$ of them ($\theta_{(N-M)},\cdots,\theta_N$) distributed following the density $\frac{1}{2\pi}$.
\end{enumerate}
In terms of the Hilbert space, this corresponds to the separation to the deconfined block and the confined block pictorially shown in Fig.~\ref{fig:matrix-partial-deconfinement}~\cite{Hanada:2020uvt,Hanada:2018zxn}.
This subdivision is fixing the residual S$_N$ permutation symmetry further to S$_M\times$S$_{N-M}$, where rearrangements inside the two separate groups are indistinguishable for gauge-invariant properties.\footnote{
In fact, such separation is not unique; 
there is residual symmetry under the exchange of $\theta$'s with the same value in the confined and deconfined sectors. 
This is not a bug, this is a feature. 
Our numerical results are consistent with the separation, including the consequence of this ambiguity. 
See Sec.~\ref{sec:scalar-Pol-correlation}, especially the description after Eq.~\eqref{eq:ev-Ki}, for details. 
}

We note that in terms of the Euclidean path integral there is still a small residual symmetry.
Namely, the same value of $\theta$ can appear in both sectors,
\footnote{
Strictly speaking, at finite $N$, because of the Faddeev-Popov term associated with the gauge fixing $S_{\rm FP}=-\sum_{i<j}\log\left|\sin^2\left(\frac{\theta_i-\theta_j}{2}\right)\right|$, $\theta$'s cannot exactly coincide. However in the large-$N$ limit neighboring $\theta$'s can come infinitesimally close.
}
and the permutation acting on them
\footnote{
When we permute $\theta_i$ and $\theta_j$, we exchange the $i$-th and $j$-th rows and columns of the scalars $X_I$ as well.
} does not change the distribution of the Polyakov line phases.
As we will see below, this is a feature rather than a bug.

In the rest of this section, we will discuss a few properties of the master field which are related to partial deconfinement.
\footnote{In principle, it should be possible to learn more detailed of the master field. See for example Ref.~\cite{Gopakumar:1994iq}.}
Having the application to the Yang-Mills matrix model in mind, we will demonstrate that such properties are visible in lattice configurations.

\subsubsection{Distribution of $X_{I,ij}$}
\hspace{0.51cm}

As a simple characterization of the master field, let us consider the distribution of $\sqrt{N}X_{I,jj}(t)$, $\sqrt{2N}{\rm Re}X_{I,jk}(t)$ and $\sqrt{2N}{\rm Im}X_{I,jk}(t)$.
These are the diagonal and off-diagonal elements of all the scalar hermitean matrices and represent a standard \emph{lattice field configuration}.
We collectively denote them as a random variable `$x$' with distribution $\rho^{\rm(X)}(x)$.
At $T=T_c$, we want to identify the contributions to this distribution coming from the confined and deconfined sectors.
We denote them by $\rho^{\rm(X)}_{\rm con}(x)$ and $\rho^{\rm(X)}_{\rm dec}(x)$, respectively.
In fact, we expect a very specific form,
\begin{eqnarray}
\rho^{\rm(X)}(x)
=
\left(1-\left(\frac{M}{N}\right)^2\right)
\cdot
\rho^{\rm(X)}_{\rm con}(x)
+
\left(\frac{M}{N}\right)^2
\cdot
\rho^{\rm(X)}_{\rm dec}(x).
\label{eq:separation_rho_scalar}
\end{eqnarray}
where $M$ is related to $P$ by Eq.~\eqref{eq:P-vs-M-in-GMM}, i.e.~$M=2PN$.
From the point of view of partial deconfinement, this relation can readily be understood: the confined and deconfined sectors coexist in the space of color degrees of freedom.
On the other hand, without introducing partial deconfinement, this is extremely nontrivial.
In order to determine this distribution from lattice configurations at fixed $M$, we can obtain many samples of $x$ and plot their histogram (for example, we can collect $dLN^2=49152$ samples from a $d=2$, $N=32$, $L=24$ configuration, where $L$ is the number of lattice points).

The distribution $\rho^{\rm(X)}(x)$ has some interesting properties.
The part of the action describing the scalar is
\begin{eqnarray}
\lefteqn{
N\int_0^\beta dt \sum_I\sum_{j,k}
\frac{1}{2}
\left(
\left|\partial_tX_{Ijk}-\frac{i(\theta_j-\theta_k)}{\beta}X_{Ijk}\right|^2
+
|X_{Ijk}|^2
\right)}\nonumber\\
&=&
N\int_0^\beta dt \sum_I\sum_{j,k}
\frac{1}{2}
\left\{
\left|\partial_t({\rm Re}X_{Ijk})\right|^2
+
\left(
1+\frac{(\theta_j-\theta_k)^2}{\beta^2}
\right)
|{\rm Re}X_{Ijk}|^2
\right.
\nonumber\\
& &
\qquad\qquad\qquad\qquad\qquad
\left.
+\left|\partial_t({\rm Im}X_{Ijk})\right|^2
+
\left(
1+\frac{(\theta_j-\theta_k)^2}{\beta^2}
\right)
|{\rm Im}X_{Ijk}|^2
\right\}.
\end{eqnarray}
From this expression, we can see that the distribution of $\sqrt{2N}{\rm Re}X_{I,jk}(t)$ and $\sqrt{2N}{\rm Im}X_{I,jk}(t)$ depends only on $\theta_j-\theta_k$.

From here on, we focus on $T=T_c$ where partial deconfinement takes place, and we consider the SU($M$)-partially-deconfined phase.
As mentioned before, it is convenient to go to a `gauge-fixed' picture, in which the S$_N$ permutation symmetry is fixed to S$_M\times$S$_{N-M}$.
It is achieved by separating $\theta_i$'s into two groups, such that
$\theta_1,\cdots,\theta_M$ and $\theta_{M+1},\cdots,\theta_N$ are distributed as
$\frac{1+\cos\theta}{2\pi}$ and $\frac{1}{2\pi}$, respectively. 
If we take the average over $j=M+1,\cdots,N$ or $k=M+1,\cdots,N$, then the distribution of $\theta_j-\theta_k$ is uniform, just as in the completely confined phase.
In other words, the distribution of $x$ is the same as in the completely confined phase.
On the other hand, if we take the average over $j,k=1,\cdots,M$ we see the difference from the confined phase, because
$\theta_j-\theta_k$ is {\it not} uniform.
In particular, the energy $E$ defined by Eq.~\eqref{eq:GMM-E-vs-M} can be directly related to the second moment of $\rho^{\rm(X)}(x)$ by
\begin{eqnarray}
E
&=&
\left\langle
\frac{N}{\beta}\int_0^\beta dt {\rm Tr}X_I^2
\right\rangle
\nonumber\\
&=&
dN^2\int x^2\rho^{\rm(X)}(x)dx
\nonumber\\
&=&
d(N^2-M^2)\int x^2\rho_{\rm con}^{\rm(X)}(x)dx
+
dM^2\int x^2\rho_{\rm dec}^{\rm(X)}(x)dx,
\end{eqnarray}
where
\begin{eqnarray}
dM^2\int x^2\rho_{\rm dec}^{\rm(X)}(x)dx
=
\left\langle
\frac{N}{\beta}\int_0^\beta dt \sum_{1\le j,k\le M}|X_{I,jk}|^2
\right\rangle
\end{eqnarray}
corresponds to the deconfined sector where $\theta_j-\theta_k$ is not uniform,
and
\begin{eqnarray}
d(N^2-M^2)\int x^2\rho_{\rm con}^{\rm(X)}(x)dx
=
\left\langle
\frac{N}{\beta}\int_0^\beta dt \sum_{j\ge M+1\ {\rm or}\ k\ge M+1}|X_{I,jk}|^2
\right\rangle
\end{eqnarray}
corresponds to the confined sector where $\theta_j-\theta_k$ is uniform.
By referring back to Eq.~\eqref{eq:energy-Gaussian}, we can see that the variances in the equation above can be computed as
\begin{eqnarray}
\int dx\ x^2\rho^{\rm(X)}_{\rm con}(x)=\frac{1}{2}
\label{eq:variance_con_GMM}
\end{eqnarray}
and
\begin{eqnarray}
\int dx\ x^2\rho^{\rm(X)}_{\rm dec}(x)=\frac{1}{2}+\frac{1}{4d}.
\label{eq:variance_dec_GMM}
\end{eqnarray}
Such increase can be understood as the deconfinement of the upper-left $M\times M$-block in Fig.~\ref{fig:matrix-partial-deconfinement}.
Due to our specific choice of the separation of the phases to two groups,
 the upper-left $M\times M$-block is identical to the completely deconfined phase of the SU($M$) theory.
In this way, the coexistence of two phases in color space can be seen manifestly.

Note that this separation is not completely unique.
As we have mentioned before, there is a small residual symmetry, namely the S$_M\times$S$_{N-M}$ permutations which do not change the distribution of the Polyakov line phases, and leave $\rho^{\rm(X)}_{\rm con}$ and $\rho^{\rm(X)}_{\rm dec}$ unchanged.

So far we took the upper-left $M\times M$ sector to be deconfining.
However, the same distributions $\rho^{\rm(X)}_{\rm dec}$ and $\rho^{\rm(X)}_{\rm con}$ can be obtained without fixing the S$_N$ permutation symmetry, as long as the static diagonal gauge is used.
If we pick up only some specific configurations with a common value of $P$ and make a histogram, we should get Eq.~\eqref{eq:separation_rho_scalar}.
Numerically, we can determine $\rho^{\rm(X)}_{\rm con}$ and $\rho^{\rm(X)}_{\rm dec}$ by using different values of $M$.
For example we can take $a=(M/N)^2$, $b=(M'/N)^2$, and then
\begin{eqnarray}
b\rho^{\rm(X)}(x;a)
-
a\rho^{\rm(X)}(x;b)
=
(b-a)
\rho^{\rm(X)}_{\rm con}(x)
\label{how-to-determine-rho-con}
\end{eqnarray}
and
\begin{eqnarray}
(1-b)\rho^{\rm(X)}(x;a)
-
(1-a)\rho^{\rm(X)}(x;b)
=
(a-b)
\rho^{\rm(X)}_{\rm dec}(x).
\label{how-to-determine-rho-dec}
\end{eqnarray}

We can confirm this separation numerically by lattice Monte Carlo simulations.
First, from a numerical simulation at fixed $d=2$, $N$ and $L$ at $T=T_c$, we sort through the field configurations with fixed Polyakov loop value $P$ such that it reflects the value of $M$ that we are interested it.
Those field configurations become the basis for constructing $\rho^{\rm(X)}(x)$ at a pair of $M$, $M'$ values.
By using several pairs of $M$ and $M'$, we can construct $\rho^{\rm(X)}_{\rm con}(x)$ and $\rho^{\rm(X)}_{\rm dec}(x)$ solving Eqs.~\eqref{how-to-determine-rho-con}-~\eqref{how-to-determine-rho-dec}.

\begin{figure}[htbp]
\centering
	\begin{subfigure}[b]{0.5\textwidth}
		\includegraphics[width=\textwidth]{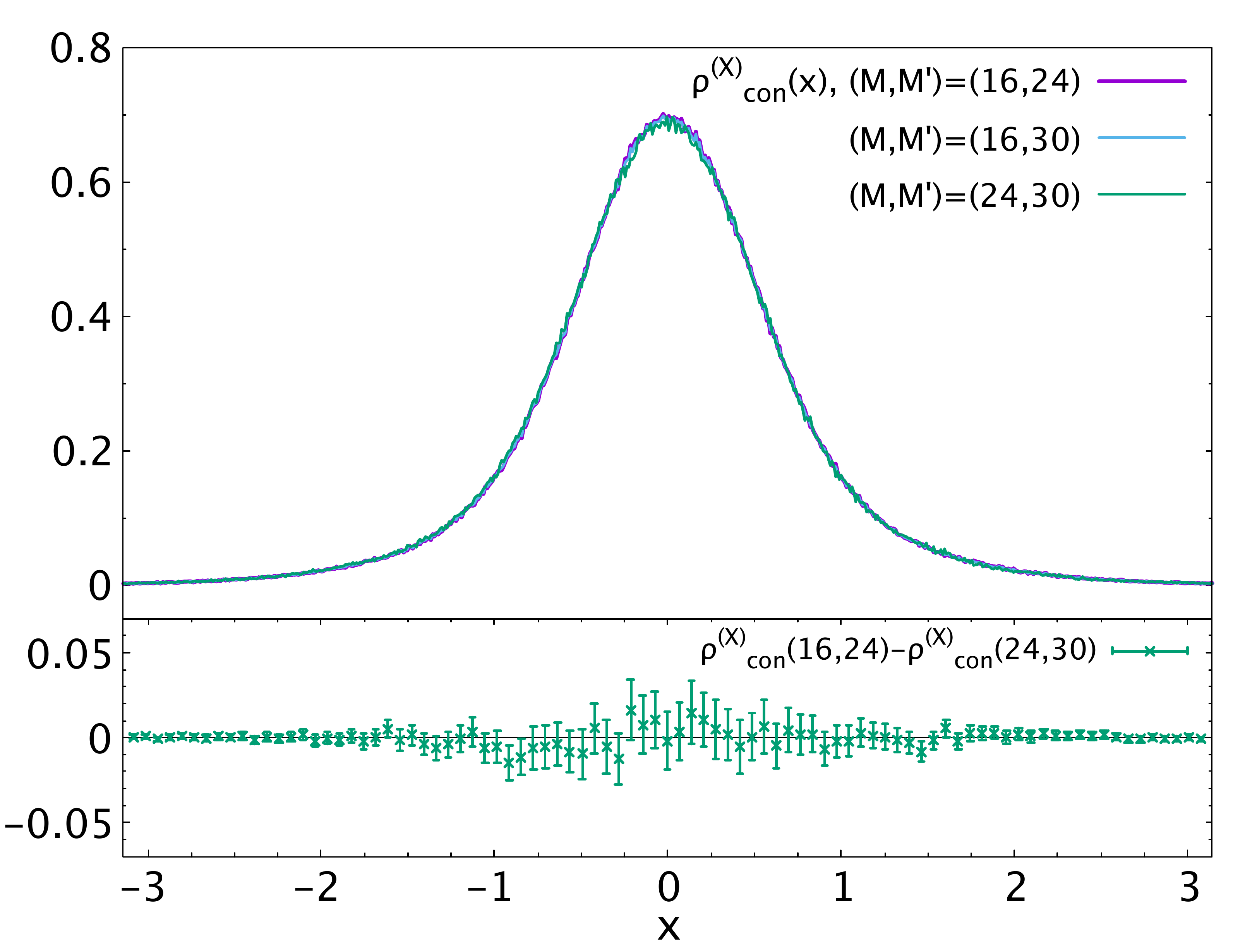}
		\caption{$\rho^{\rm(X)}_{\rm con}(x)$}
	\end{subfigure}
	\begin{subfigure}[b]{0.5\textwidth}
		\includegraphics[width=\textwidth]{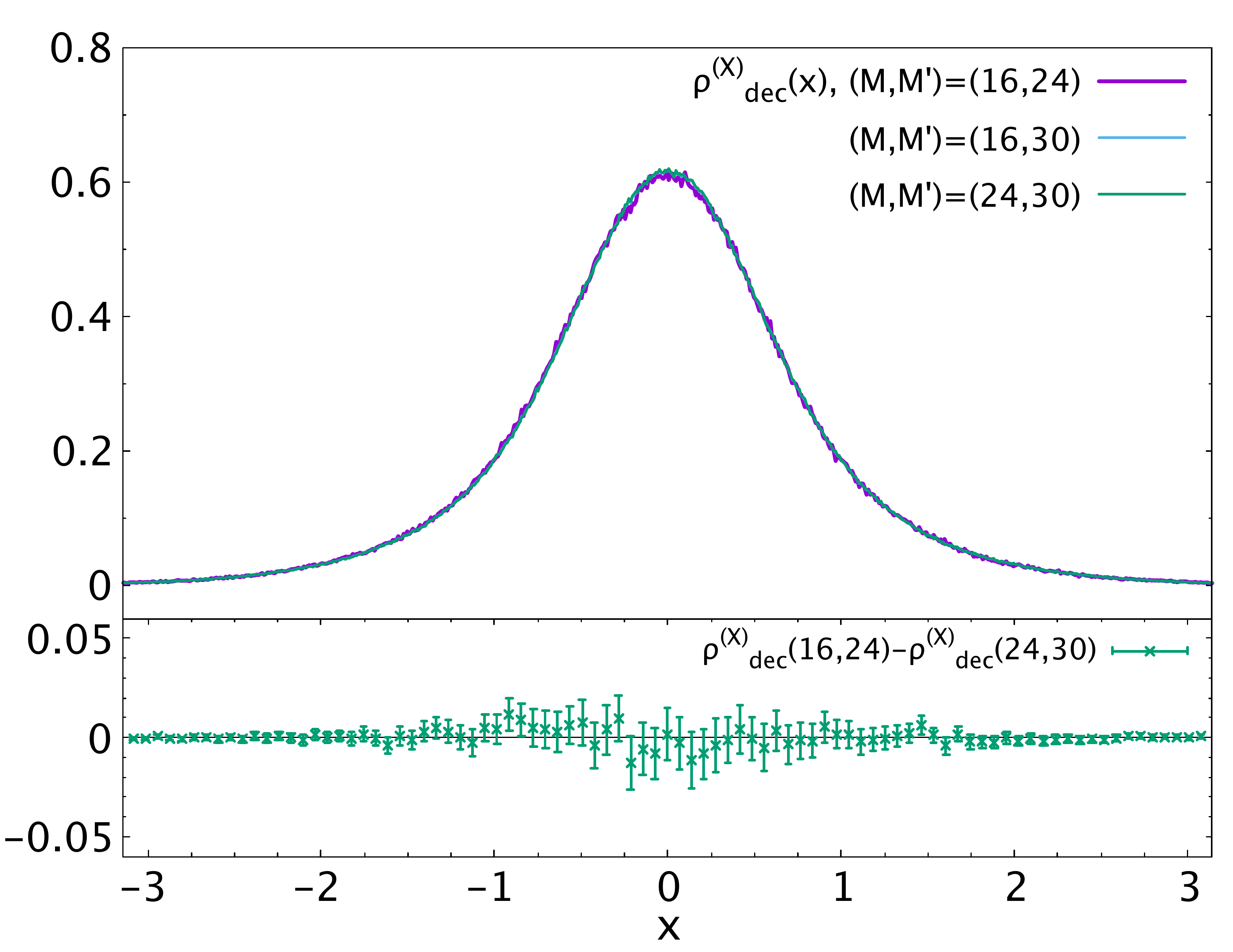}
		\caption{$\rho^{\rm(X)}_{\rm dec}(x)$}
	\end{subfigure}
	\begin{subfigure}[b]{0.5\textwidth}
		\includegraphics[width=\textwidth]{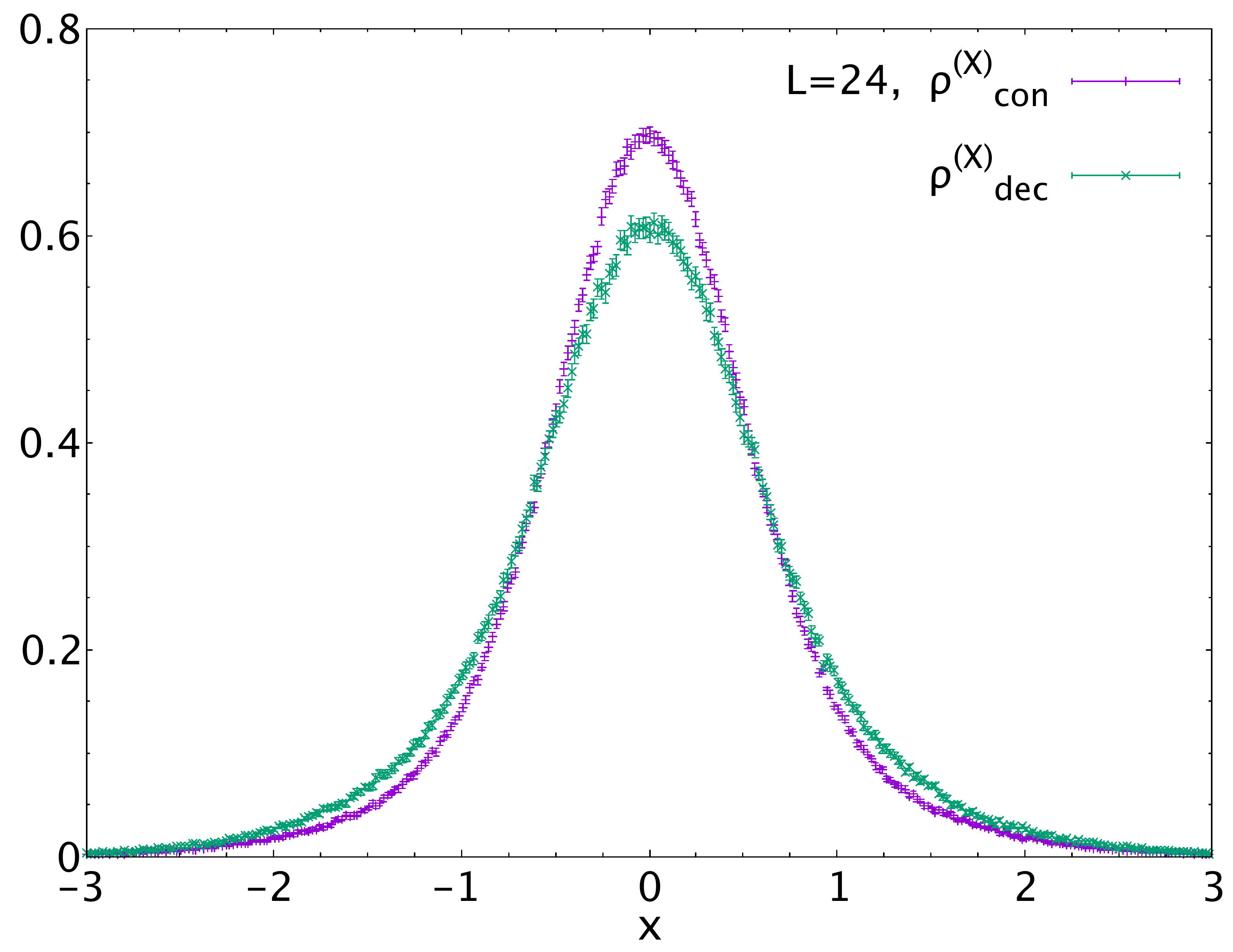}
		\caption{$(M,M')=(16,24)$}
	\end{subfigure}
	\caption{$\rho^{\rm(X)}_{\rm con}(x)$ and $\rho^{\rm(X)}_{\rm dec}(x)$ obtained from the Gaussian Matrix model, $d=2$, $N=32$, number of lattice sites $L=24$, at $T=T_c=\frac{1}{\log 2}$.
By using different combinations of $M$ and $M'$, the same distribution is obtained.
The error bars in each figure are obtained by jackknife analysis.
}\label{fig:rho_con_and_dec}
\end{figure}

\begin{figure}[htbp]
\centering
	\begin{subfigure}[b]{0.5\textwidth}
		\includegraphics[width=\textwidth]{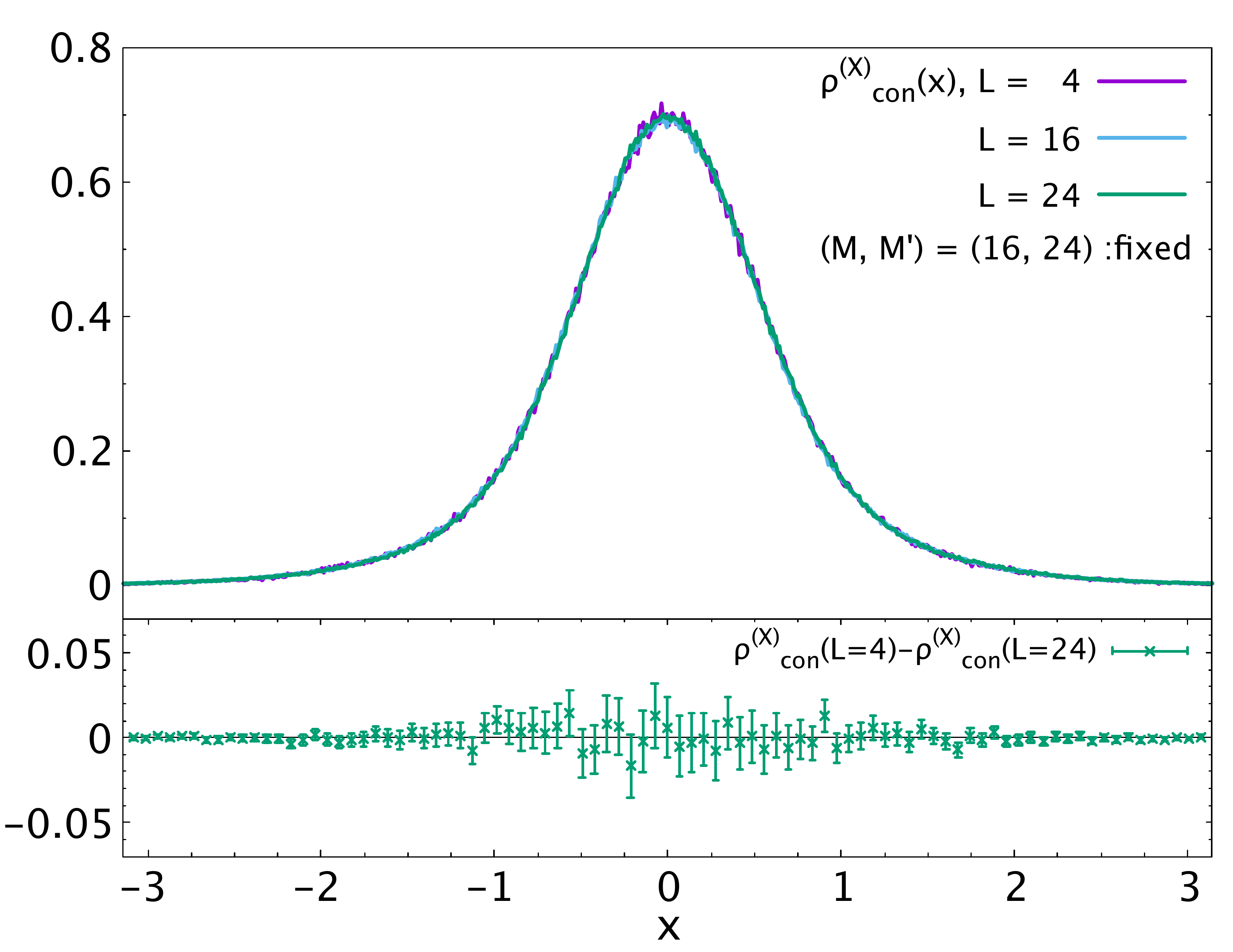}
		\caption{$\rho^{\rm(X)}_{\rm con}(x)$}
	\end{subfigure}
	\begin{subfigure}[b]{0.5\textwidth}
		\includegraphics[width=\textwidth]{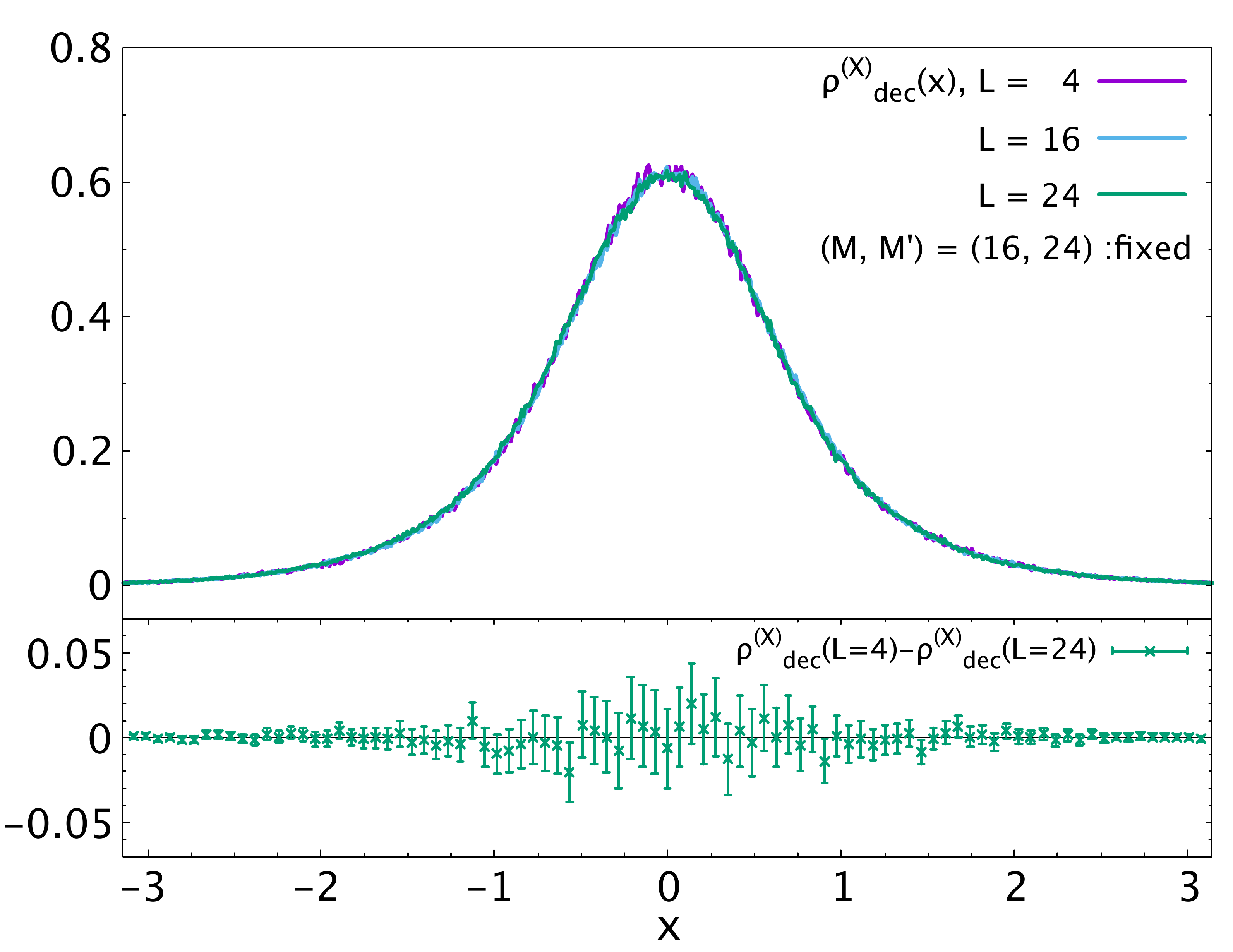}
		\caption{$\rho^{\rm(X)}_{\rm dec}(x)$}
	\end{subfigure}
	\caption{$\rho^{\rm(X)}_{\rm con}(x)$ and $\rho^{\rm(X)}_{\rm dec}(x)$ obtained from the Gaussian Matrix model, $d=2$, $N=32$, number of lattice sites $L=4, 16, 24$, at $T=T_c=\frac{1}{\log 2}$.
The finite lattice size effect is small.
}\label{fig:rho_con_and_dec_L_dependence}
\end{figure}

In Fig.~\ref{fig:rho_con_and_dec} we confirm that the extracted distributions are indistinguishable from each other, as expected from the equations above.
Moreover, the effects due to having only a finite number of lattice sites are very small: using a lattice with only 4 sites is already enough in this case, as is demonstrated in Fig.~\ref{fig:rho_con_and_dec_L_dependence}, showing good convergence to the continuum limit.
The variances in the confined and deconfined sectors calculated from the histograms of $\rho^{\rm(X)}_{\rm con}(x)$ and $\rho^{\rm(X)}_{\rm dec}(x)$ are shown in Table~\ref{tab:variance_GMM}.
The values agree with the ones at large $N$, Eqs.~\eqref{eq:variance_con_GMM}-\eqref{eq:variance_dec_GMM}, within the statistical errors and this indicates that the numerical analysis is robust.
\begin{table}[htb]
\begin{center}
\caption{The variances in the Gaussian matrix model. $d=2, N=32$ at $T=T_c = \frac{1}{\log d}$.
The number of lattice sites is $L$.
The total number of $x$ is obtained by $\#x = 32^2\times2\times L \times(\textrm{\# configs.})$.}
\label{tab:variance_GMM}
\begin{tabular}{|c|c|c|c|c|}\hline
	$L$ &	$(M,M')$		&	Confined 	&Deconfined & \# configs. of $(M,M')$\\ \hline\hline
	4	&	(16,	24)		&	0.50(3)		&	0.63(4)	&	(389, 801)\\ \hline\hline
	16	&	(16,	24)		&	0.50(4)		&	0.62(5)	&	(241, 408)	\\ \hline\hline
	24	&	(16, 24)		&	0.50(3)		&	0.63(4)	&	(165, 287)	\\ \hline
	24	&	(16,	30)		&	0.50(2)		&	0.62(1)	&	(165, 289)	\\ \hline
	24	&	(24,	30)		&	0.50(4)		&	0.62(2)	&	(287, 289)	\\ \hline
\end{tabular}
\end{center}
\end{table}

\begin{figure}[htbp]
\centering
	\includegraphics[width=0.75\textwidth]{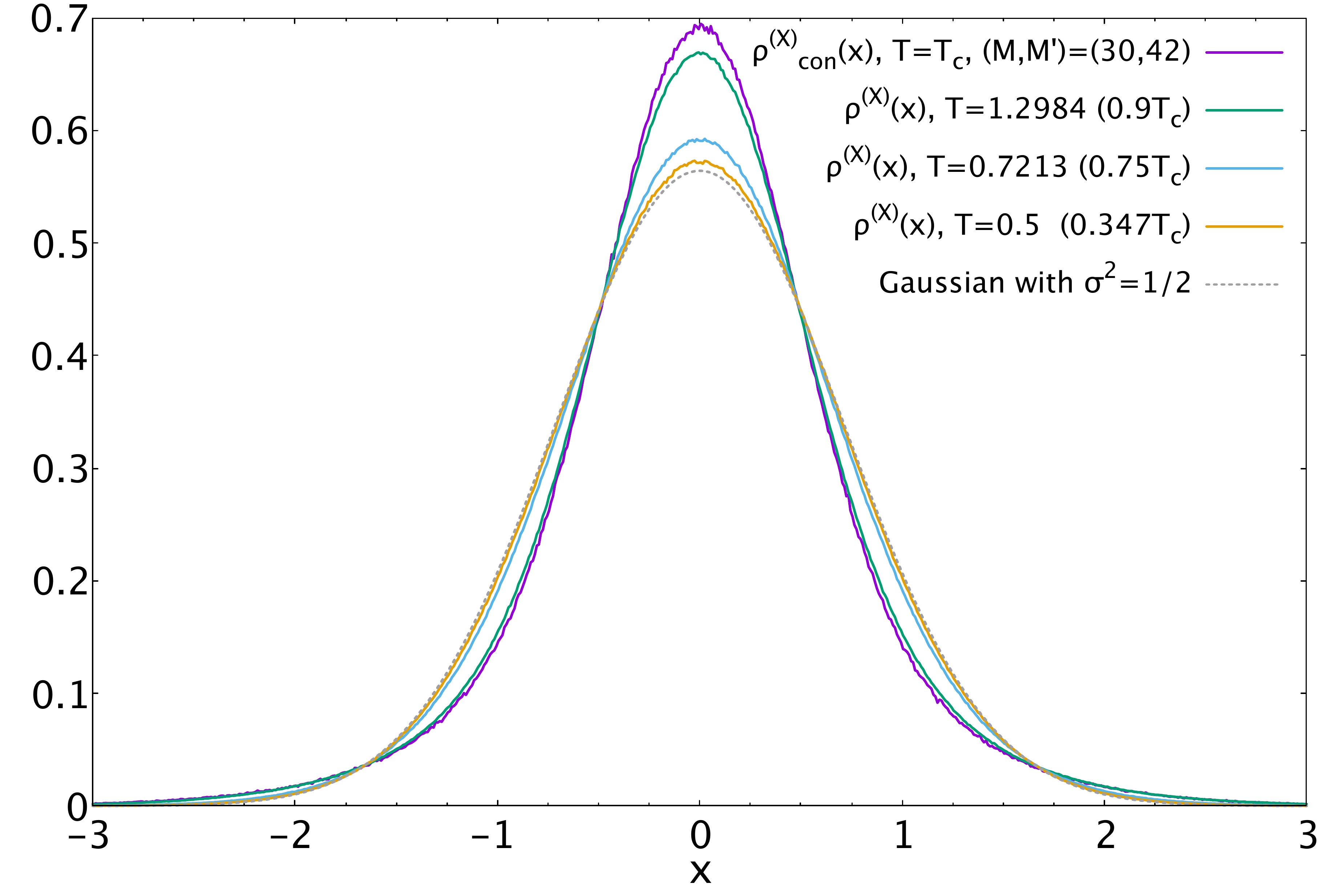}
	\caption{Gaussian matrix model, $\rho^{\rm(X)}(x)$ at low temperature, $d=2, N=48$, $L=16$. It approaches a Gaussian distribution as the temperature goes down.
}\label{fig:rho_con_low_T}
\end{figure}

Somewhat interestingly, neither $\rho^{\rm(X)}_{\rm con}(x)$ nor $\rho^{\rm(X)}_{\rm dec}(x)$ is Gaussian at $T=T_c$.
At sufficiently low temperature, deep in the completely confined phase, $\rho^{\rm(X)}_{\rm con}(x)$ (which is equivalent to $\rho^{\rm(X)}(x)$ there)
approaches a Gaussian, as shown in Fig.~\ref{fig:rho_con_low_T}.

\subsubsection{Correlation between scalars and gauge field}\label{sec:scalar-Pol-correlation}
\hspace{0.51cm}

Another relevant quantity is the correlation between the scalars $X_I$ and the Polyakov line phases $\theta_i$.
We consider the quantity $K_i$ defined by
\begin{eqnarray}
K_i\equiv\sum_{I,j}\frac{1}{\beta}\int dt|X_{I,ij}|^2.
\end{eqnarray}
For the Gaussian matrix model, at large $N$, the distribution of $|X_{I,ij}|^2$ should uniquely correspond to $\theta_i-\theta_j$,
and hence, at each value of $\frac{M}{N}$, we expect a one-to-one correspondence between $\theta_i$ and $K_i$, given by the components which are labeled by the same index $i$.

This quantity is related to the energy Eq.~\eqref{eq:GMM-E-vs-M} by $E=N\sum_{i=1}^N K_i$.
In the completely confined phase, since the entire configuration is in the ground state,
\begin{eqnarray}
\langle K_i \rangle_{\rm con}
\equiv
\frac{1}{N} \sum_{i=1}^N \left.K_i \right|_{M=0}
=
\frac {\left.E\right|_{M=0}}{N^2}
=
\frac{d}{2},
\label{eq:Ki_con}
\end{eqnarray}
and we can identify this contribution with that in the confined sector at $T=T_c$ (the light blue in Fig.~\ref{fig:matrix-partial-deconfinement}).

Next we compute the contribution from the deconfined sector, $\langle K_i \rangle_{\rm dec}$.
At the Gross-Witten-Wadia (GWW) point in SU($N$) theory, where all elements are thermally excited (the red in Fig. \ref{fig:matrix-partial-deconfinement}), namely at the point $T=T_c$ and $M=N$,
\begin{eqnarray}
\langle K_i \rangle_{{\rm GWW},N}
\equiv
\frac{1}{N} \sum_{i=1}^N \left.K_i \right|_{{\rm GWW},N}
=
\frac {\left.E\right|_{{\rm GWW},N}}{N^2}
=
\frac{d}{2}+\frac{1}{4}.
\label{eq:Ki_GWW}
\end{eqnarray}

So far in our argument, we did not fix the S$_N$ permutation symmetry.
In principle, by separating $\theta_i$'s into two groups, such that
$\theta_1,\cdots,\theta_M$ and $\theta_{M+1},\cdots,\theta_N$ are distributed as
$\frac{1+\cos\theta}{2\pi}$ and $\frac{1}{2\pi}$, respectively, we could fix the permutation symmetry
in such a way that the separation to the confined and deconfined sectors becomes manifest.
Hence each $\theta_i$ should belong to the confined or deconfined sector.
If $\theta_i$ belongs to the deconfined sector, we expect
\begin{eqnarray}
\langle K_i \rangle_{\rm dec}
=
\left(1-\frac MN \right) \langle K_i \rangle_{\rm con} + \frac MN \langle K_i \rangle_{{\rm GWW},M}
=
\frac d2+\frac M{4N}.
\label{eq:Ki_dec}
\end{eqnarray}
This is because $N-M$ and $M$ components in $|X_{ij}|^2$ behave as in the confined and deconfined phases, respectively.
For the same reason, we expect $\langle K_i \rangle_{\rm con}=\frac{d}{2}$ if $\theta_i$ belongs to the confined sector.
The average should be
\begin{eqnarray}
\langle K_i\rangle_{\rm p.d.}
=
\left(1-\frac{M}{N} \right) \langle K_i \rangle_{\rm con}
+ \frac{M}{N} \langle K_i \rangle_{\rm dec}
=
\frac d2 + \frac 14\left(\frac MN\right)^2.
\label{eq:ev-Ki}
\end{eqnarray}

When we look at the distribution of $K_i$'s, one may expect two peaks corresponding to the confined and deconfined sectors.
This naive expectation is wrong.
In fact, due to the residual symmetry under the exchange of $\theta$'s with the same value in the confined and deconfined sectors, we have to see a single peak which explains Eq.~\eqref{eq:Ki_con} and Eq.~\eqref{eq:Ki_dec} simultaneously.
This can be confirmed numerically.
Note that, due to this residual symmetry, the separation to the confined and deconfined sectors in the lattice configuration is conceptually more complicated
than in the Hamiltonian formulation.
At $\theta=\pm\pi$, all $\theta$'s belong to the confined sector,
while at $\theta\neq\pm\pi$ we can argue that each $\theta$ can belong to both sectors with
relative probability $\rho^{\rm (P)}(\pi)$ and $\rho^{\rm (P)}(\theta)-\rho^{\rm (P)}(\pi)$.

\begin{figure}[htbp]
\centering
	\begin{subfigure}[b]{0.475\textwidth}
		\includegraphics[width=\textwidth]{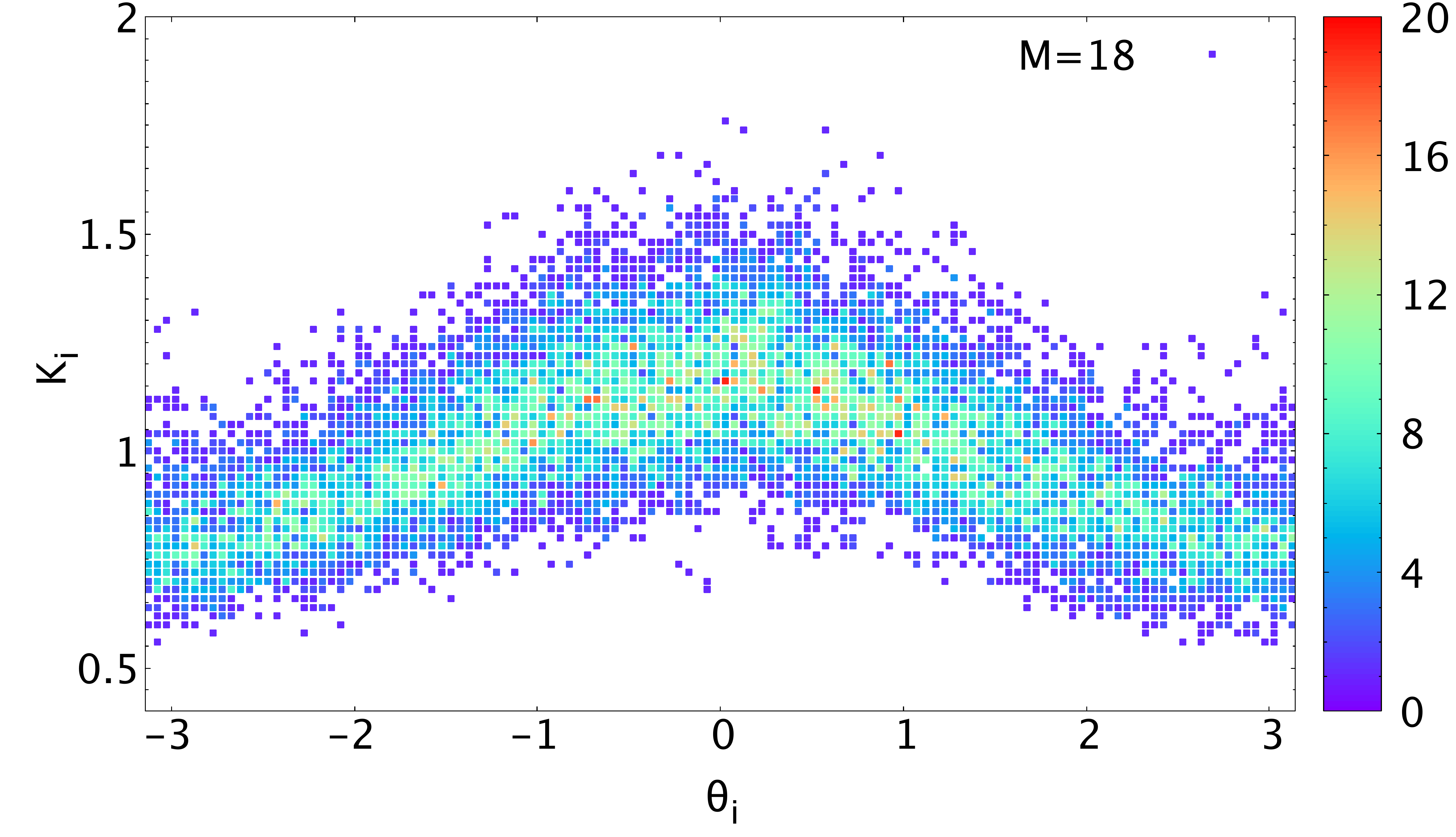}
	\end{subfigure}
	\begin{subfigure}[b]{0.475\textwidth}
		\includegraphics[width=\textwidth]{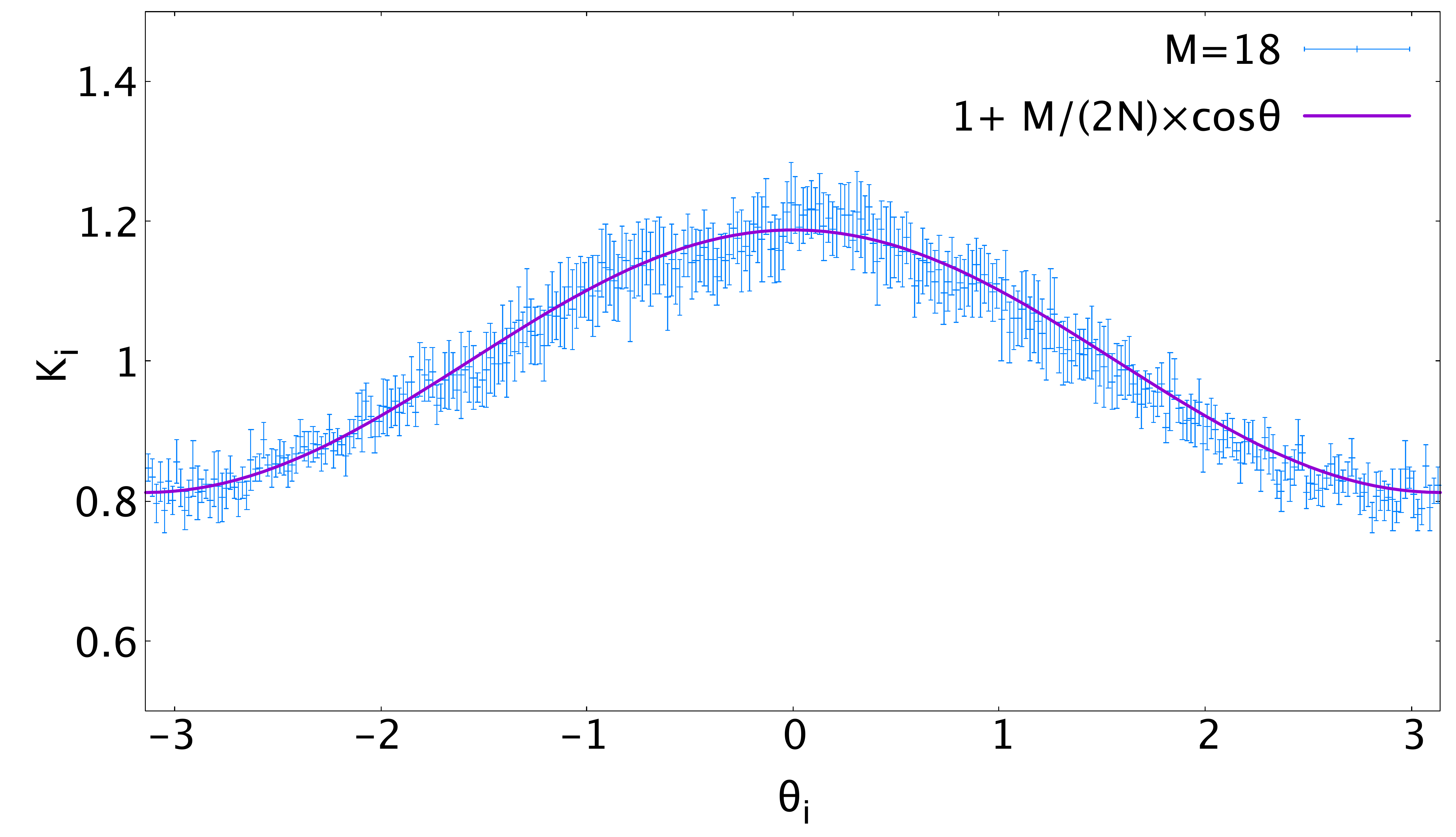}
	\end{subfigure}
	\begin{subfigure}[b]{0.475\textwidth}
		\includegraphics[width=\textwidth]{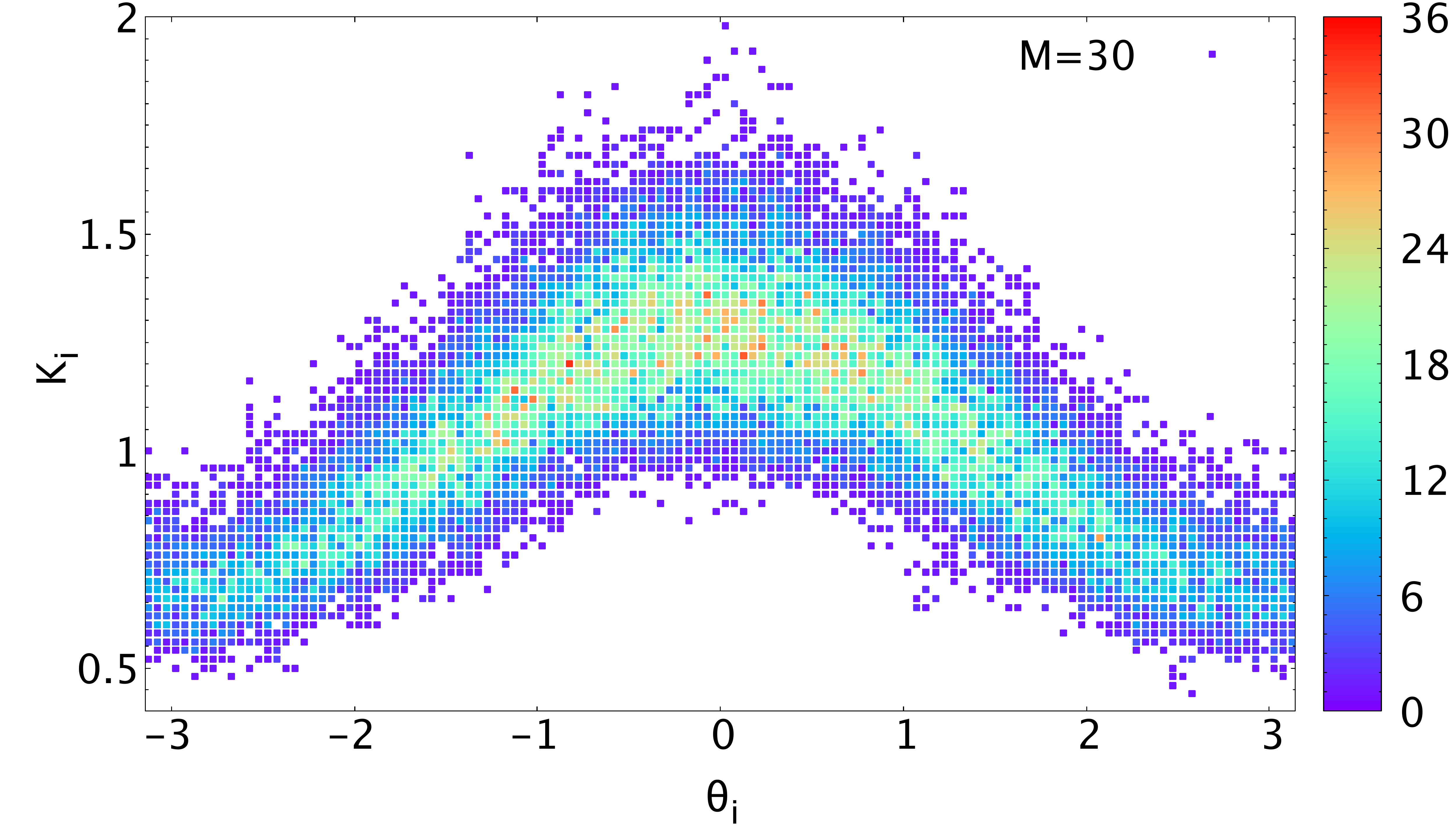}
	\end{subfigure}
	\begin{subfigure}[b]{0.475\textwidth}
		\includegraphics[width=\textwidth]{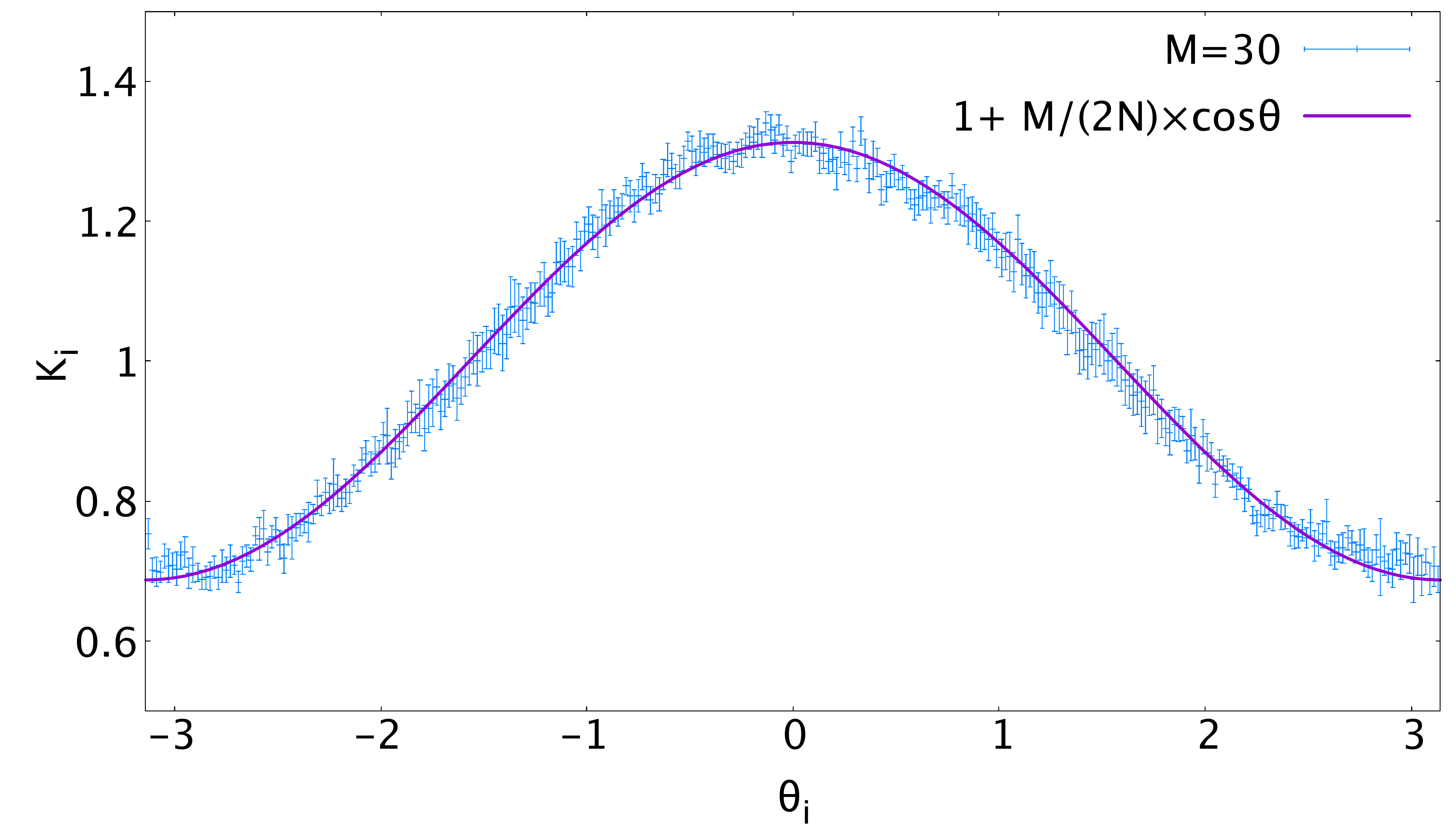}
	\end{subfigure}
	\begin{subfigure}[b]{0.475\textwidth}
		\includegraphics[width=\textwidth]{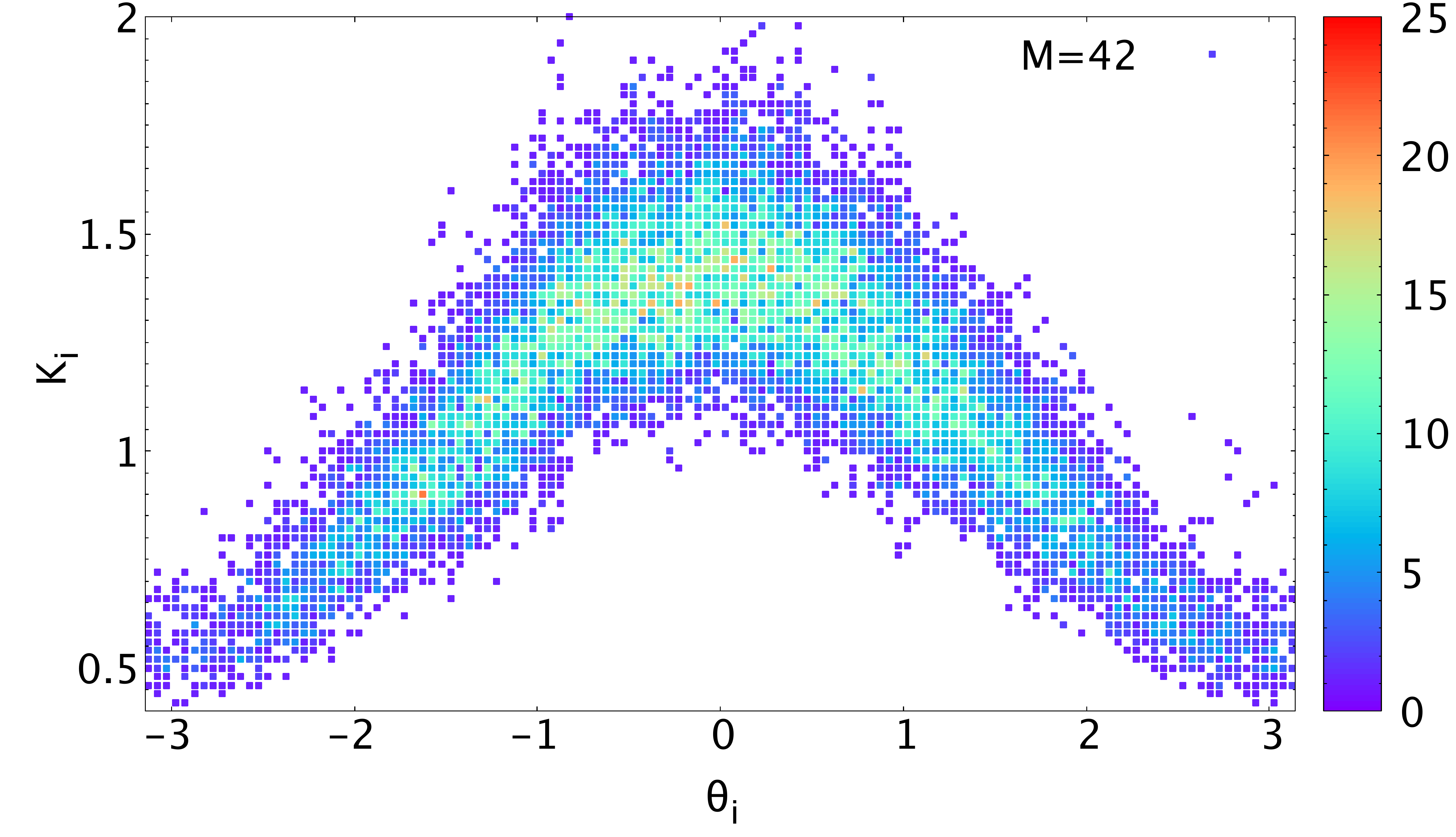}
		\caption{2D histogram}
	\end{subfigure}
	\begin{subfigure}[b]{0.475\textwidth}
		\includegraphics[width=\textwidth]{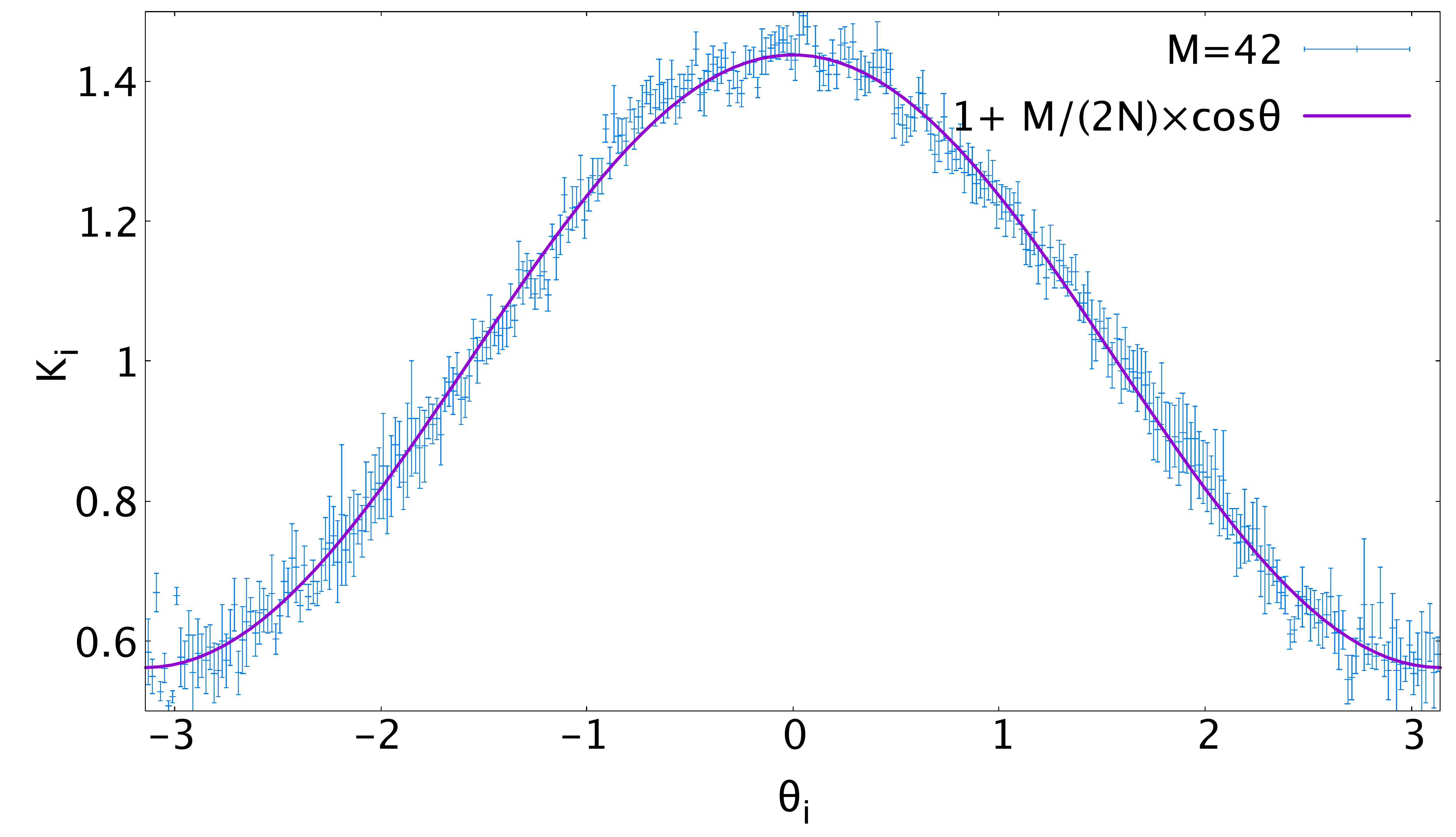}
		\caption{Binned histogram}
	\end{subfigure}
\caption{Gaussian matrix model, $\theta_i$ vs $K_i$, $N=48, d=2$, the number of lattice sites $L=16$, at $T=T_c=\frac{1}{\log 2}$.
The center symmetry is fixed sample by sample such that $P=|P|$.
(a) The two-dimensional histograms of ($\theta_i, K_i$).
(b) The averaged $K_i$ within the narrow bin $\Delta\theta=0.02$.
The magenta lines are $1+\frac{M}{2N}\cos\theta$.
The error bars are obtained by jackknife analysis.
}\label{fig:theta-vs-K}
\end{figure}

For $d=2$, $N=48$, $L=16$, at $T=T_c=\frac{1}{\log 2}$ we select configurations with three values $M=18,30,42$.
The number of sampled pairs $(\theta_i,K_i)$ are $16416, 31488, 16557$, for the different values of $M$.
In Fig.~\ref{fig:theta-vs-K} we show the two-dimensional histograms of $(\theta_i,K_i)$ at each $M$ separately.
From the two-dimensional histograms we can see only one peak at each $\theta$ (represented by a reddish hue).
Furthermore we observe that
\begin{eqnarray}
K_i
=
1+\frac{M}{2N}\cos\theta_i
\label{eq:K-vs-theta-GMM}
\end{eqnarray}
holds with good accuracy when compared to binned histograms.
These histograms are created by taking the average over the samples $K_i$ falling into a $\theta_i$ bin of size $\Delta\theta=0.02$.
The fluctuation at each fixed $\theta_i$ can be understood as the finite-$N$ effect which should be suppressed as $N$ becomes larger.
By using the distribution of $\theta_i$ in the confined and deconfined sectors ($\frac{1}{2\pi}$ and $\frac{1+\cos\theta}{2\pi}$), we obtain Eqs.~\eqref{eq:Ki_con} and \eqref{eq:Ki_dec}, and hence, also Eq.~\eqref{eq:ev-Ki}.

\section{Partial deconfinement: the Yang-Mills matrix model}\label{sec:Yang-Mills-MM}
\begin{figure}[htbp]
\centering
	\includegraphics[width=0.75\textwidth]{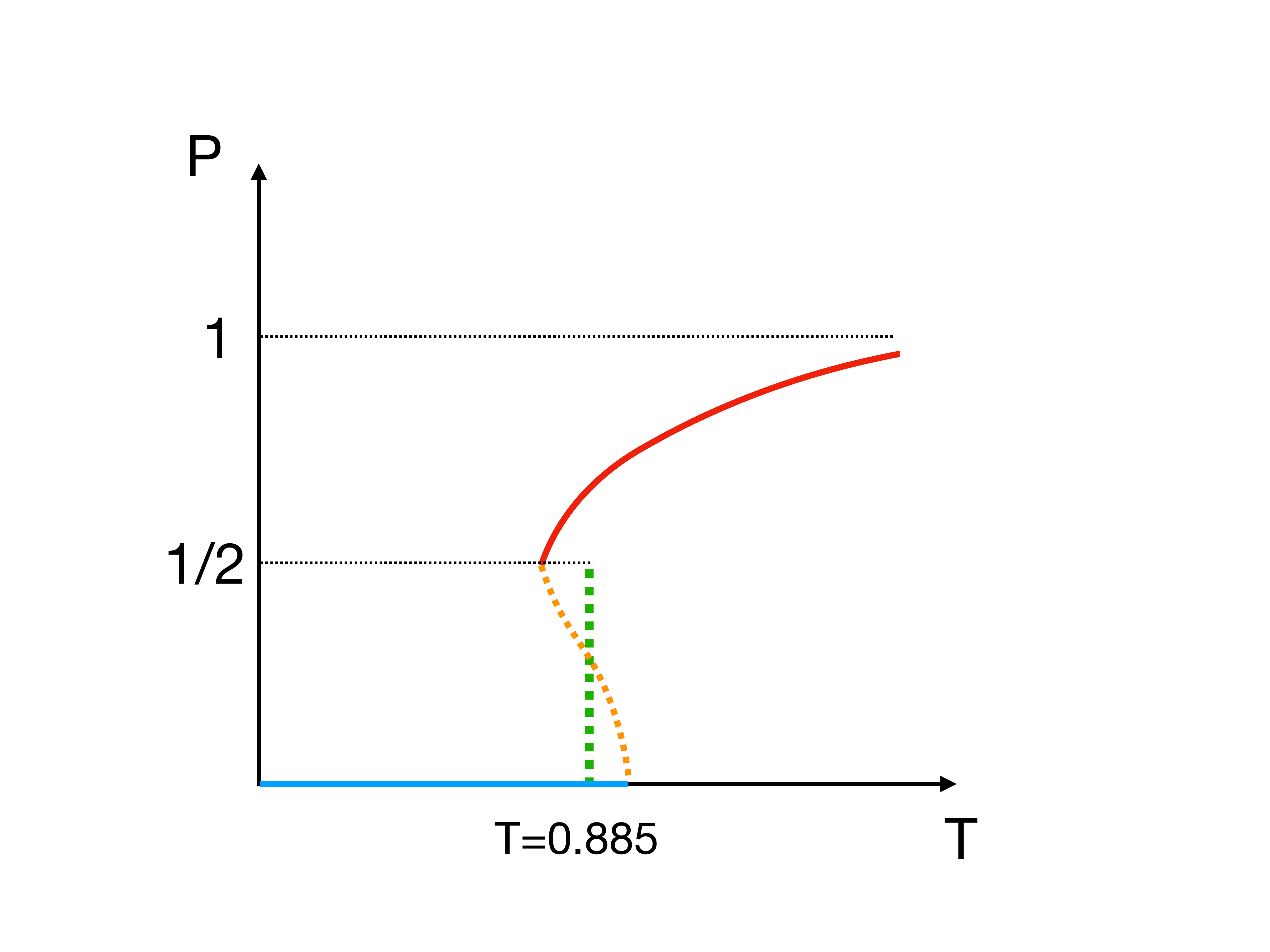}
	\caption{
A sketch of the temperature dependence of the Polyakov loop in the $d=9$ Yang-Mills matrix model \cite{Bergner:2019rca}.
Red, orange and blue lines represent the phases dominant at large $N$, in the canonical or microcanonical ensemble.
The red and blue lines represent the completely deconfined and confined phases, which minimize the free energy in the canonical ensemble.
The orange dotted line is the partially deconfined phase, which maximizes the free energy.
In order to emphasize that this phase is not favored in the canonical ensemble, we used the dotted line.
(Note however that this phase is stable in the microcanonical ensemble.)
In the canonical ensemble, there is a first order transition near $T=0.885$.
We will study the properties of the configurations at $T=0.885$, varying the value of $P$ from 0 to $\frac{1}{2}$, along the green dotted line.
Ref.~\cite{Bergner:2019rca} found that
$E$ and $R$ change as \eqref{eq:energy-YM} and \eqref{eq:R-YM}.
}\label{fig:bBFSS-simulation-cartoon}
\end{figure}

\hspace{0.51cm}
The Yang-Mills matrix model with $d=9$ exhibits a first order transition near $T=0.885$~\cite{Bergner:2019rca}, as sketched in Fig.~\ref{fig:bBFSS-simulation-cartoon}.
Slightly different from the Gaussian matrix model, there is a hysteresis
in a very narrow temperature range, which can be read off easily from the two-peak signal in the Polyakov loop distribution.
Below, we study the properties of the configurations at fixed temperature.
As a concrete example, we study $T=0.885$, varying the value of $P$ from 0 to $\frac{1}{2}$, along the green dotted line in Fig.~\ref{fig:bBFSS-simulation-cartoon}.
We expect that this fixed temperature slice is a good approximation of the partially-deconfined phase
(orange dotted line),
because the hysteresis exists in a very narrow temperature range ($0.884\lesssim T\lesssim 0.886$
for $N=64$, number of lattice sites $L=24$ \cite{Bergner:2019rca}).

At fixed temperature in the transition region, numerically we can find relations similar to
\eqref{eq:Pol-Gaussian}, \eqref{eq:energy-Gaussian} and \eqref{eq:entropy-Gaussian} for the Gaussian matrix model \cite{Bergner:2019rca}.
Namely, $\rho^{\rm(P)}(\theta)=\frac{1+2P\cos\theta}{2\pi}$ holds,\footnote{
Note that this specific form of $\rho^{\rm(P)}(\theta)$ is not a requirement, though it is observed in various theories.
Note also that we fixed the center symmetry such that $P=|P|$.
}
and by using the identification $2P=\frac{M}{N}$, we obtain
\footnote{
The fits performed in Ref.~\cite{Bergner:2019rca} were slightly different,
in that the power was not fixed to 2. The results were
$
E
=
N^2\left(a_E
+
b_E|P|^{c_E}\right)$,
where $a_E=6.1365(5)$,
$b_E=1.84(1)$ and $c_E=1.99(1)$,
and
$
R
=
N^2\left(a_R
+
b_R|P|^{c_R}\right)$,
where $a_R=2.2017(1)$,
$b_R=0.358(0)$ and $c_R=2.01(1)$. }
\begin{eqnarray}
\rho^{\rm(P)}(\theta)
=
\left(
1-\frac{M}{N}
\right)
\cdot
\frac{1}{2\pi}
+
\frac{M}{N}\cdot\frac{1+\cos\theta}{2\pi},
\label{eq:Pol-YM}
\end{eqnarray}
\begin{eqnarray}
E
=
\left\langle
-\frac{3N}{4\beta}\int dt {\rm Tr}[X_I,X_J]^2
\right\rangle
=
(N^2-M^2)\varepsilon_0
+
M^2\varepsilon_1,
\qquad
\varepsilon_0\simeq 6.14,
\quad
\varepsilon_1\simeq 6.60,
\nonumber\\
\label{eq:energy-YM}
\end{eqnarray}
and
\begin{eqnarray}
R
\equiv
\left\langle
\frac{N}{\beta}\int dt \sum_I{\rm Tr}X_I^2
\right\rangle
=
(N^2-M^2)r_0
+
M^2r_1.
\qquad
r_0\simeq 2.20,
\quad
r_1\simeq 2.29.
\label{eq:R-YM}
\end{eqnarray}

These relations can be naturally explained if we assume partial deconfinement.
The first relation \eqref{eq:Pol-YM} can be interpreted 
as showing that $N-M$ of the phases are
in the confined sector,
while the other $M$ are in the deconfined sector.
The second relation \eqref{eq:energy-YM} would mean that each degree of freedom in the deconfined sector contributes to the increment of the energy by $\varepsilon_1-\varepsilon_0$.
That it appears to be independent of $M$ would be natural because temperature $T$ is fixed.
The third relation can be interpreted in a similar manner:
$M^2$ matrix entries are excited to $|X_{ij}|^2\sim r_1$, while the rest remain $|X_{ij}|^2\sim r_0$.

Although \eqref{eq:Pol-YM}, \eqref{eq:energy-YM} and \eqref{eq:R-YM} are consistent with partial deconfinement, the separation to the SU($M$)- and SU($N-M$)-sectors shown in Fig.~\ref{fig:matrix-partial-deconfinement} has not been confirmed explicitly in previous studies.
Unlike in the Gaussian model,
whether the separation to `confined' and `deconfined' sectors can work is highly nontrivial
due to the interaction.
The explicit confirmation of this separation is the goal of this section.
Our strategy is to confirm the properties of the master field compatible with partial deconfinement,
analogous to the ones we have seen in Sec.~\ref{sec:Gaussian-matrix-model}.

Note also that we do not find a clear theoretical reason forbidding the nontrivial $M$-dependence.
Due to the interaction between the confined and deconfined sectors,
the average contribution in the confined and deconfined sectors may change depending on $M$.
We will come back to this issue later, in Sec.~\ref{sec:simulation_summary}.
In short, whether the $M$-dependence exists or not does not affect our argument significantly.

\subsection{The properties of the ensemble and the master field on the lattice}\label{sec:BFSS-unconstrained}
\hspace{0.51cm}
In this section, we use the static diagonal gauge, as we did in Sec.~\ref{sec:Gaussian-matrix-model},
i.e., the gauge field is fixed to $A_t = {\rm diag}\left(\frac{\theta_1}{\beta},\cdots,\frac{\theta_N}{\beta}\right)$.  The relations \eqref{eq:Pol-YM}, \eqref{eq:energy-YM} and \eqref{eq:R-YM} suggest
the deconfinement of the $M\times M$-block at the upper-left corner, when $\theta_1,\cdots,\theta_M$
are distributed as $\frac{1+\cos\theta}{2\pi}$ and $\theta_{M+1},\cdots,\theta_N$ are uniformly distributed.
If this is correct, then many of the arguments for the Gaussian matrix model presented in Sec.~\ref{sec:Gaussian-matrix-model} can be repeated without change.

Before showing the results confirming this expectation, let us remark a technical aspect of the simulation.
The size of the deconfined sector $M$ can change from $0$ to $N$. Therefore, in order to estimate the quantities such as $\rho^{\rm(X)}(x)$ or $K_i$ at fixed $M$, we need a very long simulation, so that the samples with that specific value of $M$ appear many times.
For the Gaussian matrix model, we took this approach, because the simulation cost was low.
For the Yang-Mills matrix model, we take a more efficient approach.
The idea is to restrict the value of the Polyakov loop $P=\frac{1}{N}\sum_{j=1}^Ne^{i\theta_j}$ by adding
\begin{eqnarray}
\Delta S
=\left\{
\begin{array}{cc}
\frac{\gamma}{2}\left(|P|-p_1\right)^2 & (|P|<p_1)\\
\frac{\gamma}{2}\left(|P|-p_2\right)^2 & (|P|>p_2)
\end{array}
\right.
\label{delta-S-unconstrained}
\end{eqnarray}
to the action.\footnote{
This deformation may look similar to Eq.~\eqref{constraint-partially-deconfined}, which will be introduced in Sec.~\ref{sec:constrained_simulation},
but actually there is a big difference.
We are fixing $|P|$ ({\it not} $|P_M|$ or $|P_{N-M}|$) between $p_1$ and $p_2$. See also Fig.~\ref{fig:Two-constrained-simulations}}
This allows us to pick up the configurations at a fixed value of $M$ effectively
by choosing $p_1$ and $p_2$ appropriately, while leaving the configurations at $p_1<|P|<p_2$
untouched.

\subsubsection{Distribution of $X_{I,ij}$}\label{sec:scalars-bBFSS-unconstrained}
\hspace{0.51cm}
The relation Eq.~\eqref{eq:R-YM} suggests the separation of the distribution of $X_{I,ij}$'s to $\rho^{\rm(X)}_{\rm dec}(x)$ and $\rho^{\rm(X)}_{\rm con}(x)$ just as in the Gaussian model, by using the same expression Eq.~\eqref{eq:separation_rho_scalar}.
Let us confirm that this is indeed the case.
Here we are assuming that $\rho^{\rm(X)}_{\rm dec}(x)$ and $\rho^{\rm(X)}_{\rm con}(x)$ are independent of $M$.
As we will see, this is valid with a reasonably good precision, although a weak $M$-dependence may exist.
In Sec.~\ref{sec:constrained_simulation}, we show a different analysis that does not assume $M$-independence.
(See Sec.~\ref{sec:simulation_summary} for further discussions.)

\begin{figure}[htbp]
\centering
	\begin{subfigure}[b]{0.475\textwidth}
		\includegraphics[width=\textwidth]{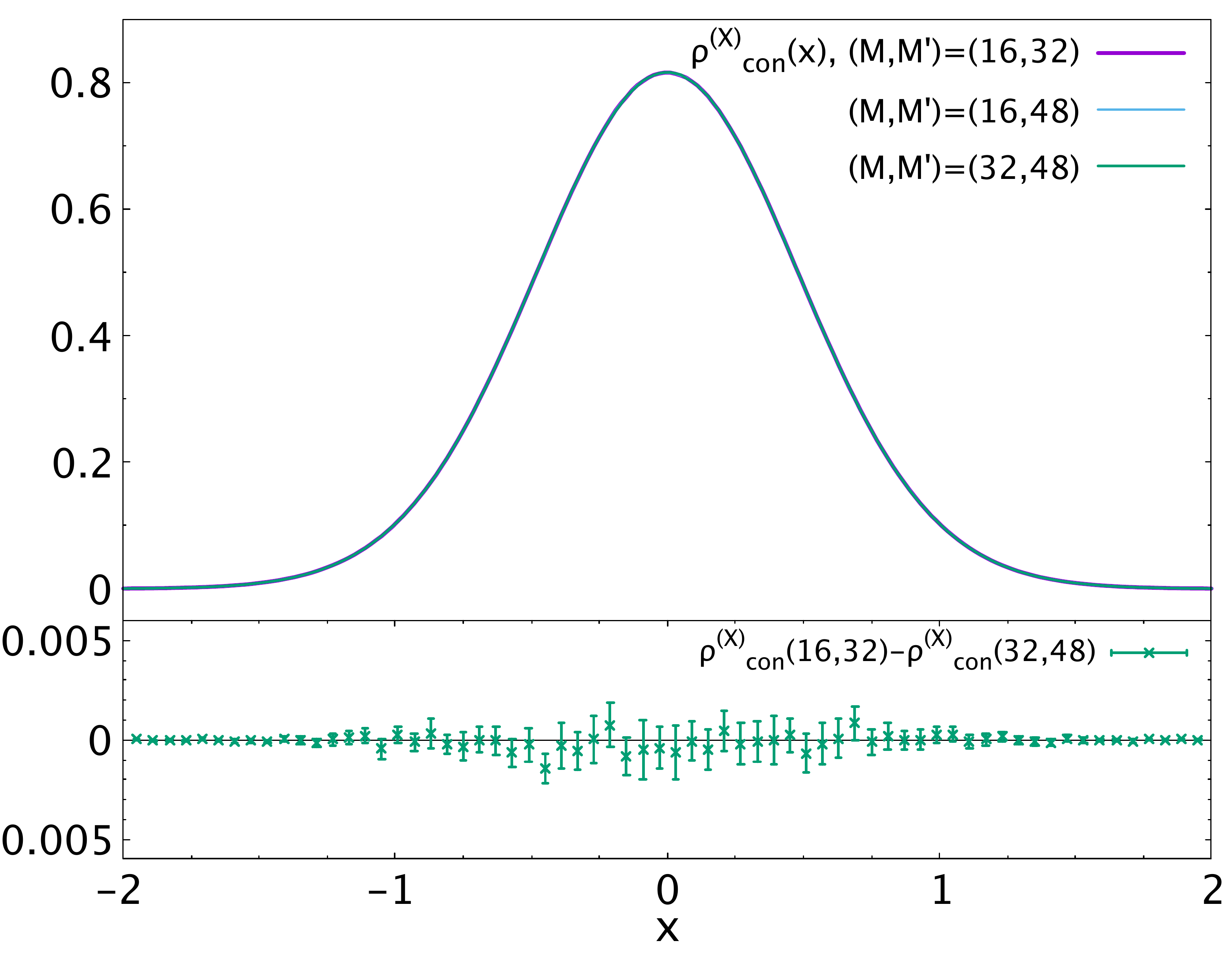}
		\caption{$\rho^{\rm(X)}_{\rm con}(x)$}
	\end{subfigure}
	\begin{subfigure}[b]{0.475\textwidth}
		\includegraphics[width=\textwidth]{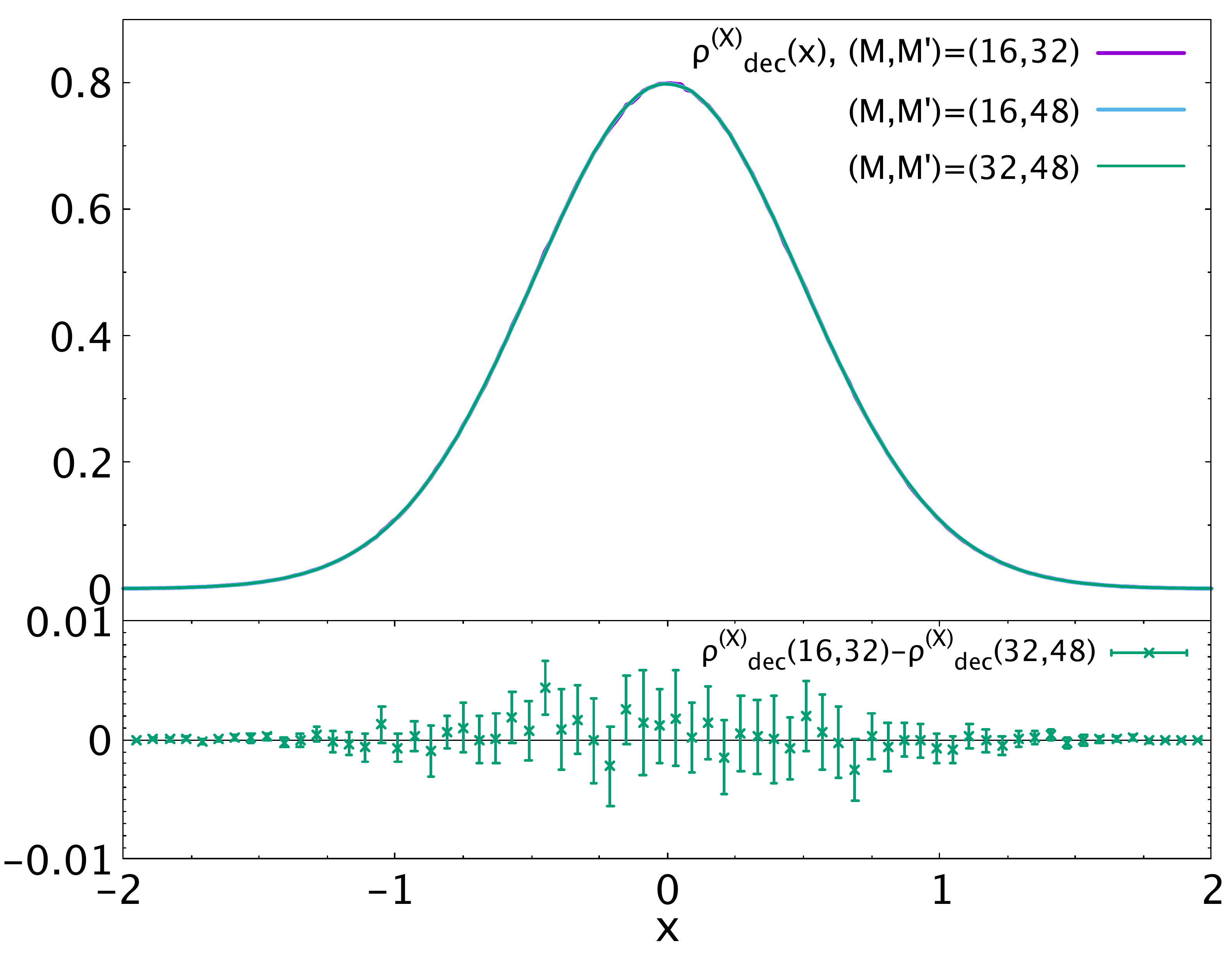}
		\caption{$\rho^{\rm(X)}_{\rm dec}(x)$}
	\end{subfigure}
\caption{$\rho^{\rm(X)}_{\rm con}(x)$ and $\rho^{\rm(X)}_{\rm dec}(x)$ in the Yang-Mills matrix model, $N=64$, $L=24$, and $T=0.885$.
In order to sample $M=16,32$ and $48$ efficiently, we used the trick explained around Eq.~\eqref{delta-S-unconstrained}.
Different combinations of $M$ and $M'$ lead to the same result within the error bars.
The error bars in each figure are obtained by jackknife analysis.
}\label{fig:P_constrained-BFSS-rho-scalar}
\end{figure}

\begin{figure}[htbp]
\centering
	\begin{subfigure}[b]{0.475\textwidth}
		\includegraphics[width=\textwidth]{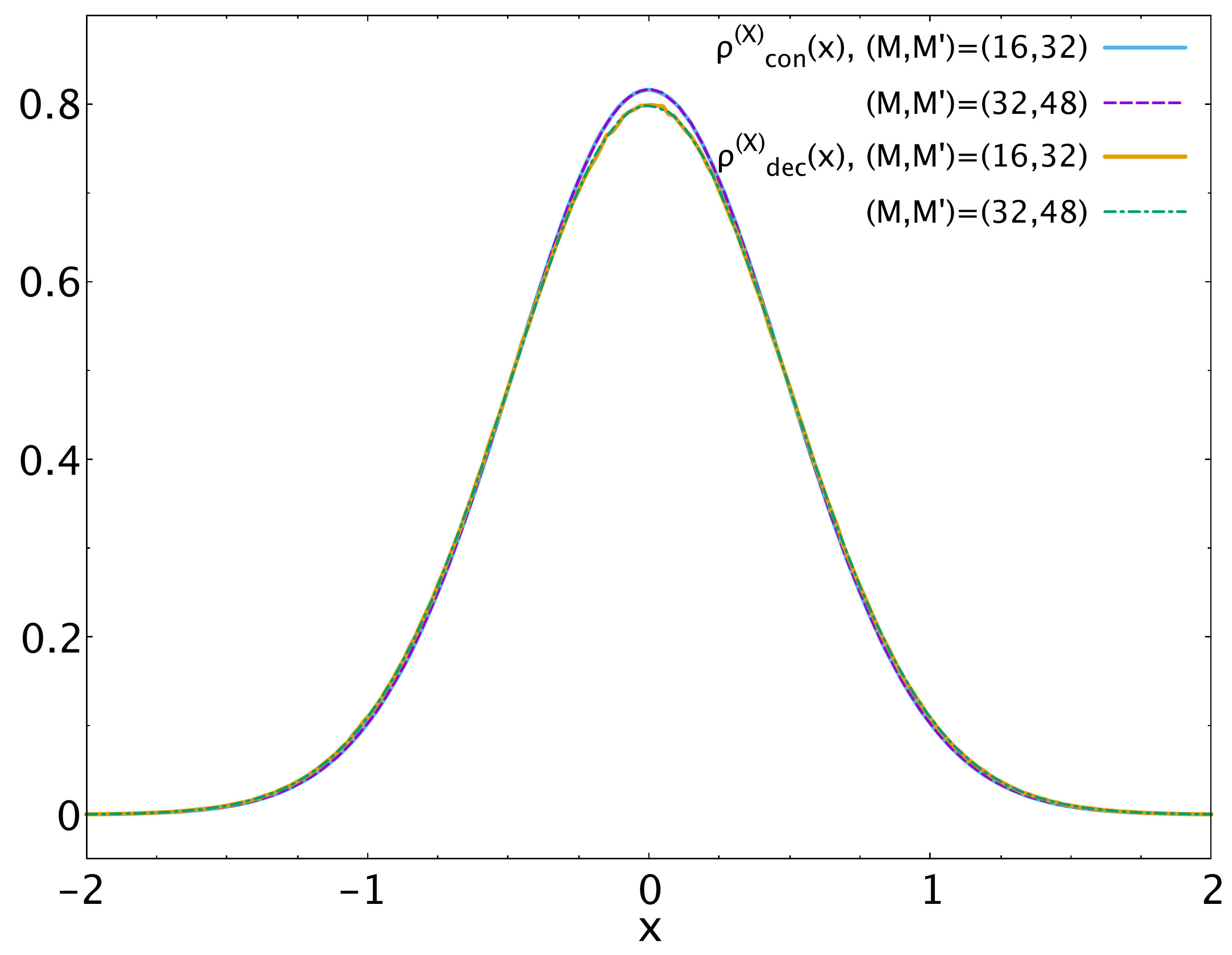}
		\caption{Distributions}
	\end{subfigure}
	\begin{subfigure}[b]{0.475\textwidth}
		\includegraphics[width=\textwidth]{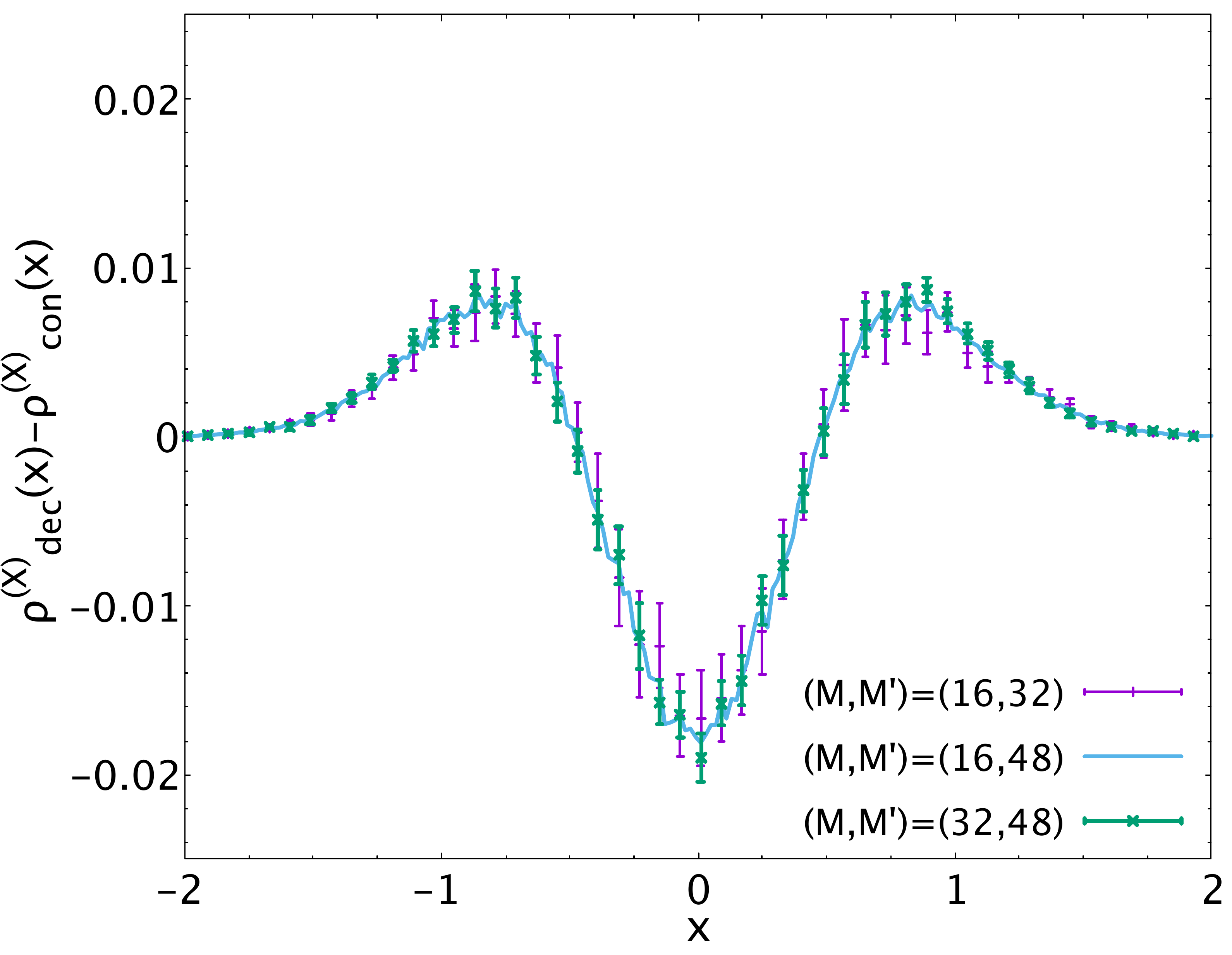}
		\caption{Difference}
	\end{subfigure}
\caption{Yang-Mills matrix model, $N=64$, $L=24$, and $T=0.885$.
Comparison between $\rho^{\rm(X)}_{\rm dec}(x)$ and $\rho^{\rm(X)}_{\rm con}(x)$ as shown in Fig.~\ref{fig:P_constrained-BFSS-rho-scalar}.
(a) A small but visible difference exists between $\rho^{\rm(X)}_{\rm dec}(x)$ and $\rho^{\rm(X)}_{\rm con}(x)$.
(b) The difference $\rho^{\rm(X)}_{\rm dec}(x)-\rho^{\rm(X)}_{\rm con}(x)$ is significantly larger than the error bars of $\rho^{\rm(X)}_{\rm dec}(x)$ and $\rho^{\rm(X)}_{\rm con}(x)$.
The error bars in each figure are obtained by jackknife analysis.
}\label{fig:P_constrained-BFSS-rho-scalar_1}
\end{figure}

In Figs.~\ref{fig:P_constrained-BFSS-rho-scalar} and \ref{fig:P_constrained-BFSS-rho-scalar_1}, we show the distributions $\rho^{\rm(X)}_{\rm dec}(x)$ and $\rho^{\rm(X)}_{\rm con}(x)$ obtained by using Eqs.~\eqref{how-to-determine-rho-con}-\eqref{how-to-determine-rho-dec}.
Different pairs $(M,M')$ lead to the same distributions, and the confined and deconfined sectors behave differently.
The variances computed from the histograms $\rho^{\rm(X)}_{\rm con}(x)$ and $\rho^{\rm(X)}_{\rm dec}(x)$ in Fig.~\ref{fig:P_constrained-BFSS-rho-scalar} are shown in Table~\ref{tab:variance_Pconst}.

\begin{table}[htbp]
\begin{center}
\caption{The variances of $\rho^{\rm(X)}_{\rm dec}$ and $\rho^{\rm(X)}_{\rm con}$ in the simulation, $N=64$, $T=0.885$, with 24 lattice sites.
The total number of $x$ is obtained by $\#x = 64^2\times9\times24\times(\textrm{\# configs.})$.}
\label{tab:variance_Pconst}
\begin{tabular}{|c|c|c|c|}\hline
	$(M,M')$		&	$\sigma^2_{\rm con}$ 	&	$\sigma^2_{\rm dec}$ &\# configs. of $(M,M')$ 	\\ \hline	\hline
	(16, 32)		&	0.2447(5)	&	0.254(3)	&	(854, 857)	\\\hline
	(16,	48)		&	0.2447(4)	&	0.254(1)	&	(854, 814)	\\	\hline
	(32,	48)		&	0.2447(8)	&	0.254(1)	&	(857, 814)	\\ \hline
\end{tabular}
\end{center}
\end{table}

If $\rho^{\rm(X)}_{\rm dec}(x)$ and $\rho^{\rm(X)}_{\rm con}(x)$ do not depend of $M$ then the variances have to be related to $r_0$ and $r_1$ according to Eq.~\eqref{eq:R-YM} and Eq.~\eqref{eq:separation_rho_scalar},
\begin{eqnarray}
\sigma^2_{\rm con}\equiv
\int dx\ x^2\rho^{\rm(X)}_{\rm con}(x)=\frac{r_0}{d}\simeq\frac{2.20}{9}\simeq 0.244,
\label{eq:sigma^2_con}
\\
\sigma^2_{\rm dec}\equiv
\int dx\ x^2\rho^{\rm(X)}_{\rm dec}(x)=\frac{r_1}{d}\simeq\frac{2.29}{9}\simeq 0.254.
\label{eq:sigma^2_dec}
\end{eqnarray}
The agreement with the values in Table~\ref{tab:variance_Pconst} is very good.
We will show more numerical results for the distributions of $x$ and discuss their properties in Appendix~\ref{sec:comments_rho(x)}.

Though highly nontrivial, this fact alone does not establish the separation into the SU($M$)- and SU($N-M$)-sectors; logically, it just implies a separation between $M^2$ and $N^2-M^2$ degrees of freedom.
Furthermore we have assumed that $\rho^{\rm(X)}_{\rm dec}(x)$ and $\rho^{\rm(X)}_{\rm con}(x)$ are independent of $M$, although, ideally, we do not want to assume it.
We will establish the separation to the SU($M$)- and SU($N-M$)-sectors more rigorously in Sec.~\ref{sec:constrained_simulation}.

\subsubsection{Correlation between scalars and gauge field}\label{sec:BFSS-theta-vs-K-unconstrained}
\hspace{0.51cm}

The correlation between $K_i\equiv\sum_{I,j}\frac{1}{\beta}\int dt|X_{I,ij}|^2$ and $\theta_i$ is very similar to the case of the Gaussian matrix model.
Firstly, because the fit \eqref{eq:R-YM} works well,
\begin{eqnarray}
\langle K_i \rangle_{\rm con}
=
r_0
=
9\sigma^2_{\rm con},
\qquad
\langle K_i \rangle_{{\rm GWW},N}
=
r_1
=
9\sigma^2_{\rm dec}
\label{eq:Ki_con_BFSS}
\end{eqnarray}
holds with good numerical precision.
(In the case of the Gaussian matrix model, we could show it analytically.)
From this, if we assume that the contributions from the confined and deconfined sectors
are always $r_0$ and $r_1$ regardless of the value of $M$,
we obtain
\begin{eqnarray}
\langle K_i \rangle_{\rm dec}
=
r_0+\frac{M}{N}\cdot (r_1-r_0)
\label{eq:Ki_dec_BFSS}
\end{eqnarray}
when $\theta_i$ is in the deconfined sector.
Because the distribution of the Polyakov line phases in the confined and deconfined sectors are the same as in the Gaussian matrix model ($\frac{1}{2\pi}$ and $\frac{1+\cos\theta}{2\pi}$), we naturally expect essentially the same form as Eq.~\eqref{eq:K-vs-theta-GMM}:
\begin{eqnarray}
K_i
=
r_0+\frac{M}{N}\cdot 2(r_1-r_0)\cos\theta_i.
\label{K-vs-theta-BFSS}
\end{eqnarray}

\begin{figure}[htbp]
\centering
	\begin{subfigure}[b]{0.475\textwidth}
		\includegraphics[width=\textwidth]{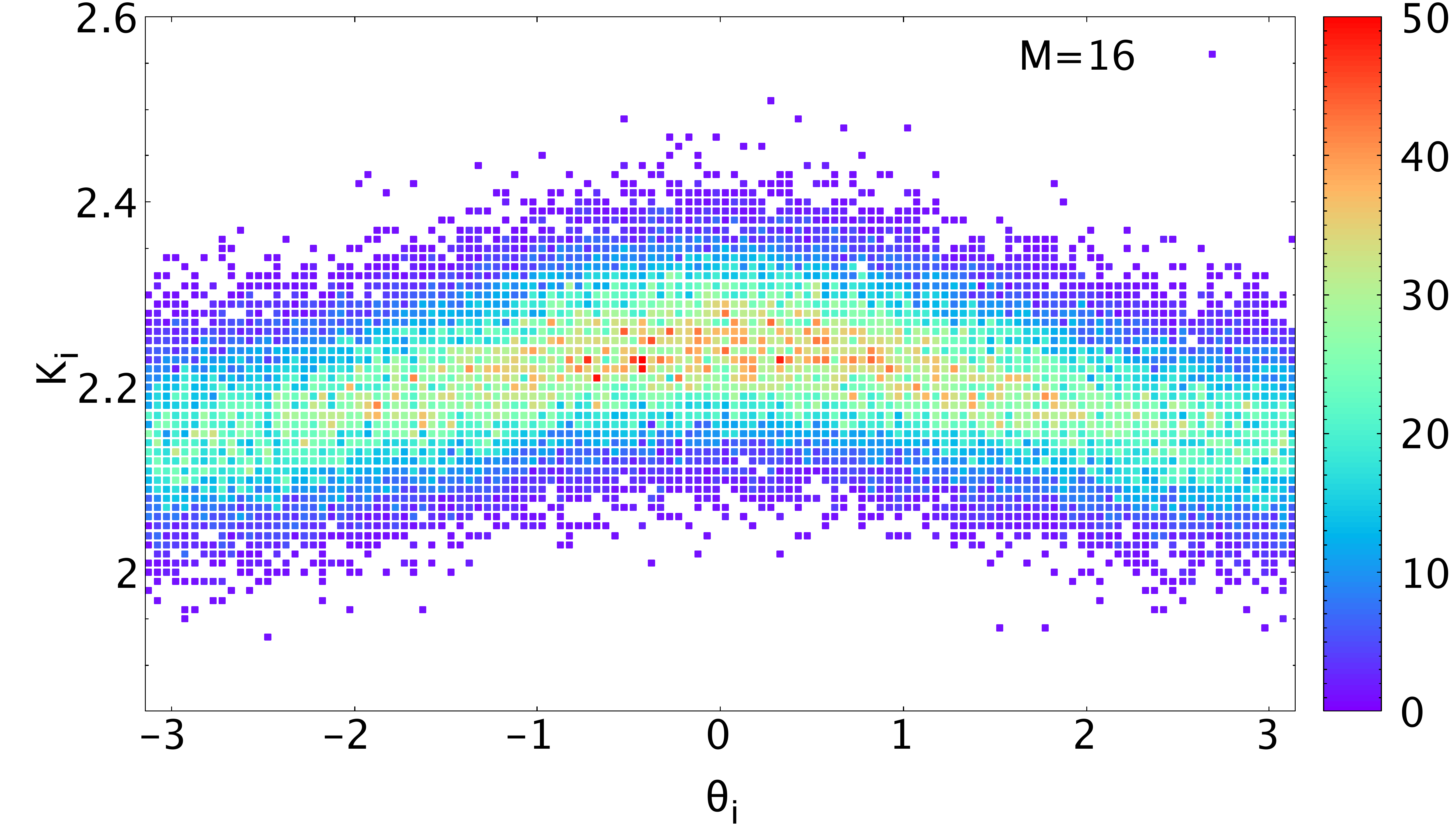}
	\end{subfigure}
	\begin{subfigure}[b]{0.475\textwidth}
		\includegraphics[width=\textwidth]{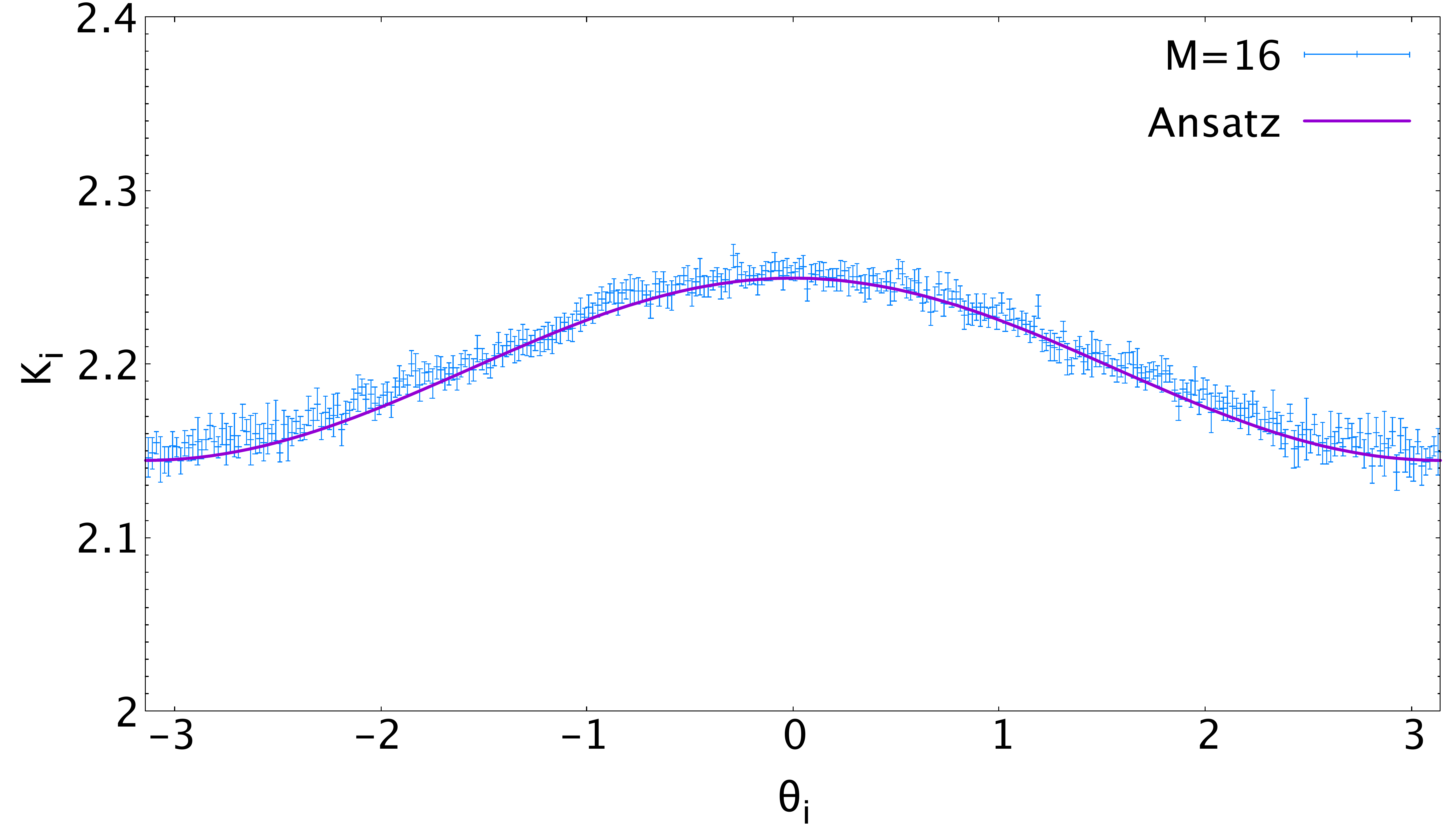}
	\end{subfigure}
	\begin{subfigure}[b]{0.475\textwidth}
		\includegraphics[width=\textwidth]{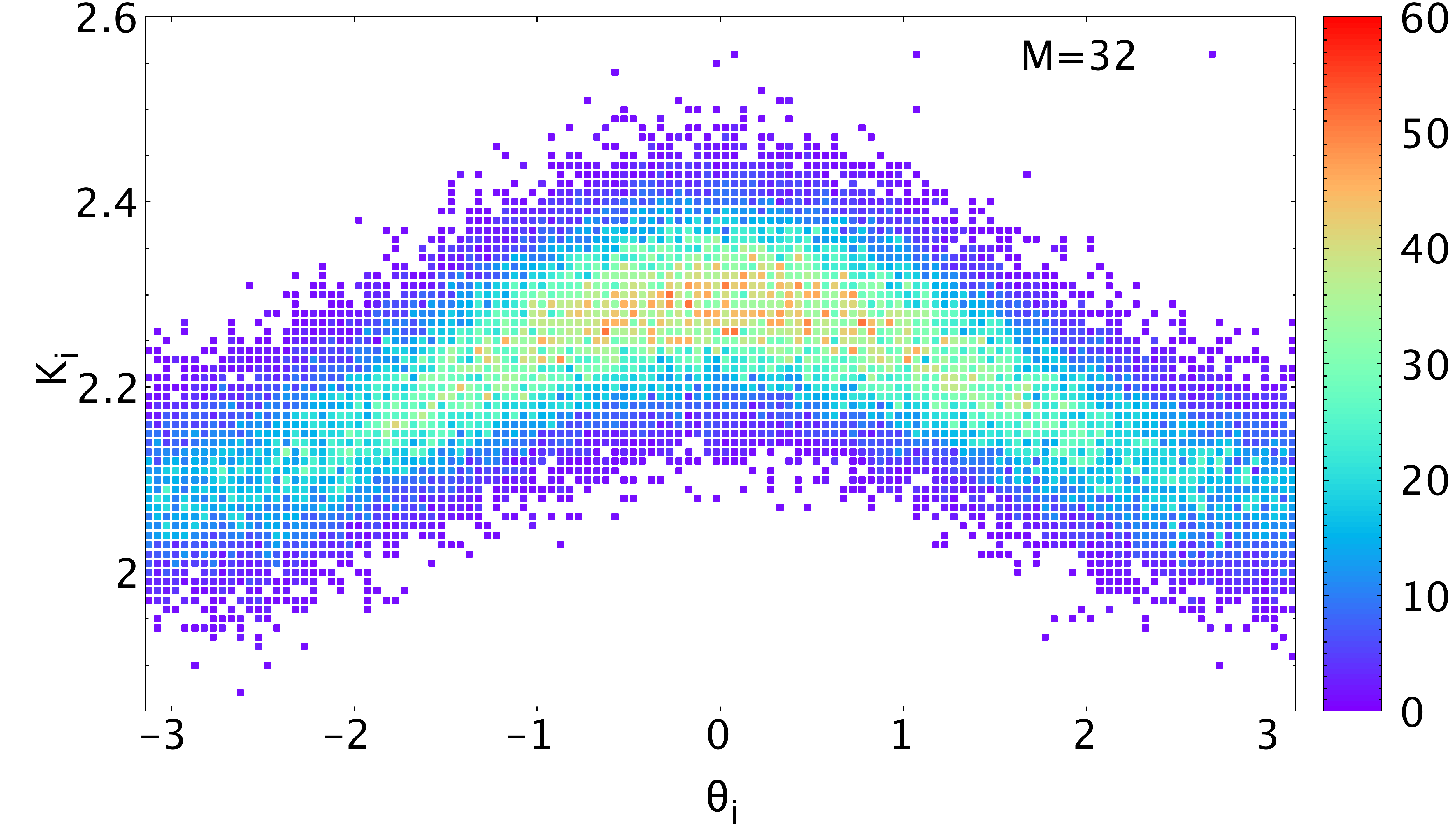}
	\end{subfigure}
	\begin{subfigure}[b]{0.475\textwidth}
		\includegraphics[width=\textwidth]{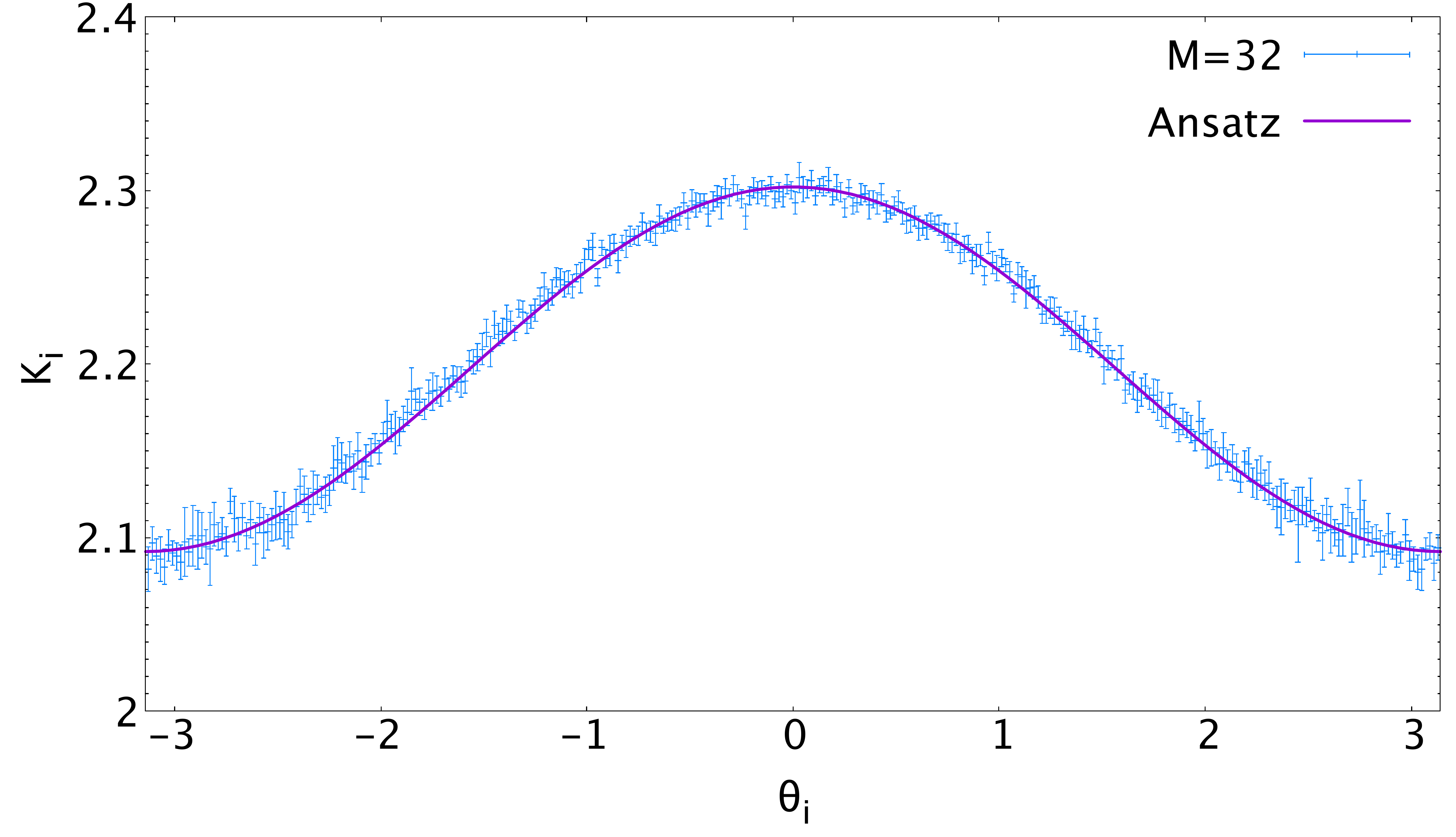}
	\end{subfigure}
	\begin{subfigure}[b]{0.475\textwidth}
		\includegraphics[width=\textwidth]{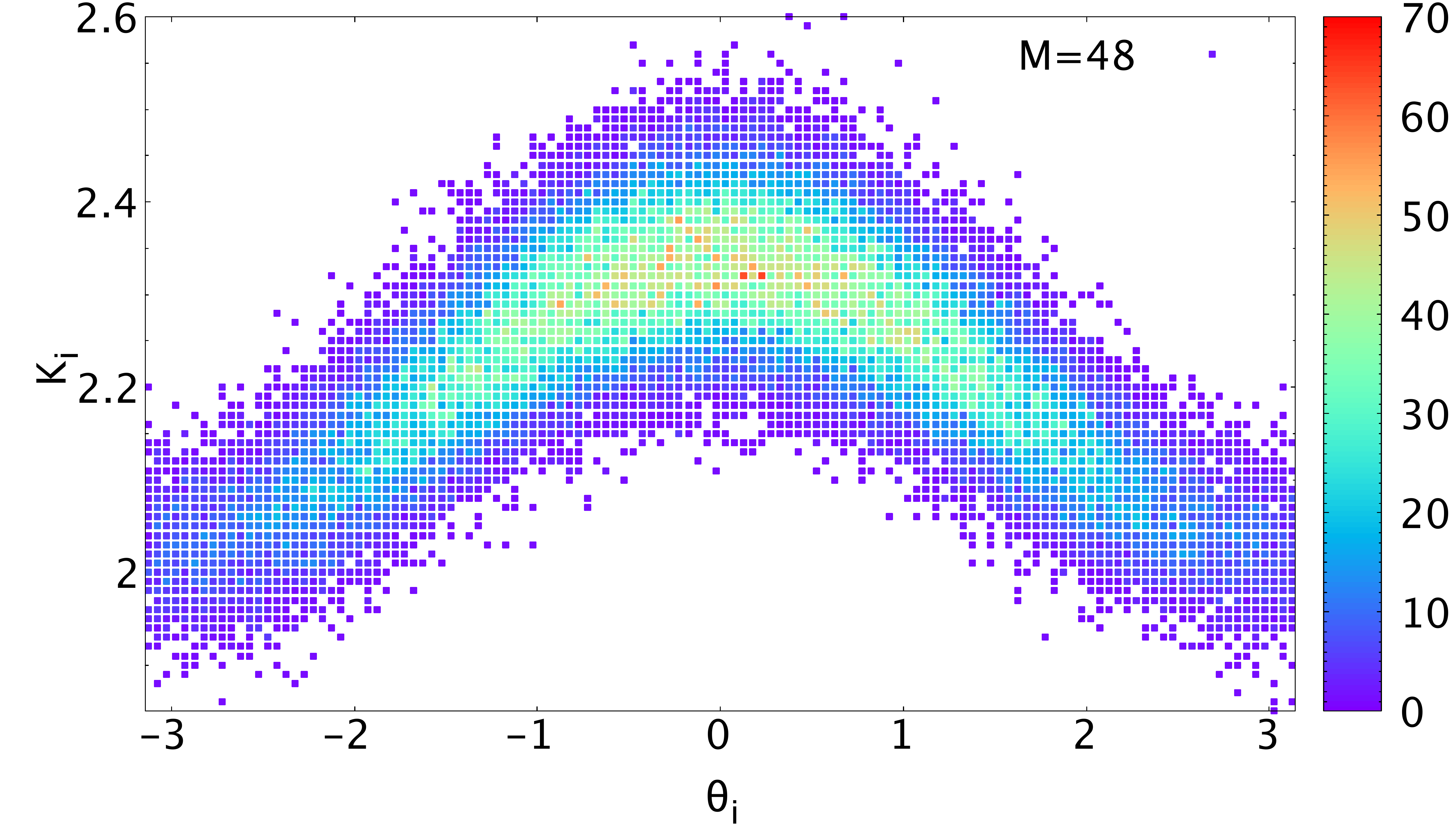}
		\caption{2D histogram}
	\end{subfigure}
	\begin{subfigure}[b]{0.475\textwidth}
		\includegraphics[width=\textwidth]{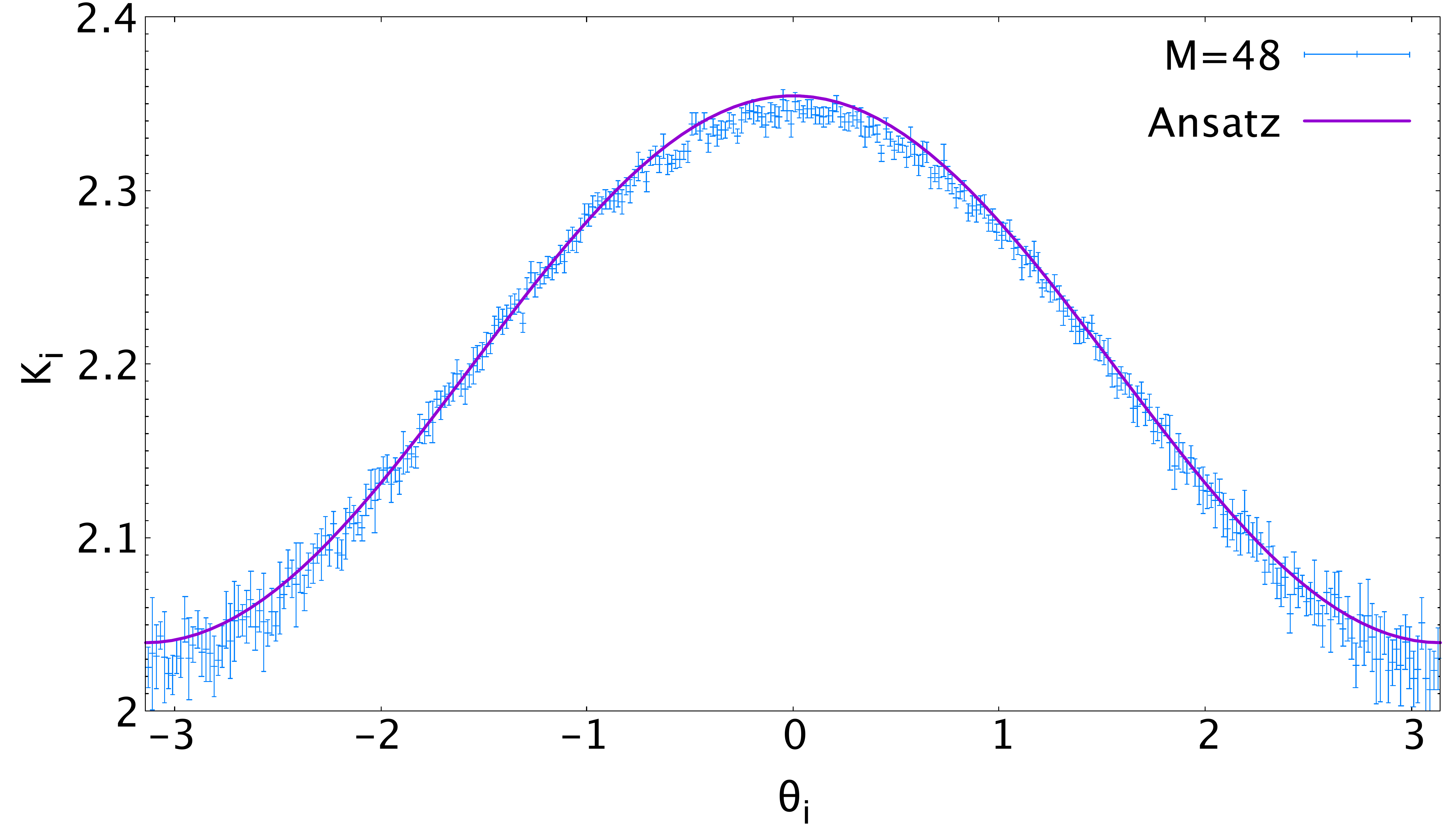}
		\caption{Binned histogram}
	\end{subfigure}
\caption{Yang-Mills matrix model, $\theta_i$ vs $K_i$, $N=64$, $L=64$, and $T=0.885$.
The total number of data points for each $M$ is obtained by $64\times \textrm{(\# config.)}$ in Table~\ref{tab:variance_Pconst}.
The center symmetry is fixed sample by sample such that $P=|P|$.
(a) The two-dimensional histograms of ($\theta_i, K_i$).
(b) The averaged $K_i$ within the narrow bin $\Delta\theta=0.02$.
The magenta lines are fit based on Eq.~\eqref{K-vs-theta-BFSS}, by using the best-fit values $r_0=2.197$ and $r_1=2.302$.
The values obtained by the fits are consistent with the values of the variances in Table~\ref{tab:variance_Pconst}.
The error bars in each figure are obtained by jackknife analysis.
}\label{fig:theta-vs-K_bBFSS-Pconst}
\end{figure}

In Fig.~\ref{fig:theta-vs-K_bBFSS-Pconst}, we show the correlation between $\theta_i$ and $K_i$ obtained by numerical simulations.
The values of $r_0$ and $r_1$ obtained by using Eq.~\eqref{K-vs-theta-BFSS} as fit ansatz are consistent with the values of the variances in Table~\ref{tab:variance_Pconst}.

\subsection{Constrained simulation}\label{sec:constrained_simulation}
\hspace{0.51cm}
In this section, we introduce a practically useful approach which makes the $M\times M$-block-structure manifest.
Again, we adopt the lattice regularization in the static diagonal gauge.
We separate $\theta_1,\cdots,\theta_N$ to two groups, and define Polyakov loops separately, as
\begin{eqnarray}
P_M=\frac{1}{M}\sum_{j=1}^Me^{i\theta_j},
\qquad
P_{N-M}=\frac{1}{N-M}\sum_{j=M+1}^Ne^{i\theta_j}.
\end{eqnarray}
Then we add the constraint term to the action,
\begin{eqnarray}
\Delta S
=\left\{
\begin{array}{cc}
\frac{\gamma}{2}\left(|P_M|-\frac{1+\delta}{2}\right)^2 & (|P_M|>\frac{1+\delta}{2})\\
\frac{\gamma}{2}\left(|P_M|-\frac{1-\delta}{2}\right)^2 & (|P_M|<\frac{1-\delta}{2})\\
\frac{\gamma}{2}\left(|P_{N-M}|-\delta\right)^2 & (|P_{N-M}|>\delta)
\end{array}
\right.
\label{constraint-partially-deconfined}
\end{eqnarray}
By taking $\gamma$ sufficiently large and $\delta$ sufficiently small, 
$|P_M|$ and $|P_{N-M}|$ can be constrained to be close to $\frac{1}{2}$ and 0.
We used $\gamma\sim 10^5$ and $\delta=0.002$ in our simulations.
More details about the simulation method will be explained in Appendix~\ref{sec:lattice_setup}.

If our scenario regarding partial deconfinement is correct, this constraint should fix the S$_N$ permutation symmetry and make the upper-left SU($M$) block deconfined while keeping the rest confined, as in Fig.~\ref{fig:matrix-partial-deconfinement}.

\subsubsection{Sanity checks}\label{sec:sanity-check-constrained-simulation}
\hspace{0.51cm}
At finite $N$, this constraint can change the theory slightly,
because $P_{M}=\frac{1}{2}$ and $P_{N-M}=0$ are valid only when $M$ and $N-M$ are sufficiently large.
It is easy to check that this effect is not large
at the values of $N$ and $M$ we study below ($N=48, 64$ and 128, $\frac{M}{N}=0.25, 0.50$ and $0.75$).

\begin{figure}[htbp]
\centering
	\begin{subfigure}[b]{0.475\textwidth}
		\includegraphics[width=\textwidth]{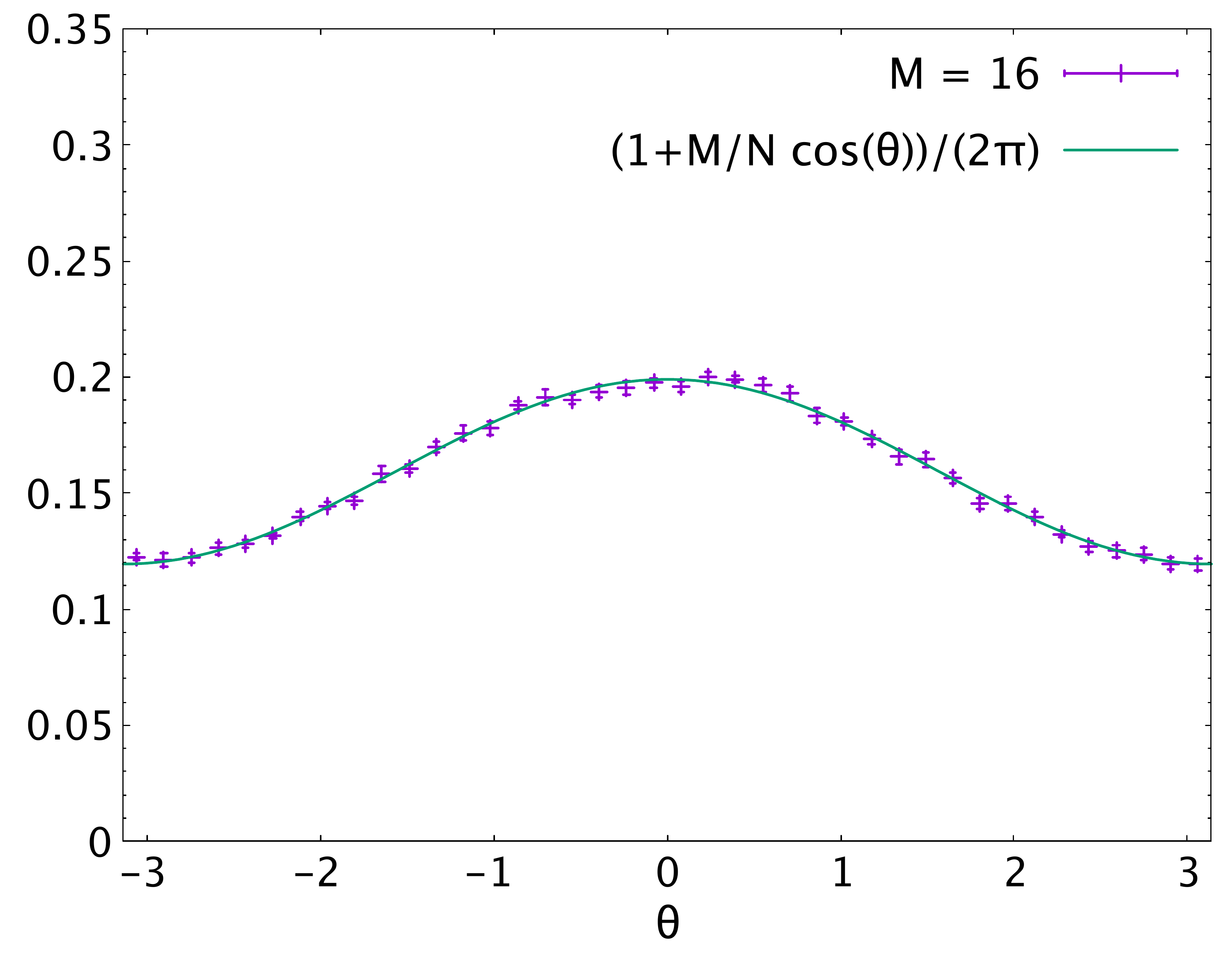}
	\end{subfigure}
	\begin{subfigure}[b]{0.475\textwidth}
		\includegraphics[width=\textwidth]{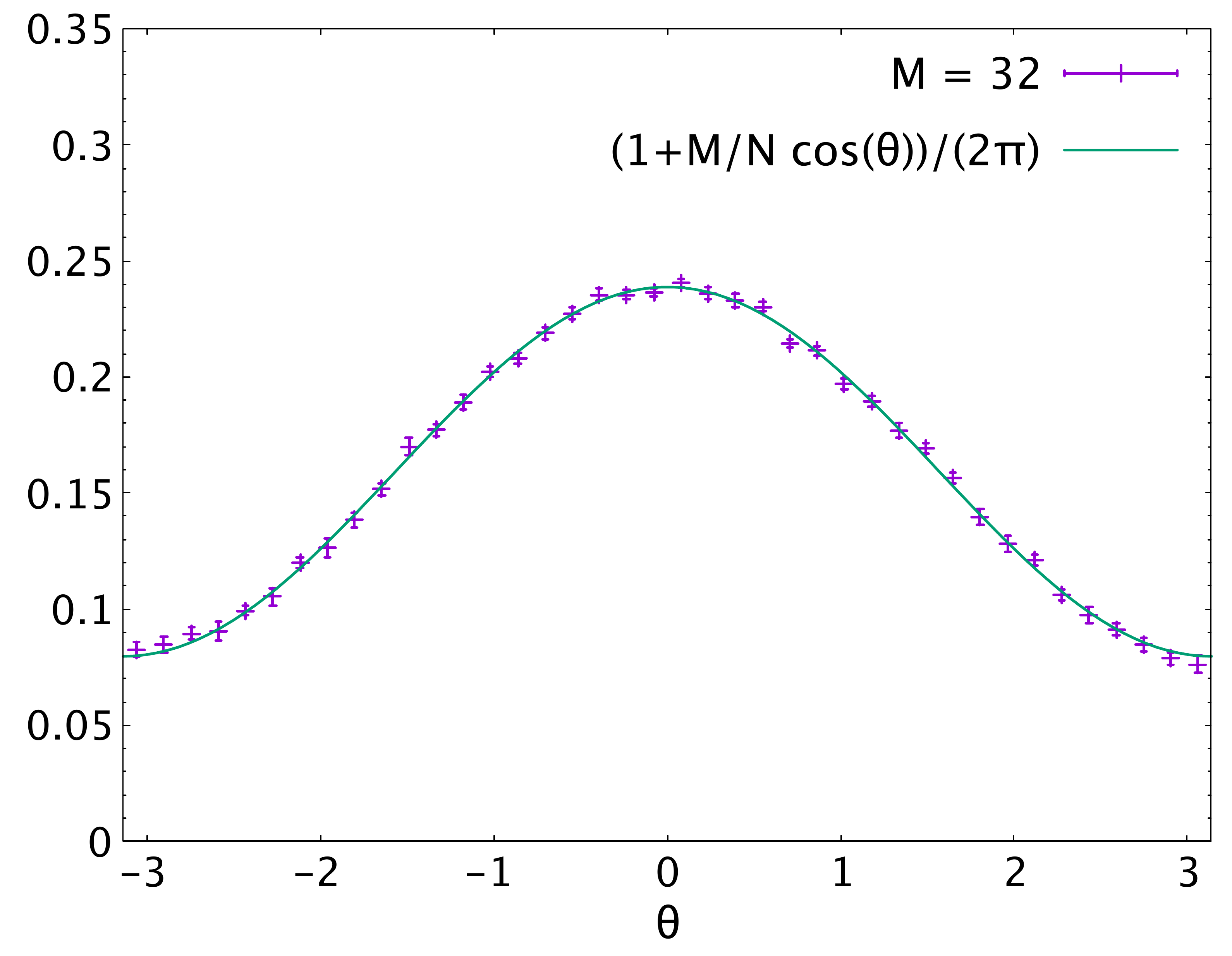}
	\end{subfigure}
	\begin{subfigure}[b]{0.475\textwidth}
		\includegraphics[width=\textwidth]{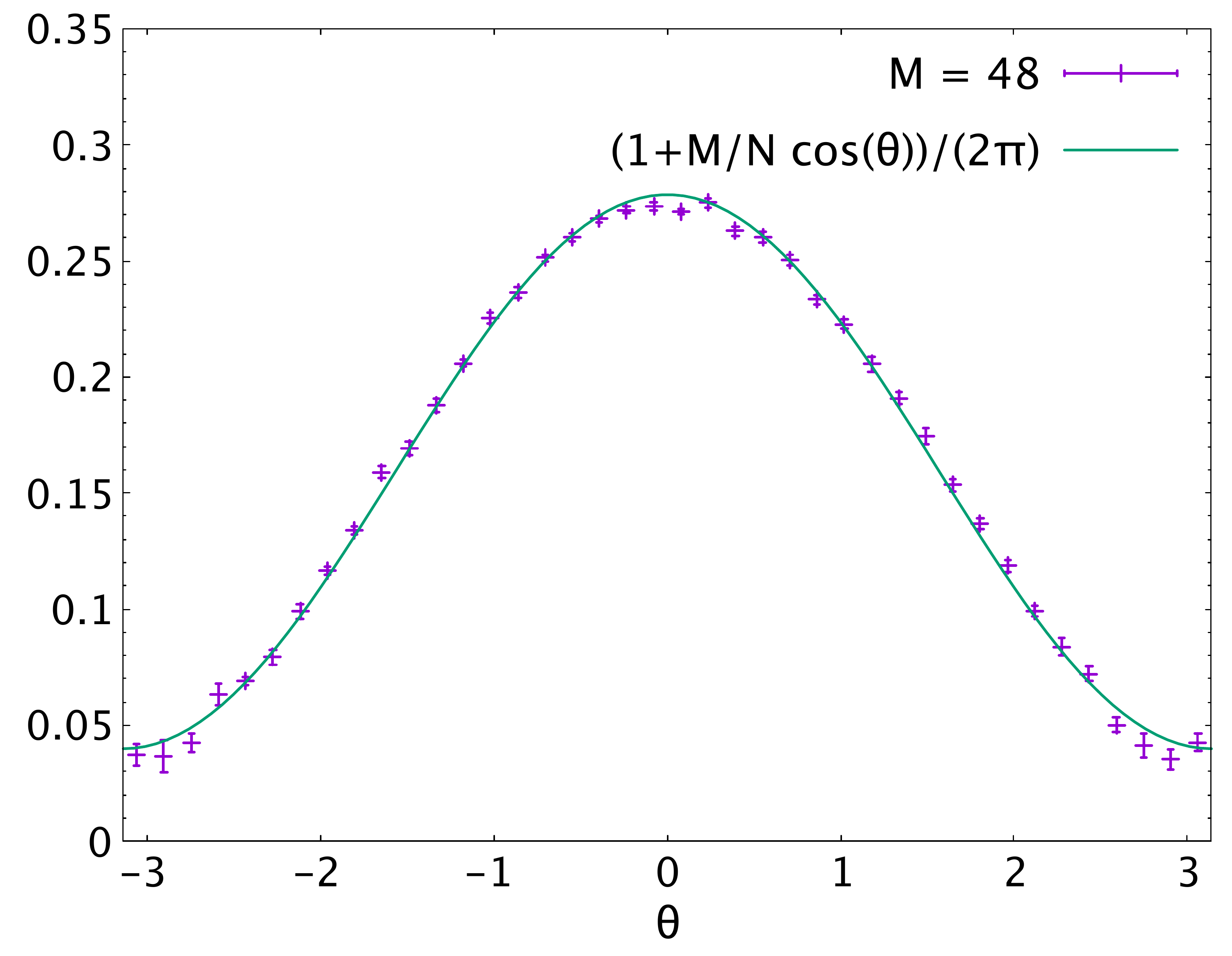}
	\end{subfigure}
	\caption{Yang-Mills matrix model, the distribution of the phases of the Polyakov loop $\rho^{\rm(P)}(\theta)$ obtained from constrained simulations, $N=64$, $L=24$, $M=16, 32,48$, and $T=0.885$.
The green lines are $\frac{1}{2\pi}\left(1+\frac{M}{N}\cos\theta\right)$.
The center symmetry is fixed sample by sample such that $P=|P|$.
The error bars in each figure are obtained by jackknife analysis.
}\label{fig:Pol-constrained}
\end{figure}

\begin{figure}[htbp]
\centering
	\begin{subfigure}[b]{0.475\textwidth}
		\includegraphics[width=\textwidth]{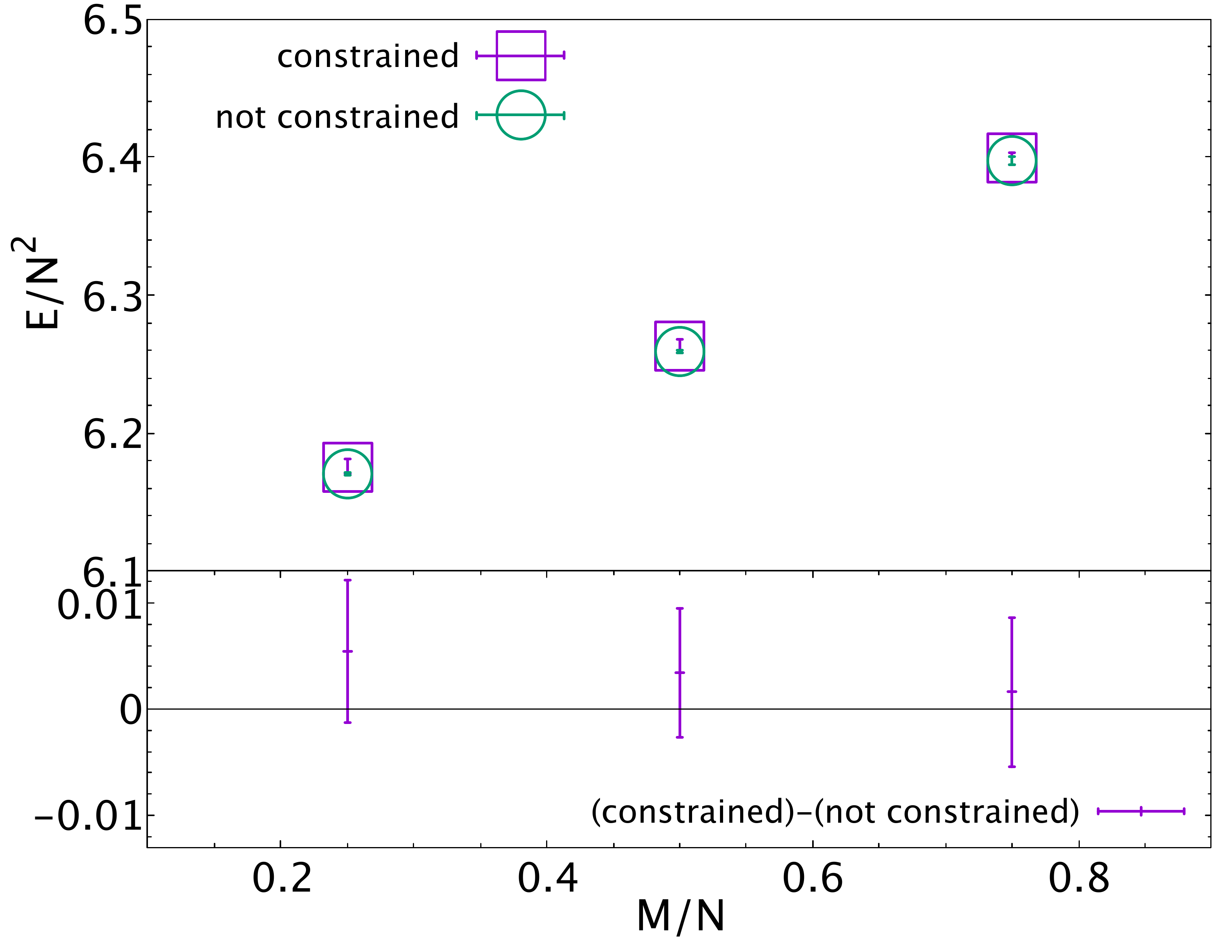}
		\caption{Energy}
	\end{subfigure}
	\begin{subfigure}[b]{0.475\textwidth}
		\includegraphics[width=\textwidth]{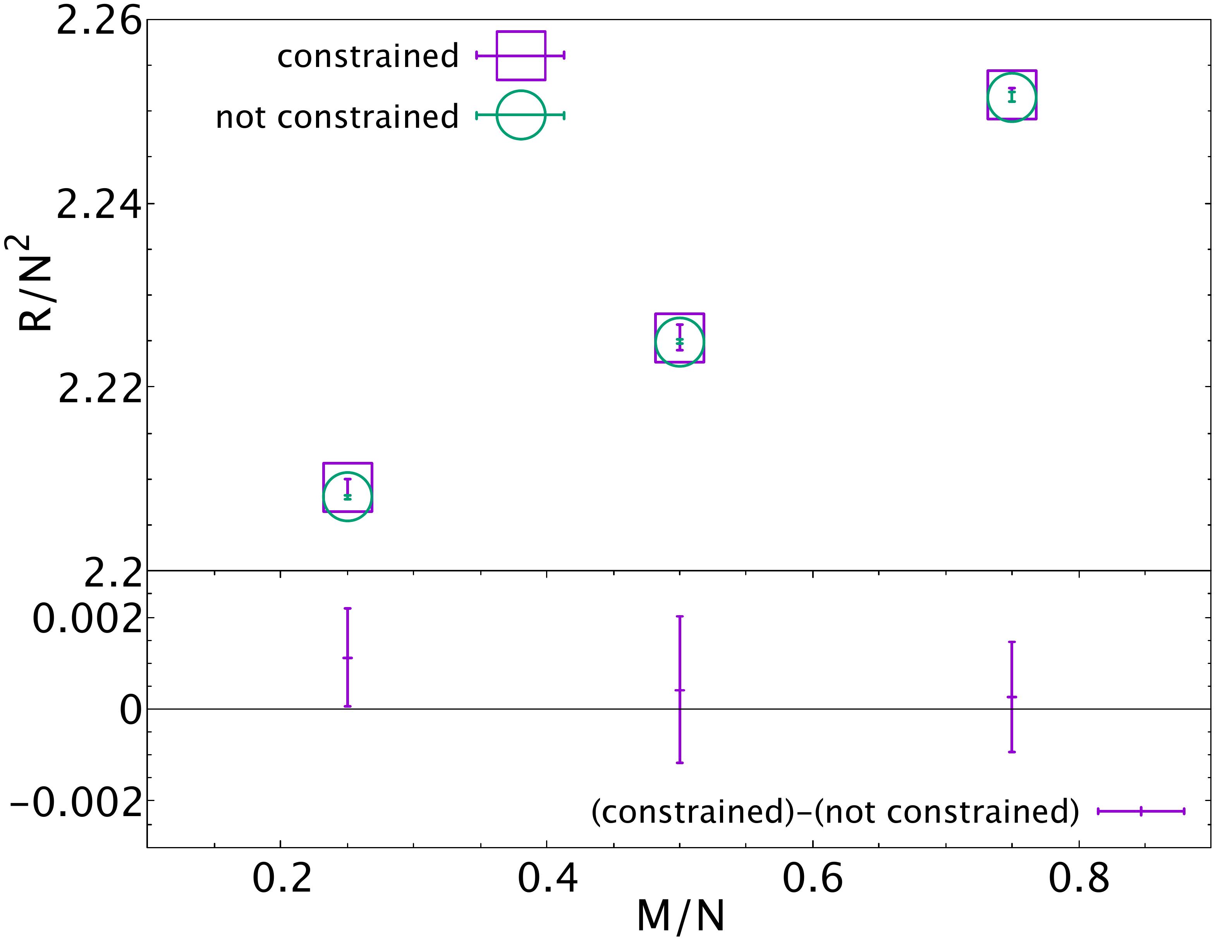}
		\caption{Extent of space}
	\end{subfigure}
	\caption{Yang-Mills matrix model, $N=64$, $L=24$, and $T=0.885$. Comparison between observables from simulations with and without the constraint term \eqref{constraint-partially-deconfined}.
The error bars in each figure are obtained by jackknife analysis.
}\label{fig:constrained-vs-unconstrained}
\end{figure}

Let us discuss $N=64$ as an example.
The distribution of the phases of the Polyakov loop $\rho^{\rm(P)}(\theta)$
obtained from the constrained simulations at $T=0.885$ with 24 lattice sites
is plotted in Fig.~\ref{fig:Pol-constrained}.
The agreement with $\frac{1}{2\pi}\left(1+\frac{M}{N}\cos\theta\right)$ is very good.
In Fig.~\ref{fig:constrained-vs-unconstrained}, $E/N^2$ and $R/N^2$ calculated with and without the constraint are compared.
We can see good agreement between them.
These observations support the expectation that the constraint term in Eq.~\eqref{constraint-partially-deconfined} does not alter the theory.

\subsubsection{Distribution of $X_{I,ij}$}\label{sec:constrained-X-distribution}
\hspace{0.51cm}

\begin{figure}[htbp]
\centering
	\begin{subfigure}[b]{0.475\textwidth}
		\includegraphics[width=\textwidth]{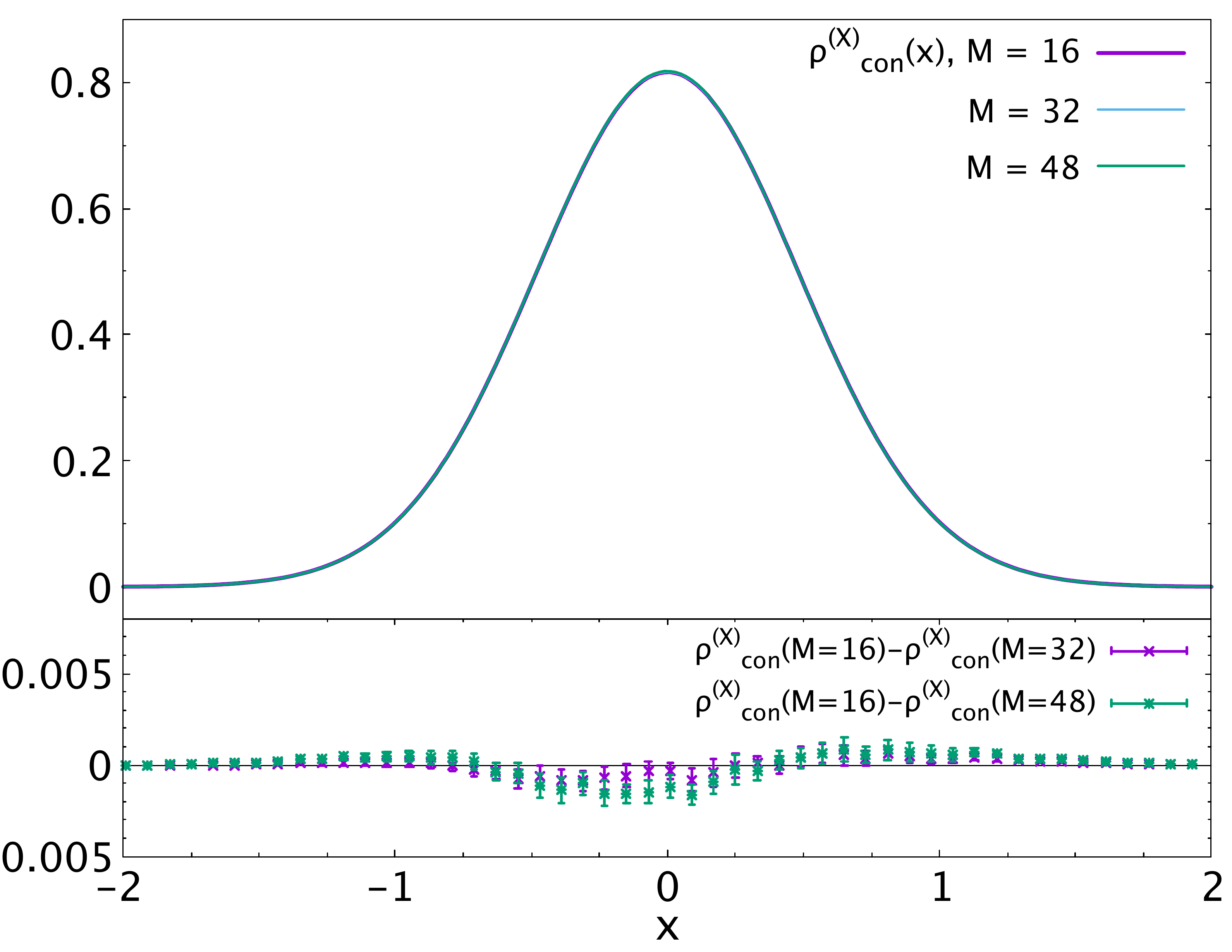}
		\caption{$\rho^{\rm(X)}_{\rm con}(x)$}
	\end{subfigure}
	\begin{subfigure}[b]{0.475\textwidth}
		\includegraphics[width=\textwidth]{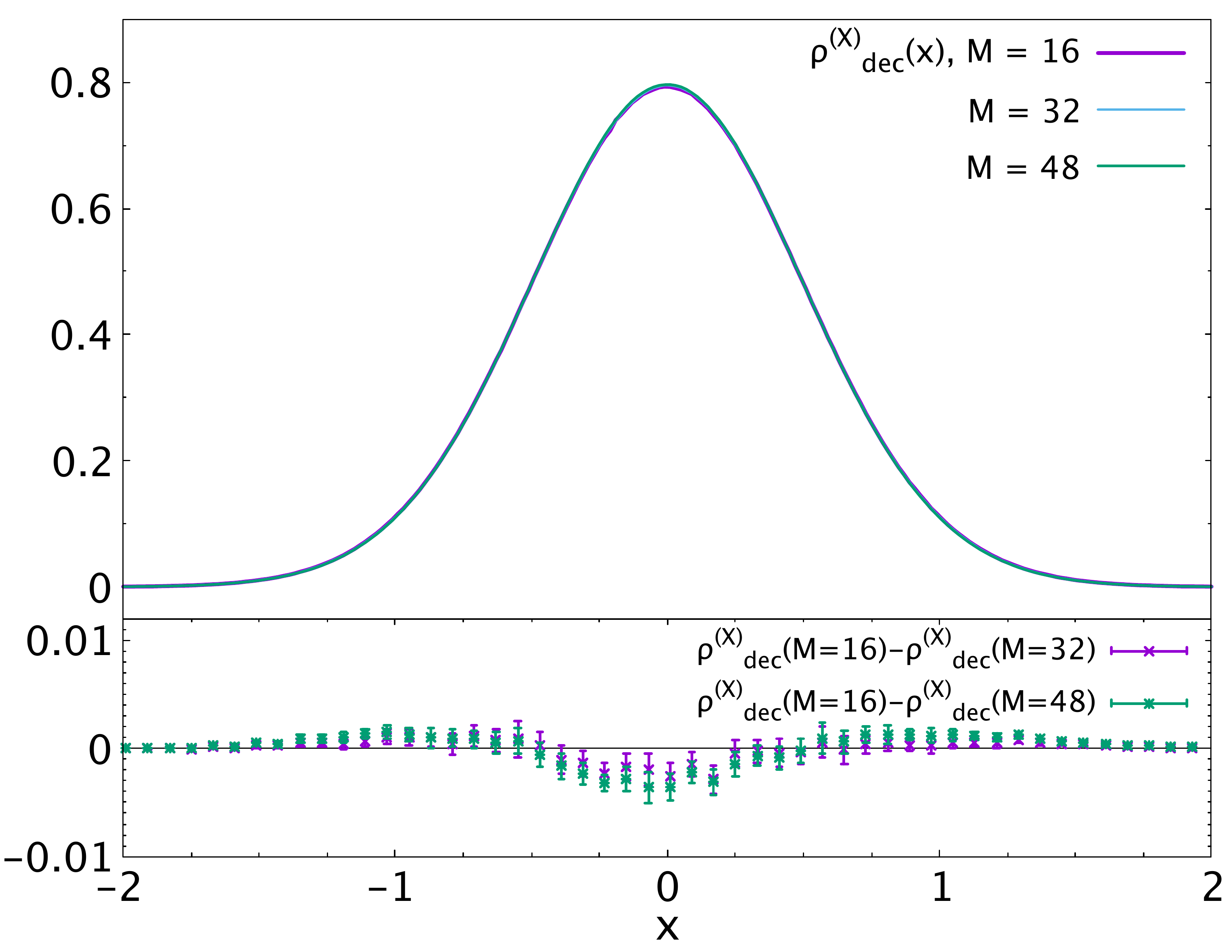}
		\caption{$\rho^{\rm(X)}_{\rm dec}(x)$}
	\end{subfigure}
	\caption{Yang-Mills matrix model, constrained simulations, $N=64$, $L=24$, $T=0.885$.
$\rho^{\rm(X)}_{\rm con}(x)$ and $\rho^{\rm(X)}_{\rm dec}(x)$ obtained from different values of $N$ and $M$.
The error bars in each figure are obtained by jackknife analysis.
}\label{fig:constrained_rho_con_dec}
\end{figure}

\begin{figure}[htbp]
\centering
	\begin{subfigure}[b]{0.475\textwidth}
		\includegraphics[width=\textwidth]{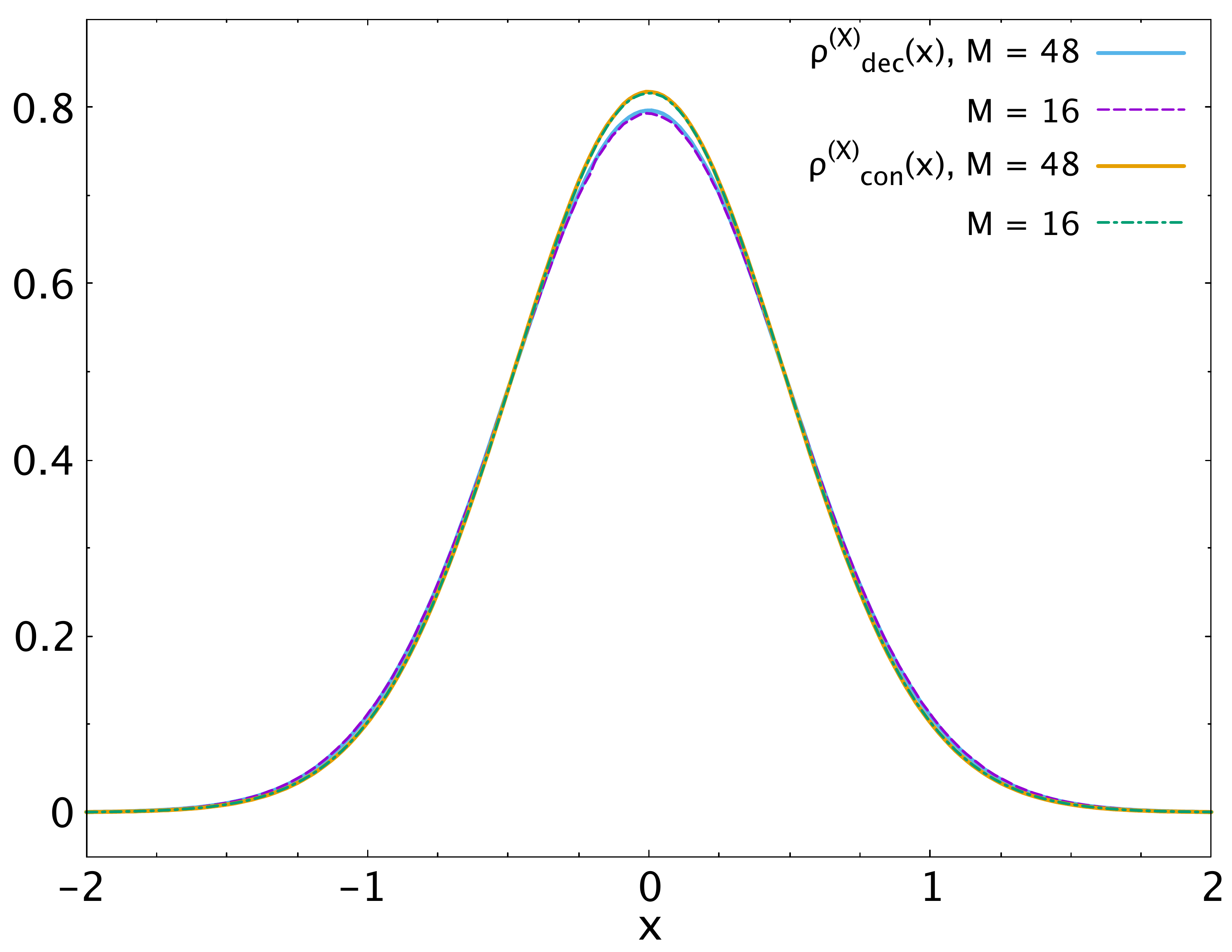}
		\caption{Distributions}
	\end{subfigure}
	\begin{subfigure}[b]{0.475\textwidth}
		\includegraphics[width=\textwidth]{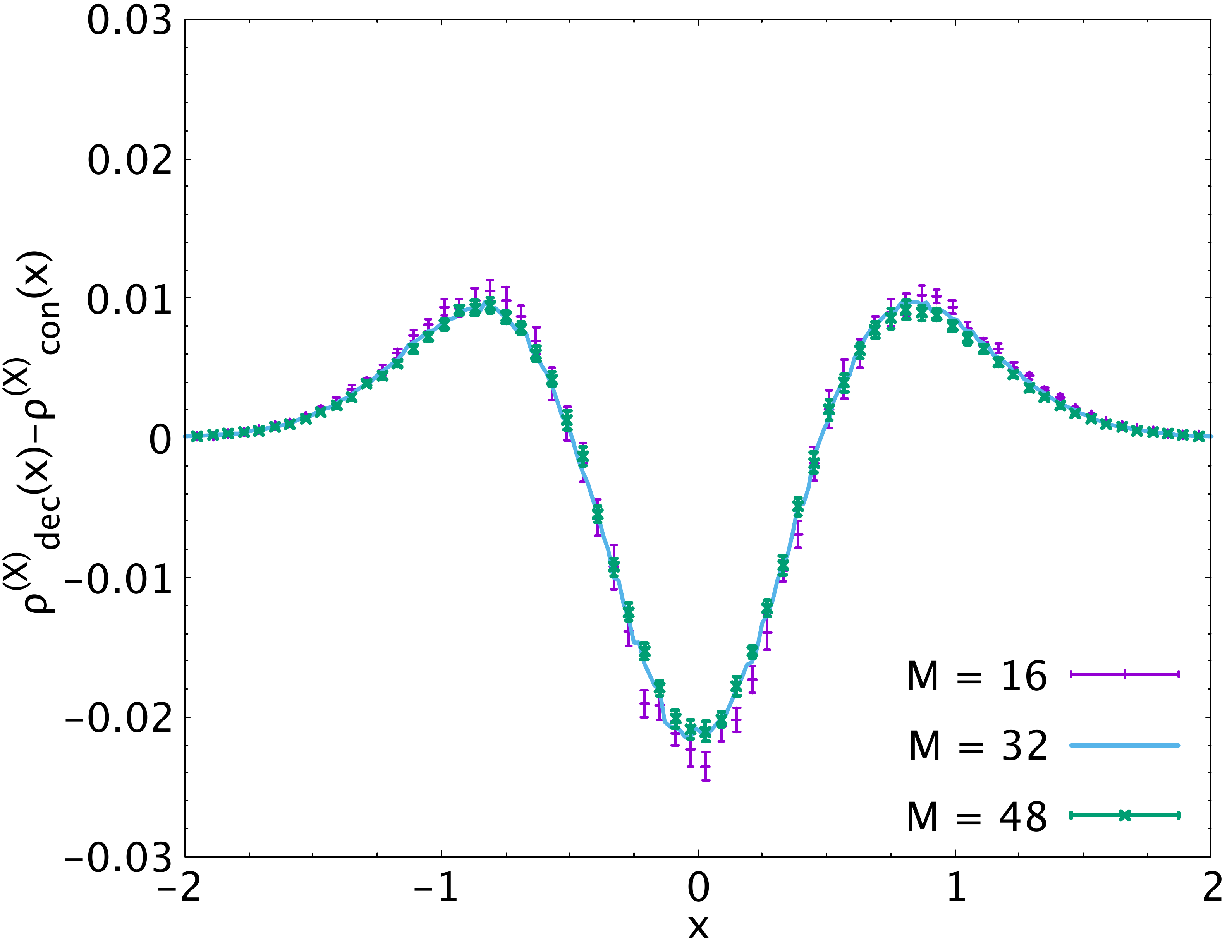}
		\caption{Difference}
	\end{subfigure}
	\caption{Yang-Mills matrix model, constrained simulations, $N=64$, $L=24$, $T=0.885$.
Comparison between $\rho^{\rm(X)}_{\rm dec}(x)$ and $\rho^{\rm(X)}_{\rm con}(x)$ showed in Fig.~\ref{fig:constrained_rho_con_dec}.
(a) A small difference between $\rho^{\rm(X)}_{\rm dec}(x)$ and $\rho^{\rm(X)}_{\rm con}(x)$ is visible.
(b) The difference $\rho^{\rm(X)}_{\rm dec}(x)-\rho^{\rm(X)}_{\rm con}(x)$ is significantly larger than the error bars of $\rho^{\rm(X)}_{\rm dec}(x)$ and $\rho^{\rm(X)}_{\rm con}(x)$.
The error bars in each figure are obtained by jackknife analysis.
}\label{fig:constrained_rho_con_dec_1}
\end{figure}

In Sec.~\ref{sec:scalars-bBFSS-unconstrained}, we determined $\rho^{\rm(X)}_{\rm dec}(x)$ and $\rho^{\rm(X)}_{\rm con}(x)$ by using Eq.~\eqref{eq:separation_rho_scalar}, assuming they are independent of $M$.
Now we can determine those distributions much more easily, without assuming $M$-independence: $\rho^{\rm(X)}_{\rm dec}(x)$ can be determined from $X_{I,jk}$ with $1\le j,k\le M$, and $\rho^{\rm(X)}_{\rm con}(x)$ can be determined from the rest.
The results are shown in Figs.~\ref{fig:constrained_rho_con_dec}, \ref{fig:constrained_rho_con_dec_1} and Table~\ref{tab:variance_const}.
\begin{table}[htb]
\begin{center}
\caption{The variances of $\rho^{\rm(X)}_{\rm dec}$ and $\rho^{\rm(X)}_{\rm con}$ in the constrained simulation, $N=48, 64$ and 128, $T=0.885$, with 24 lattice sites.
The last column is the variance of the distribution of the off-diagonal block of the confined sector.
}
\label{tab:variance_const}
\begin{tabular}{|c|c||c|c|c|c|}\hline
	$N$ & $M$	&	$\sigma^2_{\rm dec}$ 	&	$\sigma^2_{\rm con}$ & $\sigma^2_{\rm con,\ off\ diagonal}$	&\# configs. \\ \hline	\hline
	48	& 12 	& 0.2588(5) 	& 0.2442(2) 	& 0.2438(8)  & 1500\\\hline
	64	& 16	& 0.2581(2)	& 0.2446(1)	& 0.2439(3)& 1500	\\\hline
	128	& 32	& 0.2582(2) 	& 0.2445(1)	& 0.2439(1)	& 500  	\\\hline\hline
	48 	& 24  	& 0.2568(1) 	& 0.2439(1)	& 0.2433(2) & 1500\\\hline
	64	& 32	& 0.2567(1)	& 0.2441(1)	& 0.2434(2)  & 1500\\\hline
	128	& 64	& 0.2566(1)	& 0.2438(1) 	& 0.2431(1)	&500 	\\\hline\hline
	48 	& 36  	& 0.2557(2)	& 0.2438(3) 	& 0.2433(5) & 1500\\\hline
	64	& 48	& 0.2555(1)	& 0.2434(1)	& 0.2430(2)	& 1500	\\\hline
	128	& 96	& 0.2556(1) 	& 0.2433(1)	& 0.2428(1)	&500 \\\hline
\end{tabular}
\end{center}
\end{table}

\begin{figure}[htbp]
	\centering
		\begin{subfigure}[b]{0.475\textwidth}
			\includegraphics[width=\textwidth]{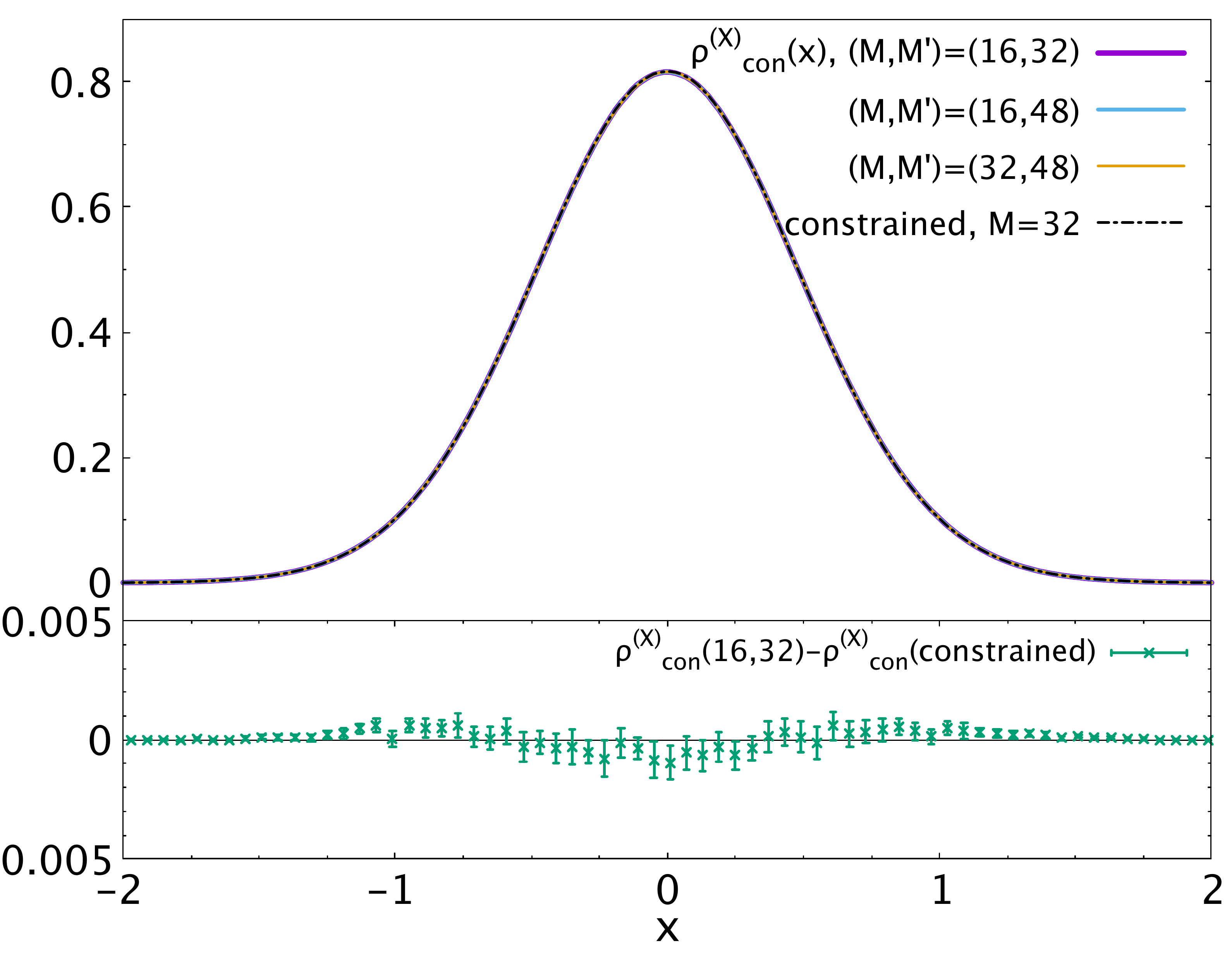}
			\caption{$\rho^{\rm(X)}_{\rm con}(x)$}
		\end{subfigure}
		\begin{subfigure}[b]{0.475\textwidth}
			\includegraphics[width=\textwidth]{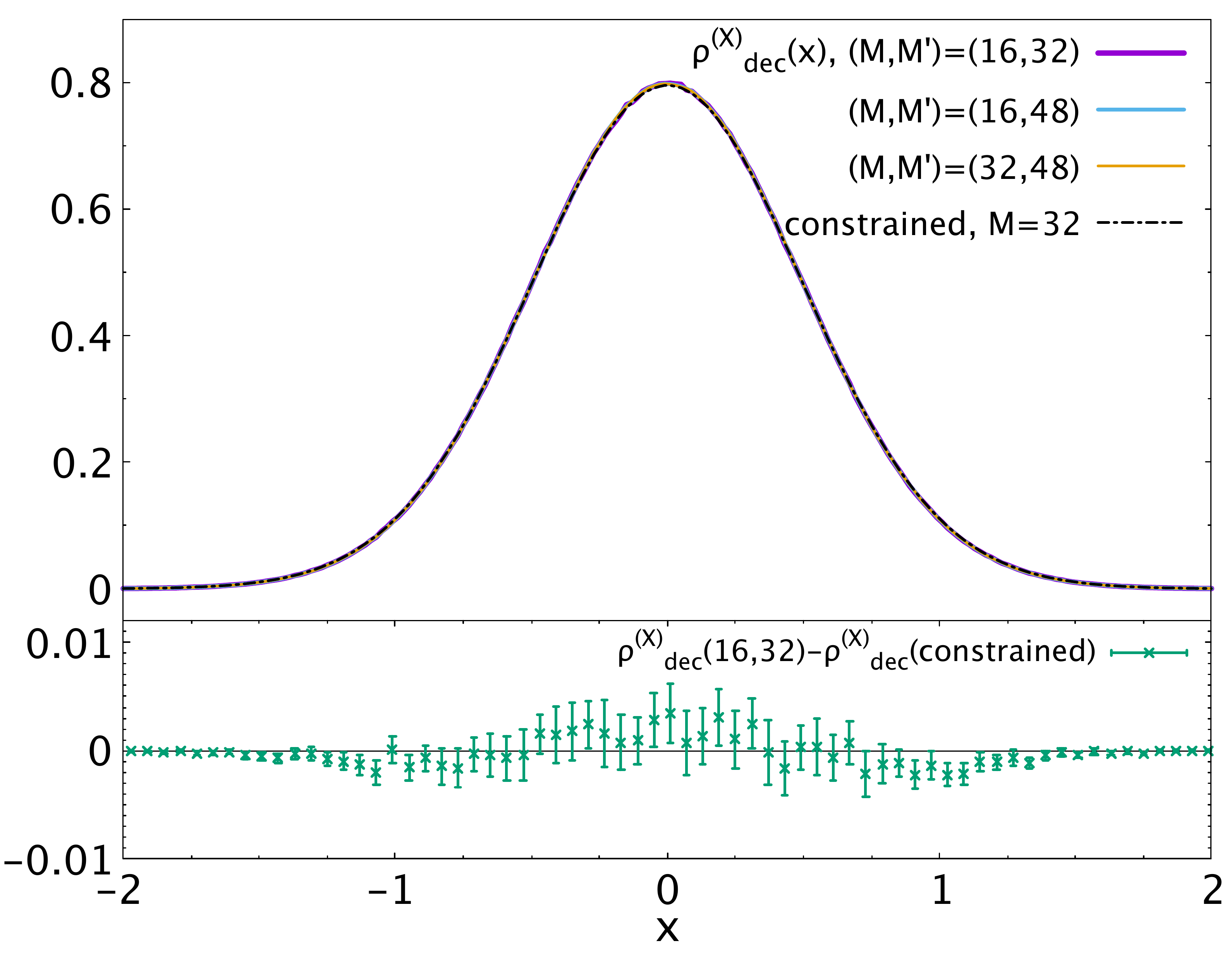}
			\caption{$\rho^{\rm(X)}_{\rm dec}(x)$}
		\end{subfigure}
\caption{Yang-Mills matrix model, the comparison between constrained and unconstrained simulations; $N=64$, $L=24$, and $T=0.885$.
The error bars in each figure are obtained by jackknife analysis.
}\label{fig:tentative-rho-constrained-unconstrained-comparison}
\end{figure}

%
%
The error bars are well under control, and we can see a clear difference between $\rho^{\rm(X)}_{\rm dec}(x)$ and $\rho^{\rm(X)}_{\rm con}(x)$.
In Fig.~\ref{fig:tentative-rho-constrained-unconstrained-comparison}, the distributions obtained from the constrained and unconstrained simulations are compared.
We can see reasonably good agreement.
These observations provide us with an explicit confirmation of the $M\times M$-block structure.

By looking at the values of the variances in Table~\ref{tab:variance_const} closely, we can see a weak $M$-dependence.
This $M$-dependence may or may not survive in the continuum limit.
We also studied the off-diagonal blocks in the confined sector separately.
The result is shown in the same table.
We can see a similar $M$-dependence.
In Sec.~\ref{sec:simulation_summary}, we will 
discuss this observation.
We will explain that our conclusions do not change, even if such an $M$-dependence actually exists.

\subsubsection{Correlation between scalars and gauge field}\label{sec:theta-vs-K-constrained}
\hspace{0.51cm}
Next let us study the correlation between $\theta_i$ and $K_i$.
We consider two options, with and without taking into account a possible $M$-dependence of
$\rho^{\rm(X)}_{\rm con}(x)$ and $\rho^{\rm(X)}_{\rm dec}(x)$.
Both options describe the data rather precisely.

\begin{figure}[htbp]
\centering
	\begin{subfigure}[b]{0.475\textwidth}
		\includegraphics[width=\textwidth]{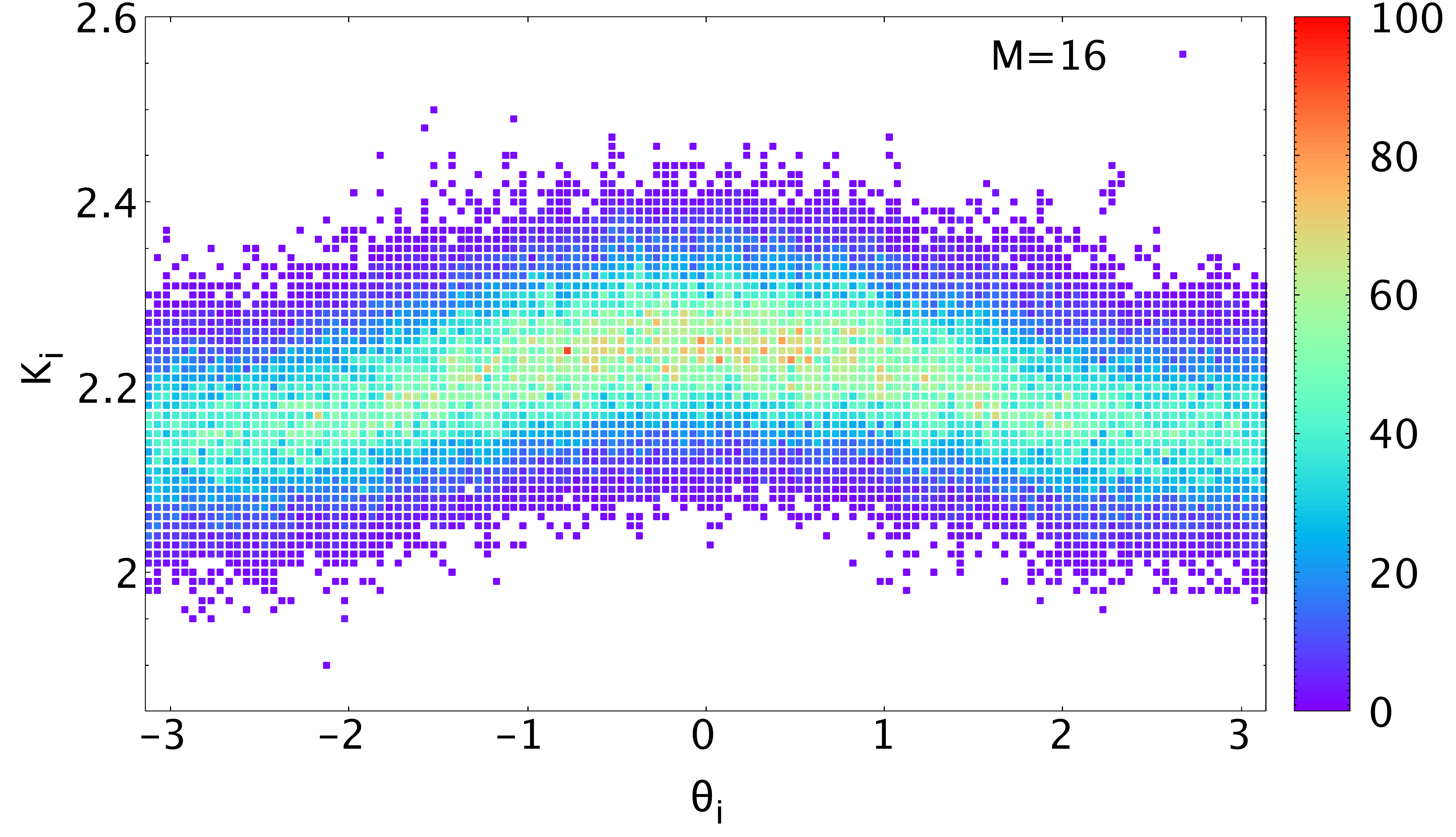}
	\end{subfigure}
	\begin{subfigure}[b]{0.475\textwidth}
		\includegraphics[width=\textwidth]{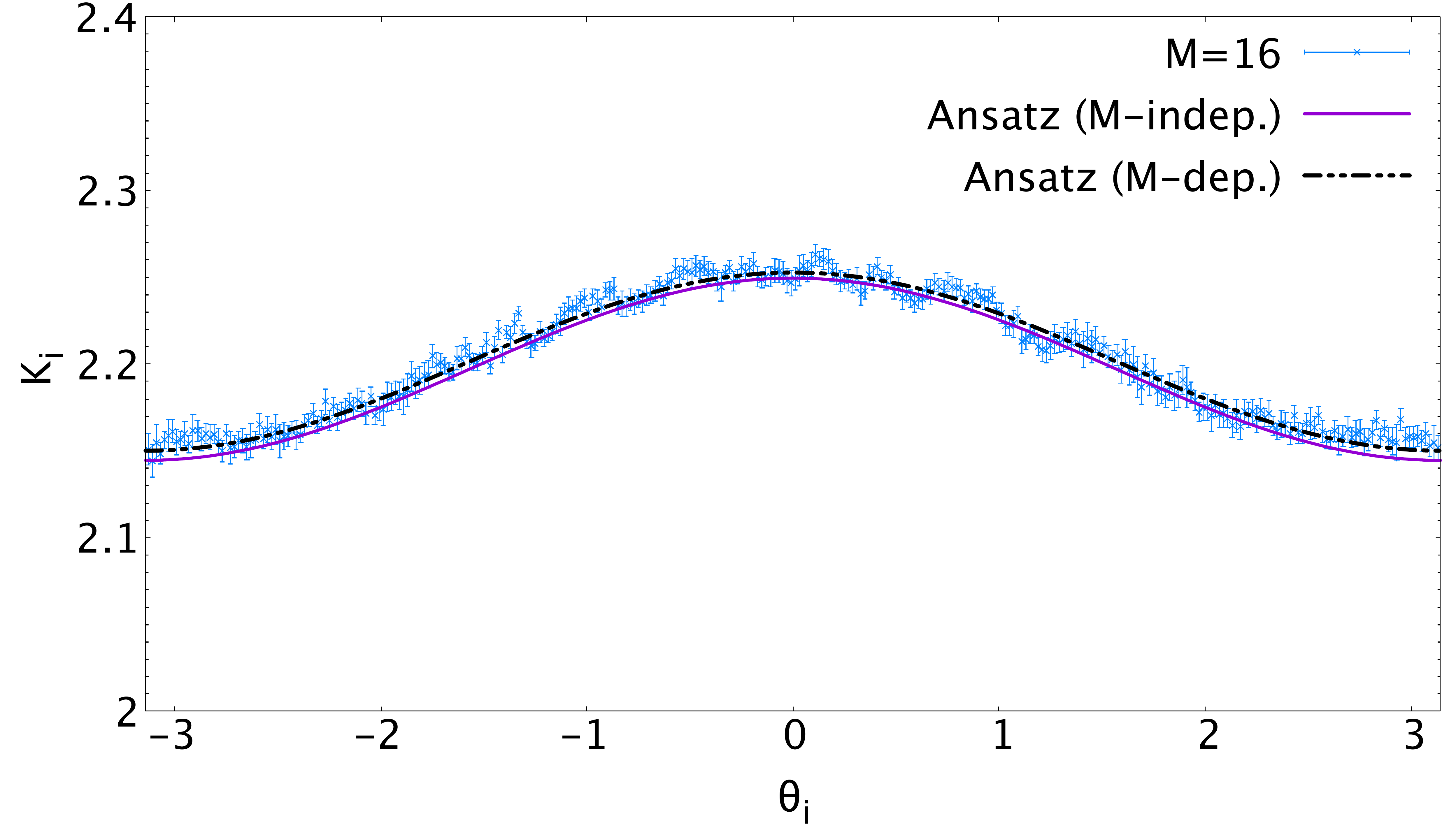}
	\end{subfigure}
	\begin{subfigure}[b]{0.475\textwidth}
		\includegraphics[width=\textwidth]{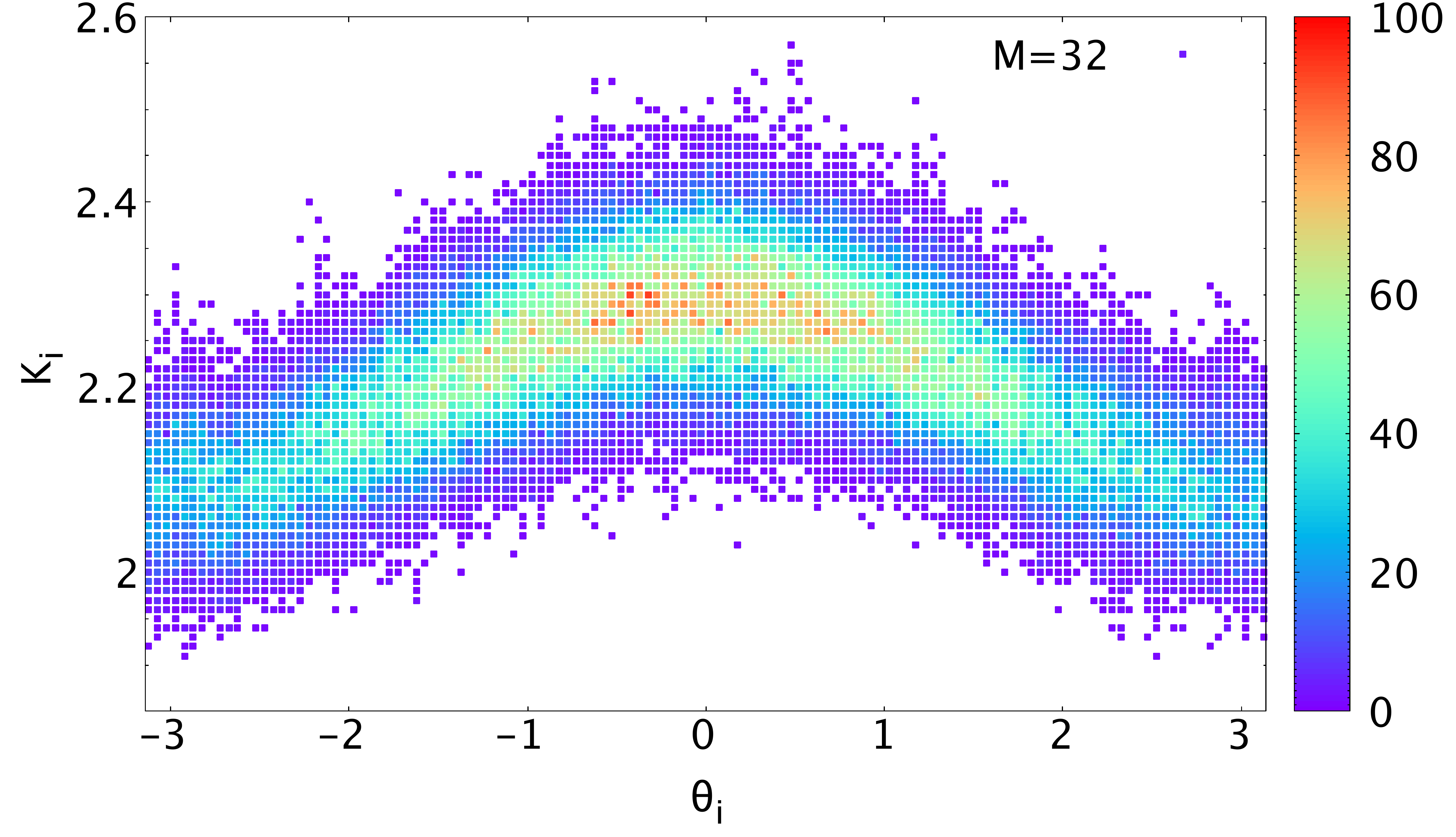}
	\end{subfigure}
	\begin{subfigure}[b]{0.475\textwidth}
		\includegraphics[width=\textwidth]{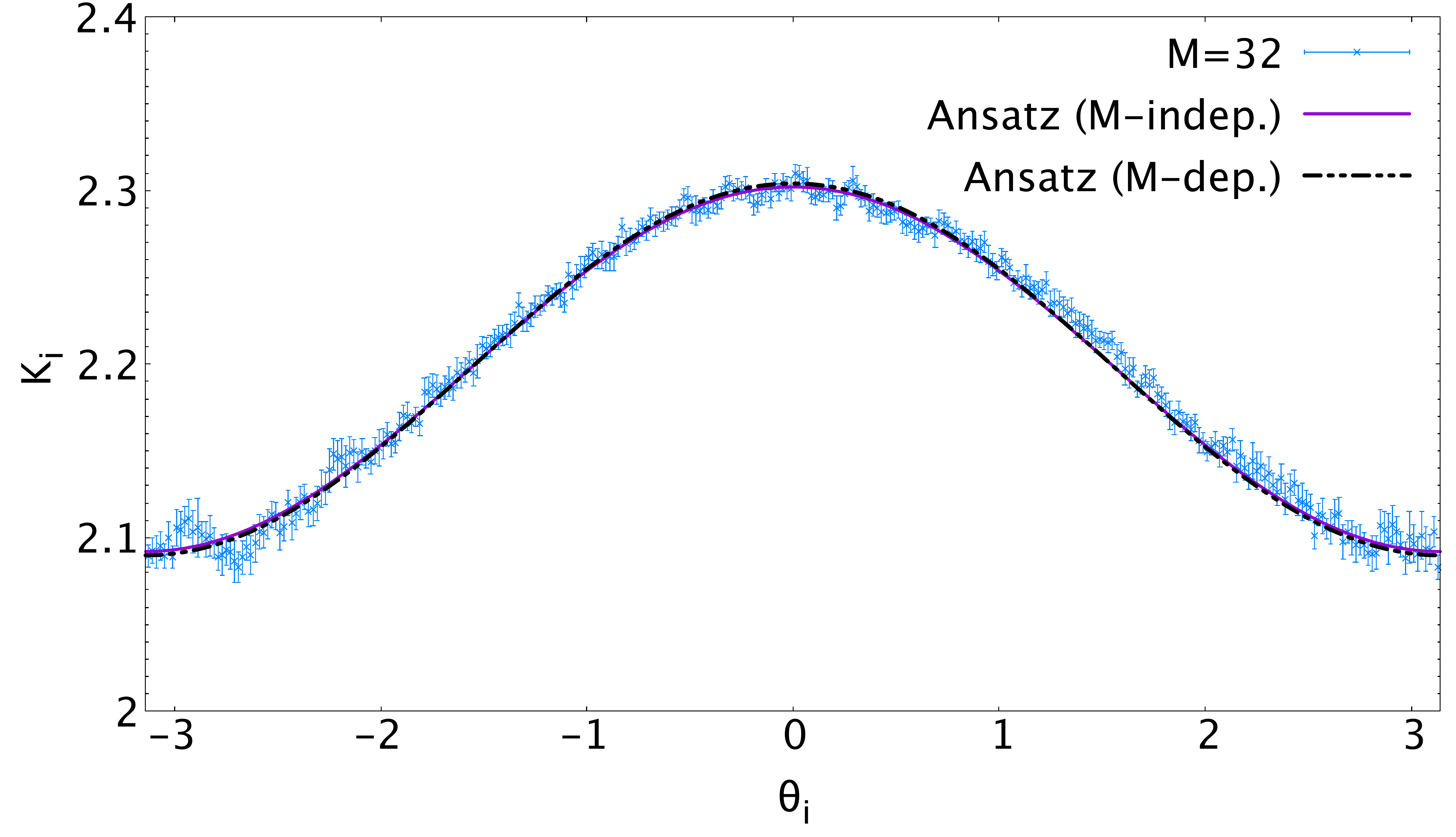}
	\end{subfigure}
	\begin{subfigure}[b]{0.475\textwidth}
		\includegraphics[width=\textwidth]{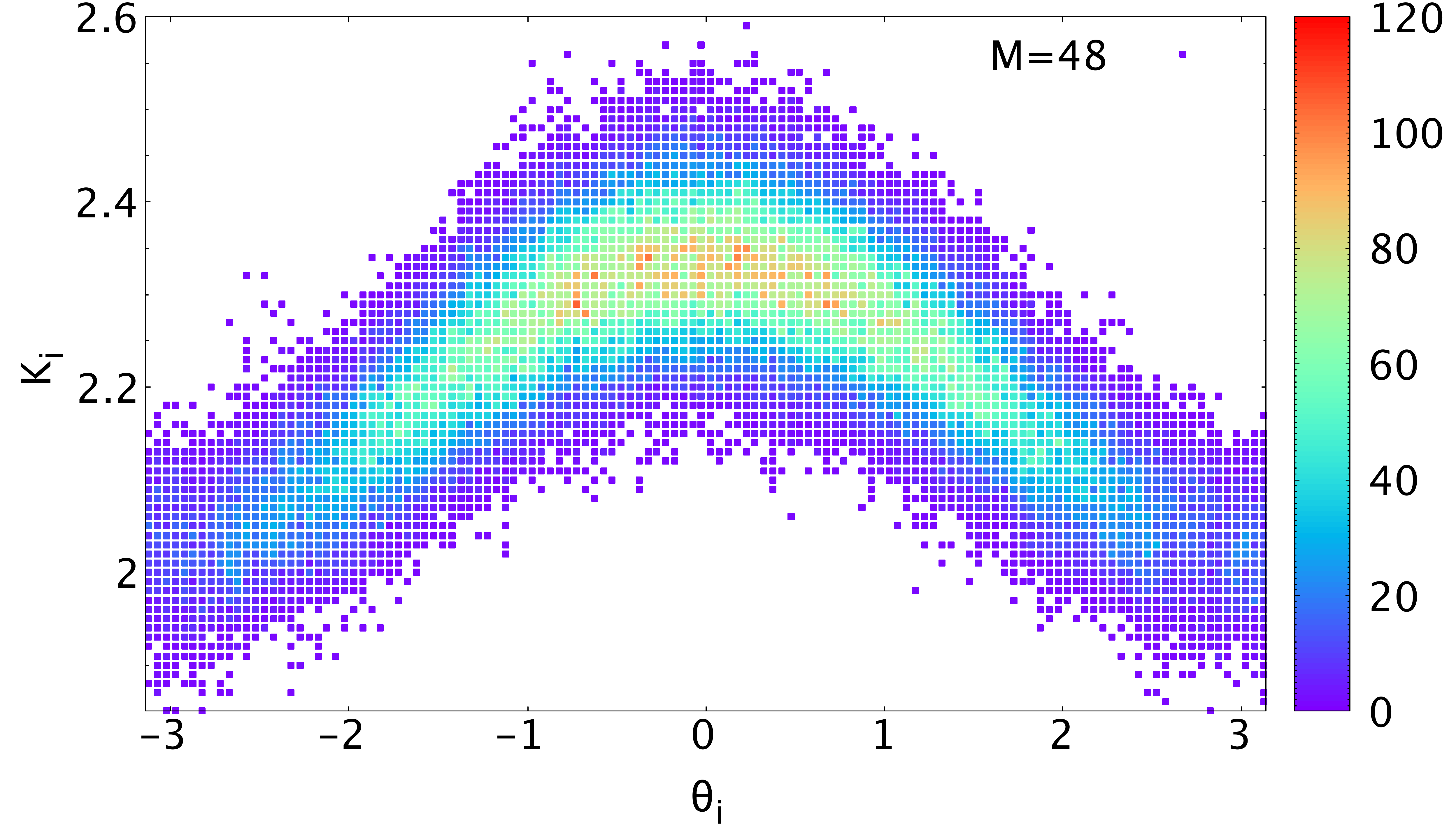}
		\caption{2D histogram}
	\end{subfigure}
	\begin{subfigure}[b]{0.475\textwidth}
		\includegraphics[width=\textwidth]{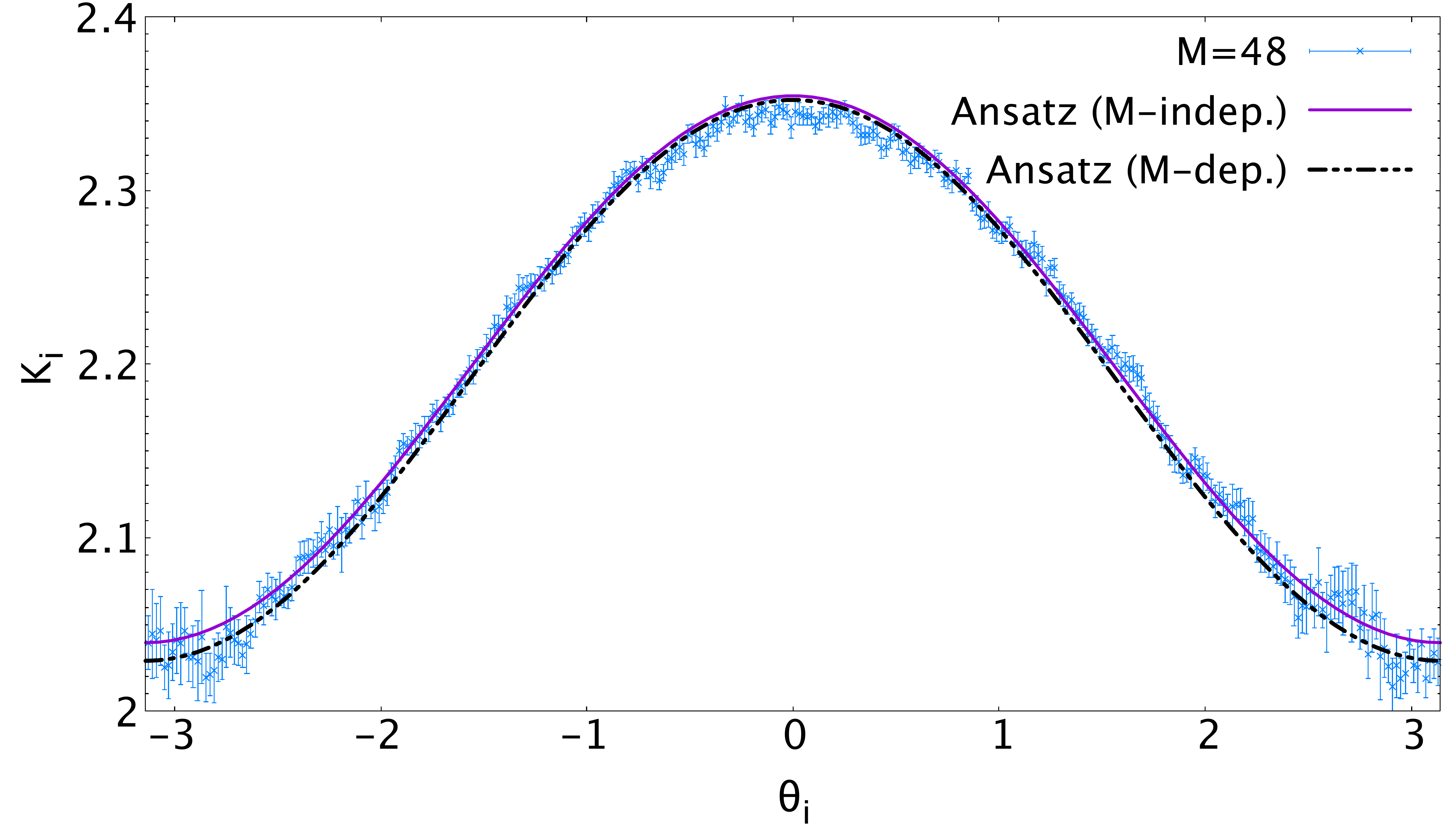}
		\caption{Binned histogram}
	\end{subfigure}
\caption{Yang-Mills matrix model, $\theta_i$ vs $K_i$, in a constrained simulation with $N=64$, $L=24$, $M=16, 32$ and $48$ at $T=0.885$.
We used 1500 configurations for each $M$ (the total number of points is $64\times 1500$).
The center symmetry is fixed sample by sample such that $P=|P|$.
(a) The two-dimensional histograms of ($\theta_i, K_i$).
Data points in the confined and deconfined sectors are shown together.
(b) The averaged $K_i$ within the narrow bin $\Delta\theta=0.02$.
The magenta lines are Eq.~\eqref{K-vs-theta-BFSS}, with $r_0$ and $r_1$ from Fig.~\ref{fig:theta-vs-K_bBFSS-Pconst}.
The black lines are Eq.~\eqref{K-vs-theta-with-M-dependence}, with variances from Table~\ref{tab:variance_const}.
The error bars in each figure are obtained by jackknife analysis.
}\label{fig:theta-vs-K-BFSS-constrained}
\end{figure}

\begin{figure}[htbp]
\centering
	\begin{subfigure}[b]{0.475\textwidth}
		\includegraphics[width=\textwidth]{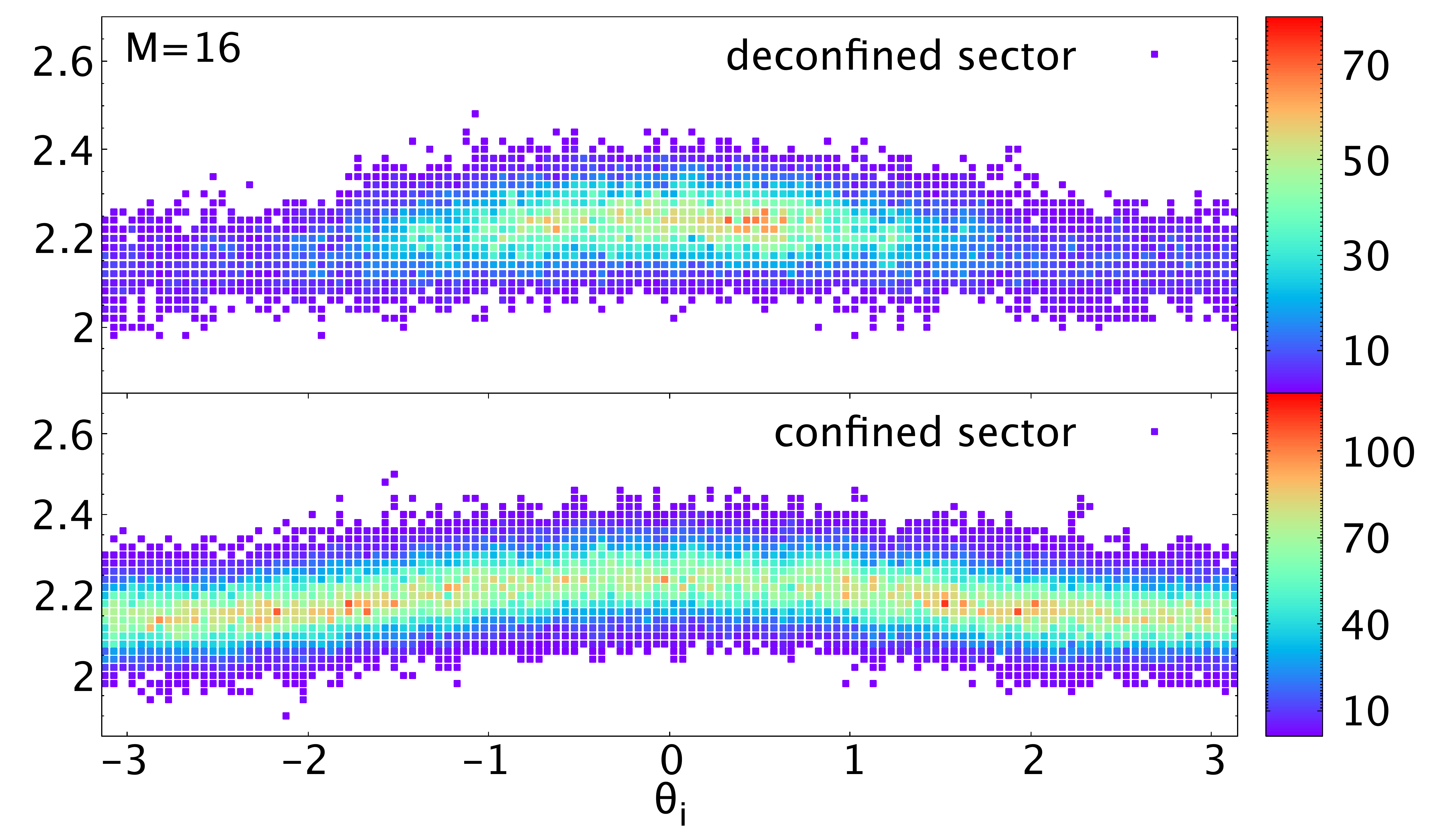}
	\end{subfigure}
	\begin{subfigure}[b]{0.475\textwidth}
		\includegraphics[width=\textwidth]{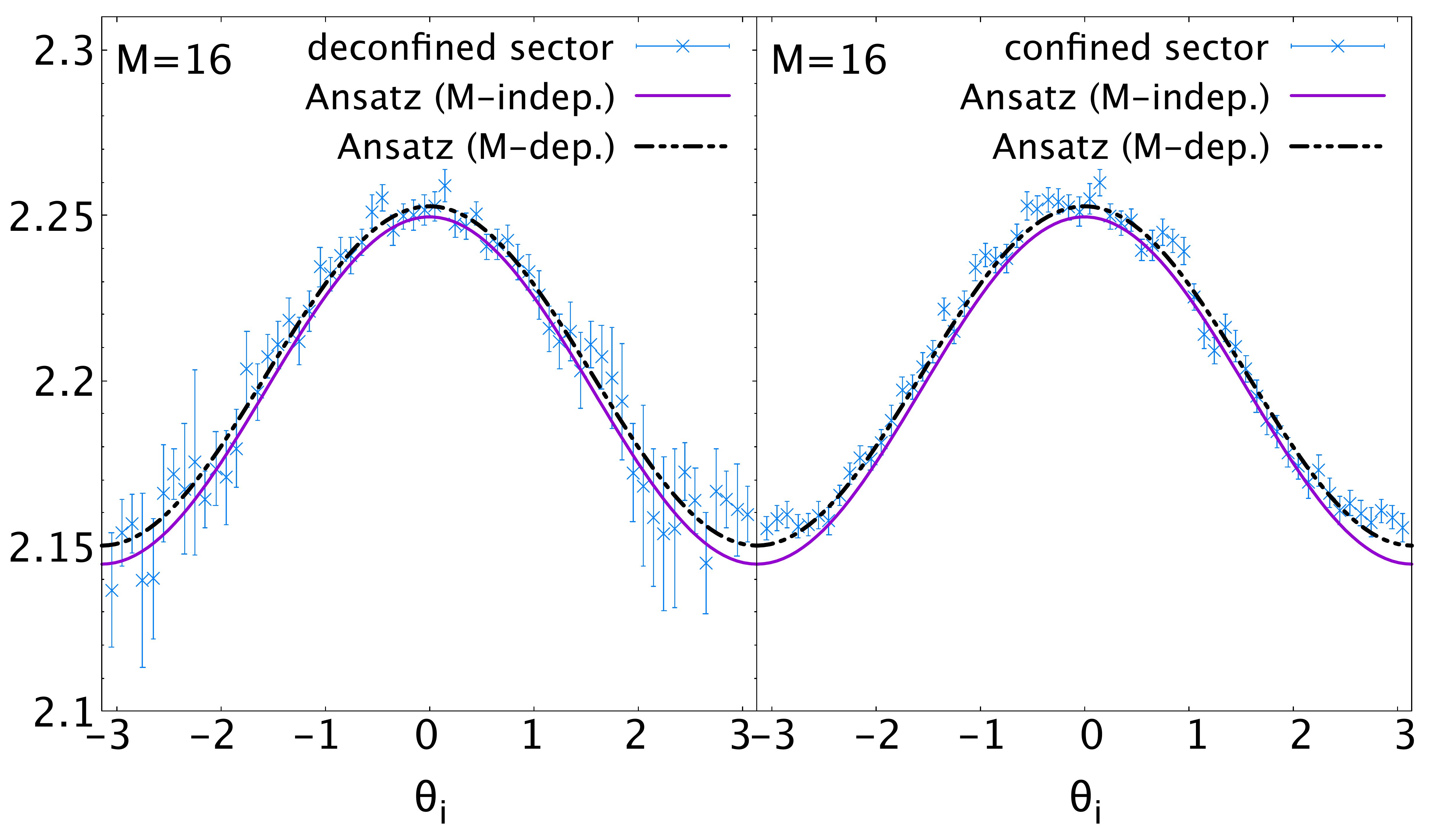}
	\end{subfigure}
	\begin{subfigure}[b]{0.475\textwidth}
		\includegraphics[width=\textwidth]{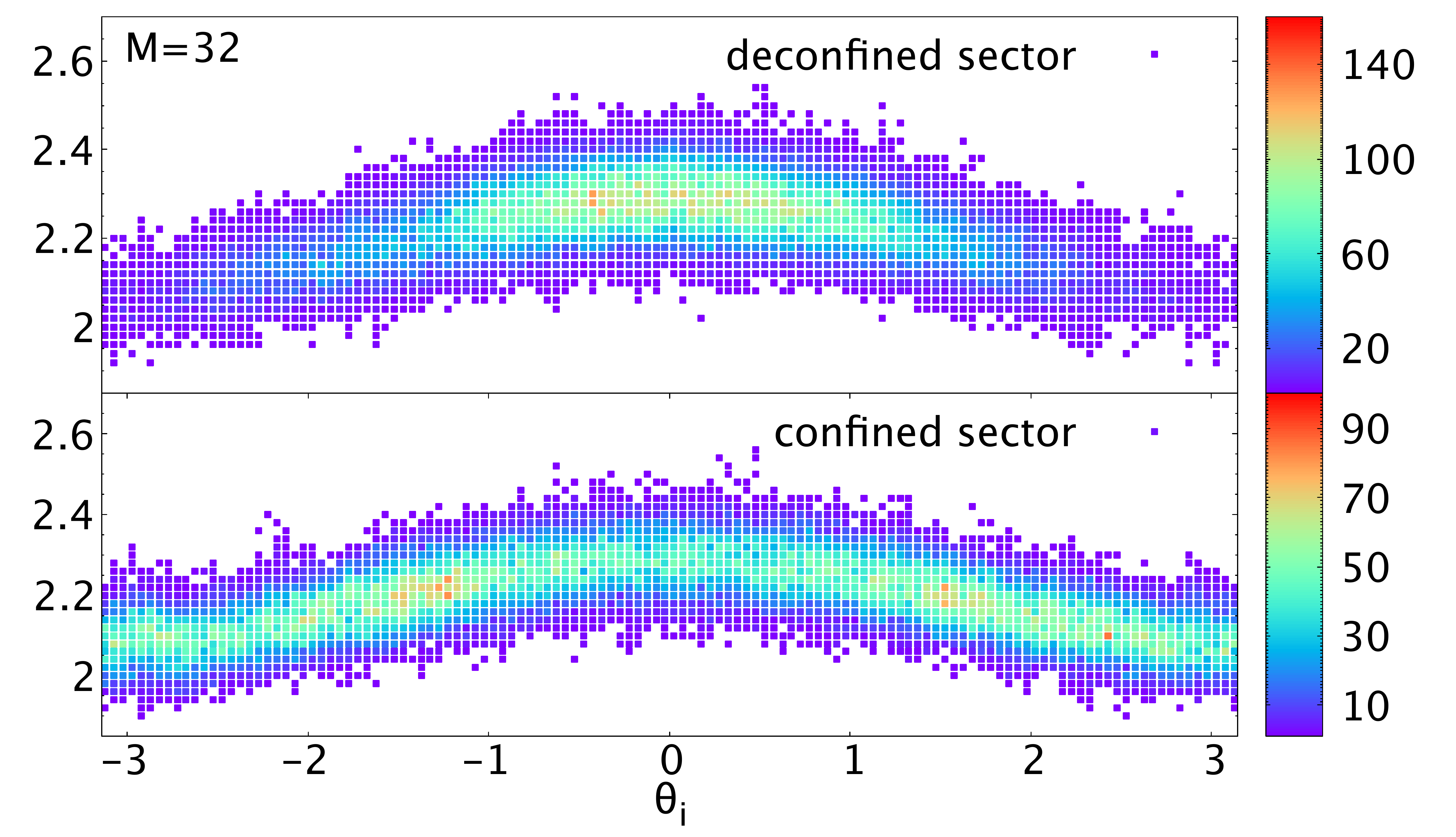}
	\end{subfigure}
	\begin{subfigure}[b]{0.475\textwidth}
		\includegraphics[width=\textwidth]{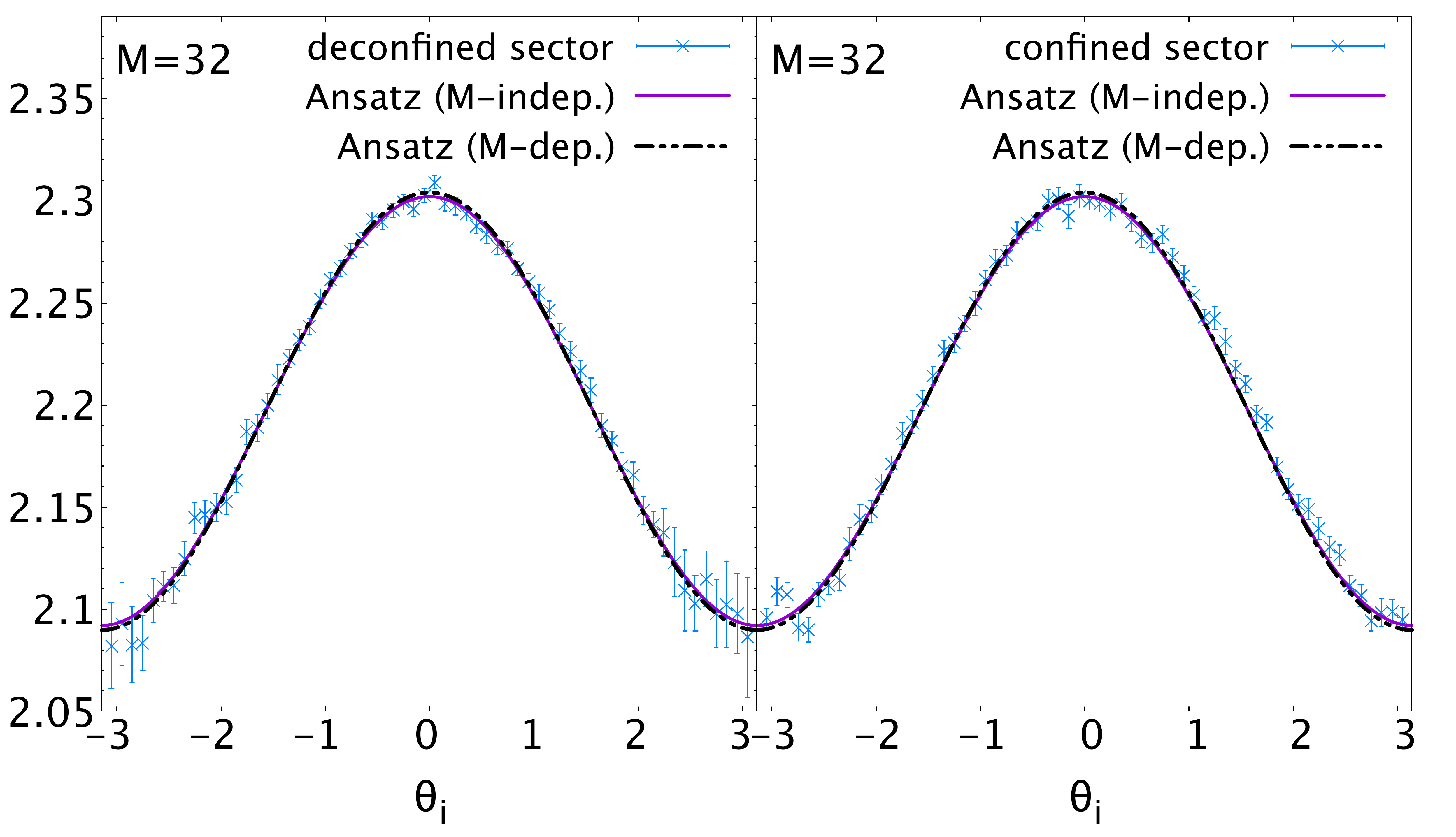}
	\end{subfigure}
	\begin{subfigure}[b]{0.475\textwidth}
		\includegraphics[width=\textwidth]{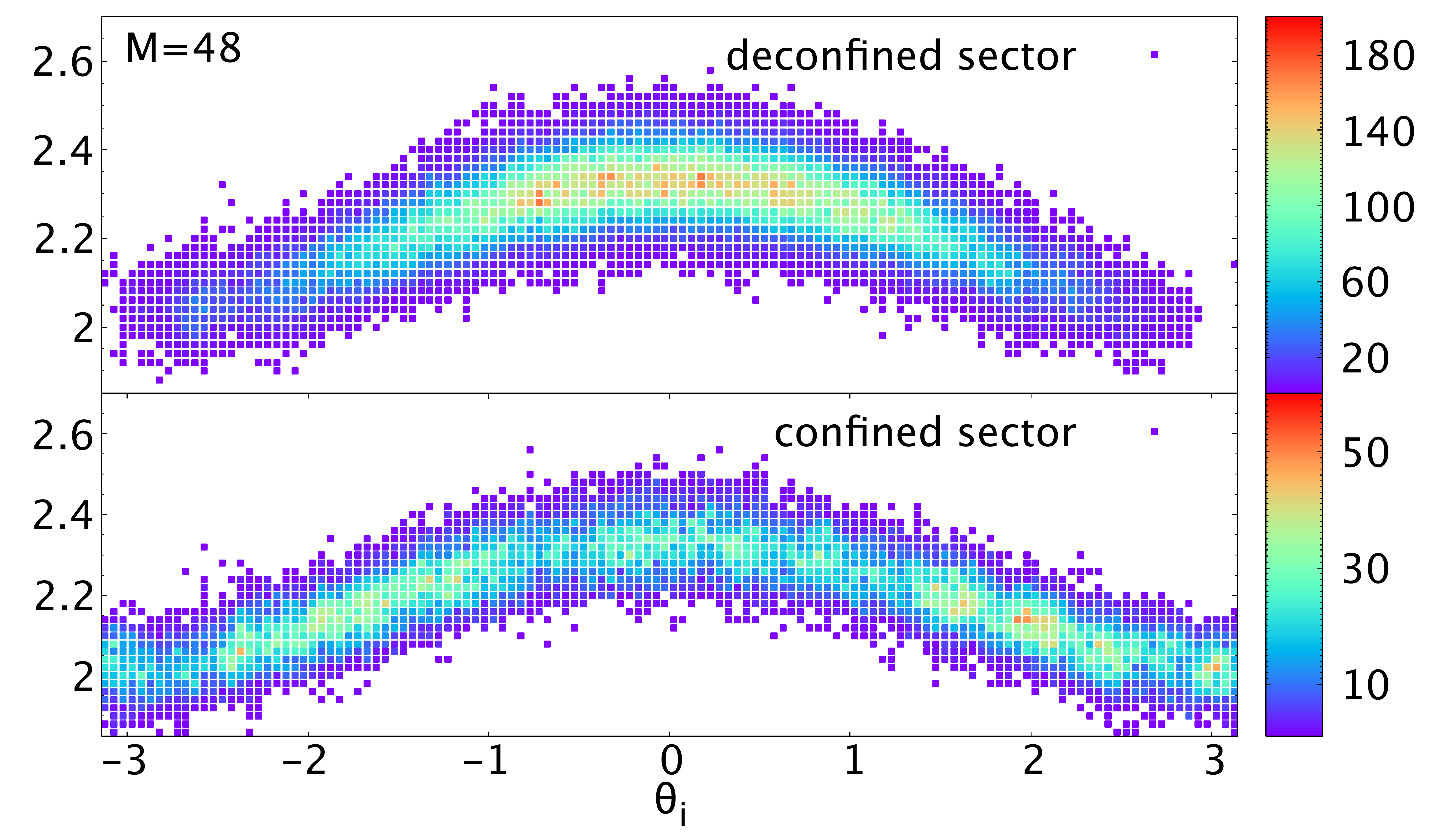}
		\caption{2D histogram}
	\end{subfigure}
	\begin{subfigure}[b]{0.475\textwidth}
		\includegraphics[width=\textwidth]{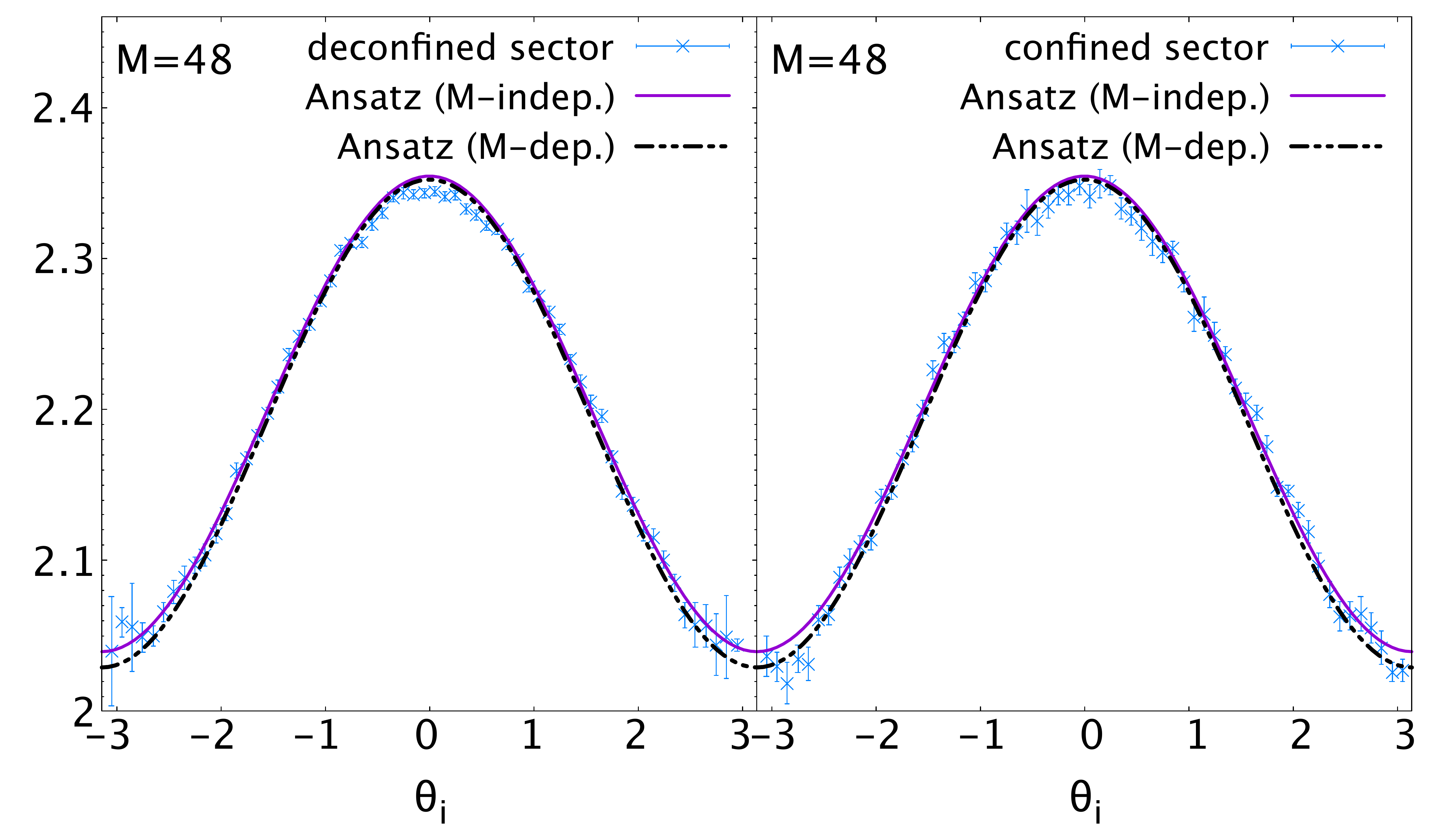}
		\caption{Binned histogram}
	\end{subfigure}
	\caption{Yang-Mills matrix model, $\theta_i$ vs $K_i$, in a constrained simulation with $N=64$, $L=24$, $M=16, 32$ and $48$ at $T=0.885$.
%
Same as Fig.~\ref{fig:theta-vs-K-BFSS-constrained}, except that the confined and deconfined sectors are shown separately, and $\Delta \theta = 0.1$ is used for (b).
The total number of the data points are $M\times 1500$ and $(64-M)\times 1500$ for the deconfined and confined sectors, respectively.
}\label{fig:theta-vs-K-BFSS-constrained_split}
\end{figure}

Firstly let us ignore a possible $M$-dependence.
We use the ansatz for the $\theta$-dependence of $K_i$ given by \eqref{K-vs-theta-BFSS}
and the values of  $r_0$ and $r_1$ used in Fig.~\ref{fig:theta-vs-K_bBFSS-Pconst}.
The results are shown in Fig.~\ref{fig:theta-vs-K-BFSS-constrained},
with the magenta lines.
Of course, Fig.~\ref{fig:theta-vs-K-BFSS-constrained} is essentially the same as Fig.~\ref{fig:theta-vs-K_bBFSS-Pconst},
except that the constraint term \eqref{constraint-partially-deconfined} is added in the simulations.
However this time we can do more: we can easily separate the confined and deconfined sectors and confirm
\eqref{K-vs-theta-BFSS} separately
in each sector; see Fig.~\ref{fig:theta-vs-K-BFSS-constrained_split}.
This illuminates the residual symmetry in the master field.

Next let us consider possible $M$-dependence as well.
This time we can calculate $\langle K_i \rangle_{\rm dec}$ and $\langle K_i \rangle_{\rm con}$
directly as
\begin{eqnarray}
\langle K_i \rangle_{\rm dec}
=
\left\langle\frac{1}{M}\sum_{i=1}^M K_i\right\rangle,
\qquad
\langle K_i \rangle_{\rm con}
=
\left\langle\frac{1}{N-M}\sum_{i=M+1}^N K_i\right\rangle.
\end{eqnarray}
These values can be obtained from Table~\ref{tab:variance_const}, as
\begin{eqnarray}
\langle K_i \rangle_{\rm dec}
=
\left(
\frac{M}{N}\cdot\sigma^2_{\rm dec}
+
\left(
1-\frac{M}{N}
\right)
\cdot\sigma^2_{\rm con,\ off\ diagonal}
\right)\times 9
\end{eqnarray}
and
\begin{eqnarray}
\langle K_i \rangle_{\rm con}
=
9\sigma^2_{\rm con}.
\end{eqnarray}

By using them, without assuming the $M$-independence of $\rho^{\rm(X)}_{\rm con}(x)$ and $\rho^{\rm(X)}_{\rm dec}(x)$, we can write down a reasonable ansatz:
\begin{eqnarray}
K_i
=
\langle K_i \rangle_{\rm con}
+
2\left(
\langle K_i \rangle_{\rm dec}
-
\langle K_i \rangle_{\rm con}
\right)
\cos\theta_i.
\label{K-vs-theta-with-M-dependence}
\end{eqnarray}
The results are shown in the right panels of Fig.~\ref{fig:theta-vs-K-BFSS-constrained} and Fig.~\ref{fig:theta-vs-K-BFSS-constrained_split},
with the dotted-blue lines.
They explain the data very well, and the difference from \eqref{K-vs-theta-BFSS} is very small.

\subsubsection{Energy}\label{sec:bBFSS-constrained-energy}
\hspace{0.51cm}

\begin{figure}[htbp]
	\centering
		\includegraphics[width=0.75\textwidth]{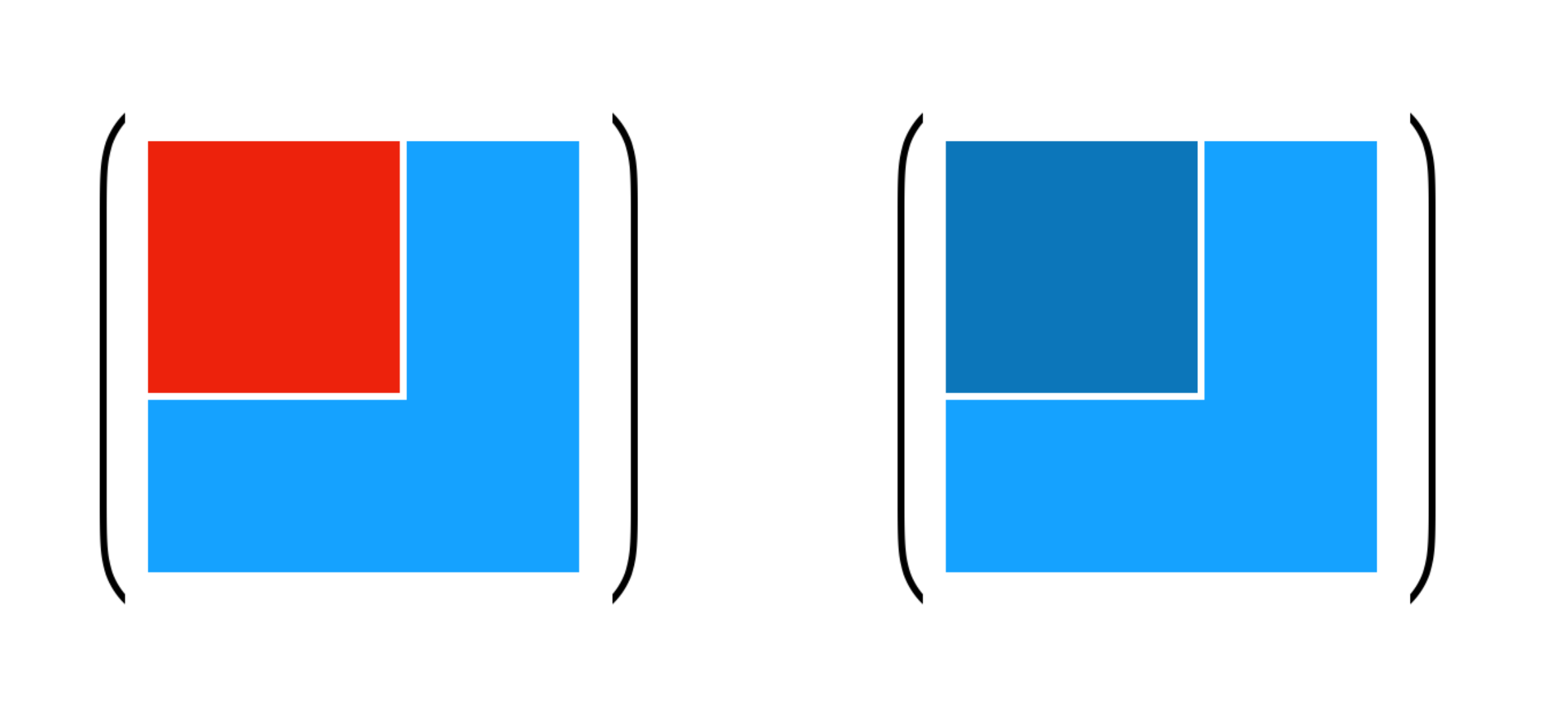}
		\caption{Two different kinds of constrained simulations. [Left] Partially deconfined phase,
with the constraint term Eq.~\eqref{constraint-partially-deconfined}.
[Right] Completely confined phase, with the constraint term Eq.~\eqref{constraint-completely-confined}.
}\label{fig:Two-constrained-simulations}
\end{figure}

Let us use $X_{\rm dec}$ and $X_{\rm con}$ to denote the deconfined and confined sectors, respectively. By definition, $X=X_{\rm dec}+X_{\rm con}$.
The energy $E=\left\langle-\frac{3N}{4\beta}\int_0^\beta dt{\rm Tr}[X_I,X_J]^2\right\rangle$ consists of the contribution from the purely confined part,
\begin{eqnarray}
E_{\rm con}
\equiv
\left\langle
-\frac{3N}{4\beta}\int_0^\beta dt\ {\rm Tr}[X_{{\rm con}I},X_{{\rm con}J}]^2
\right\rangle,
\end{eqnarray}
and the terms involving $X_{\rm dec}$,
\begin{eqnarray}
E_{\rm dec}
\equiv
E-E_{\rm con}.
\end{eqnarray}
$E_{\rm dec}$ would be interpreted as the contribution from the deconfined part and the interaction between
confined and deconfined part. A natural guess is that $E_{\rm con}$ does not know whether the SU($M$)-sector is deconfined or not.
To check it, we construct the counterpart of $X_{\rm con}$ in the completely confined phase.
Namely, as shown in Fig.~\ref{fig:Two-constrained-simulations}, we replace the SU($M$)-sector
with the confined configuration, by setting both $P_M$ and $P_{N-M}$ to zero.
If we take a generic configuration in the completely confined phase, neither $P_M$ nor $P_{N-M}$ is zero, although $P$ is zero.
Hence we perform another kind of constrained simulation
by adding
\begin{eqnarray}
\Delta S
=\left\{
\begin{array}{cc}
\frac{\gamma}{2}\left(|P_{M}|-\delta\right)^2 & (|P_{M}|>\delta)\\
\frac{\gamma}{2}\left(|P_{N-M}|-\delta\right)^2 & (|P_{N-M}|>\delta)
\end{array}
\right.
\label{constraint-completely-confined}
\end{eqnarray}
We calculate the counterparts of $E_{\rm con}$ and $E_{\rm dec}$, which we denote by $E_{\rm con}^{(0)}$ and $E_{\rm dec}^{(0)}$, for this constraint.

\begin{figure}[htbp]
\centering
	\includegraphics[width=0.65\textwidth]{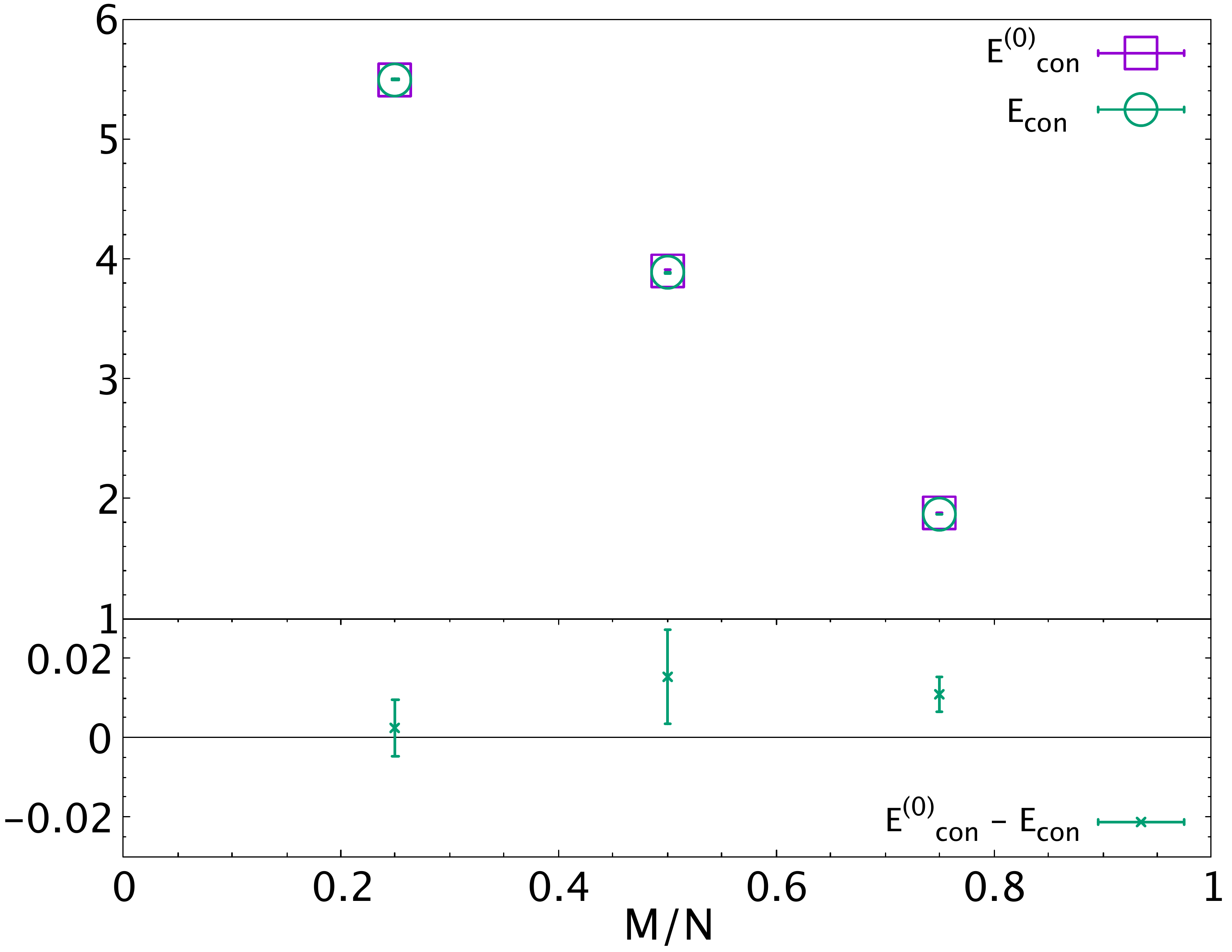}
	\caption{Yang-Mills matrix model, $E_{\rm con}$ and $E_{\rm con}^{(0)}$ for $N=64$,
$M=16,32$ and $48$, with 24 lattice points.
The error bars in each figure are obtained by jackknife analysis.
}\label{fig:Xcon_energy}
\end{figure}

The important observation is that $E_{\rm con}$ and $E_{\rm con}^{(0)}$ are very close, as shown in Fig.~\ref{fig:Xcon_energy}.
Therefore, with good numerical precision, the difference of the energy compared to the completely confined phase comes only from the increment of $E_{\rm dec}$.

\subsection{Summary of the numerical results}\label{sec:simulation_summary}
\hspace{0.51cm}
Let us summarize the simulation results and see how they are related to partial deconfinement.

In Sec.~\ref{sec:scalars-bBFSS-unconstrained},
we confirmed the separation of the distribution of $X_{I,ij}$'s to
$\rho^{\rm(X)}_{\rm dec}(x)$ and $\rho^{\rm(X)}_{\rm con}(x)$ just as in the Gaussian model,
by using the same expression \eqref{eq:separation_rho_scalar},
and assuming that they are independent of $M$.
While this is consistent with partial deconfinement,
this fact alone does not establish the separation to the SU($M$)- and SU($N-M$)-sectors;
logically, it just means the separation to $M^2$ and $N^2-M^2$ degrees of freedom.
(The results explained in Sec.~\ref{sec:constrained_simulation} establish
the separation to the SU($M$)- and SU($N-M$)-sectors.)
In Sec.~\ref{sec:BFSS-theta-vs-K-unconstrained},
we studied the correlation between $\theta_i$ and $K_i\equiv\sum_{I,j}\frac{1}{\beta}\int dt|X_{I,ij}|^2$.
We saw the same kind of correlation as in the Gaussian matrix model,
which naturally fits to the partial-deconfinement scenario.
In summary, the numerical results shown in Sec.~\ref{sec:BFSS-unconstrained}
are consistent with partial deconfinement,
but they are not yet rock-solid evidence,
due to the missing demonstration of the SU($M$)$\times$SU($N-M$)-structure.
This is the reason why we needed the constrained simulation introduced in Sec.~\ref{sec:constrained_simulation}.

In Sec.~\ref{sec:sanity-check-constrained-simulation} we confirmed that the constraint term $\Delta S$ does not change the theory, except that it fixes the ordering of $\theta$'s such that
$\theta_1,\cdots,\theta_M$ (resp. $\theta_{M+1},\cdots,\theta_N$) are distributed with the probability function $\frac{1+\cos\theta}{2\pi}$
(resp. $\frac{1}{2\pi}$), which are the form expected for the deconfined sector (resp. confined sector).
This means that, if partial deconfinement is actually taking place, then the specific embedding of
SU($M$) shown in Fig.~\ref{fig:matrix-partial-deconfinement} should be realized,
although we did not touch the scalar fields.
In the following subsections, we provided evidence supporting this expectation.
In Sec.~\ref{sec:constrained-X-distribution}, we studied the distributions of $X_{I,ij}$
in the upper-left $M\times M$ sector (the red sector in Fig.~\ref{fig:matrix-partial-deconfinement}) and the rest (the blue sector in Fig.~\ref{fig:matrix-partial-deconfinement}). The results were consistent with
$\rho^{\rm(X)}_{\rm dec}$ and $\rho^{\rm(X)}_{\rm con}$ obtained in Sec.~\ref{sec:scalars-bBFSS-unconstrained}, and there was only weak $M$-dependence.
Indeed, all the excitations are coming from the red sector, consistent with  SU($M$)-partial-deconfinement with the specific embedding of SU($M$) shown in Fig.~\ref{fig:matrix-partial-deconfinement}.
In Sec.~\ref{sec:theta-vs-K-constrained}, we studied the correlation between $\theta_i$ and $K_i\equiv\sum_{I,j}\frac{1}{\beta}\int dt|X_{I,ij}|^2$.
We looked at the statistical features of $i=1,\cdots,M$ and $i=M+1,\cdots,N$ separately,
and confirmed that the results are consistent with the specific embedding of SU($M$)
shown in Fig.~\ref{fig:matrix-partial-deconfinement}.
In Sec.~\ref{sec:bBFSS-constrained-energy}, we separated the energy to two parts: $E_{\rm dec}$, which involves the SU($M$)-sector, and $E_{\rm con}$, which does not involve the SU($M$)-sector.
Then we showed that the increment of the energy compared to the ground state comes solely from $E_{\rm dec}$.
Again, this is consistent with SU($M$)-partial-deconfinement with the specific embedding of SU($M$) shown in Fig.~\ref{fig:matrix-partial-deconfinement}.
Based on these observations in Sec.~\ref{sec:constrained_simulation}, we conclude that partial deconfinement is taking place in the Yang-Mills matrix model.

Let us close this section by discussing a possibility of small, additional $M$-dependence.
As we have mentioned before, we do not find a theoretical reason that $\rho^{\rm(X)}_{\rm con}$ and $\rho^{\rm(X)}_{\rm dec}$ have to be completely independent of $M$.
Due to the interaction between the confined and deconfined sectors, they might change depending on $M$.

Via constrained simulations, it is easier to see the $M$-dependence, if it exists.
Actually, as we have seen in Sec.~\ref{sec:constrained-X-distribution}, $\rho^{\rm(X)}_{\rm con}$ and $\rho^{\rm(X)}_{\rm dec}$ appear to have a small $M$-dependence.
In Table~\ref{tab:variance_const}, we can see that $N=48, 64$ and $N=128$ exhibit almost the same dependence on $\frac{M}{N}$, and hence, this $M$-dependence is unlikely to be a finite-$N$ artifact.
Somewhat miraculously, the changes of $r_0$ and $r_1$ cancel and ${\rm Tr}X_I^2$ can be fit as in \eqref{eq:R-YM} by using the $M$-independent values.
Note also that the analysis in Sec.~\ref{sec:theta-vs-K-constrained} was compatible with this $M$-dependence.
Such intricate $M$-dependence may be a finite-lattice-size artifact.

In constrained simulations, we could confirm the separation between two sectors taking into account a possible $M$-dependence.
We did not assume $M$-independence, and the difference between the confined and deconfined sector turned out to be much larger than a possible $M$-dependence.
Therefore, even in case the small $M$-dependence observed in Sec.~\ref{sec:constrained-X-distribution} survives in the continuum limit, it does not invalidate our conclusions.

\section{Conclusion and discussion}\label{sec:conclusion}
\hspace{0.51cm}
In this paper, we presented numerical evidence for partial deconfinement in the Yang-Mills matrix model at strong coupling.
In order to establish the numerical methods we have studied the Gaussian matrix model as well.
We identified a few nontrivial properties of the master field which are consistent with partial deconfinement, and confirmed that those properties are visible in lattice simulations.
Because the master field is unique only up to gauge transformations, we used the static diagonal gauge, which drastically simplified the analysis.
We expect that other strongly coupled theories exhibit partial deconfinement in the same manner;
as discussed in Refs.~\cite{Hanada:2018zxn,Hanada:2019kue,Hanada:2020uvt}, heuristic arguments supporting partial deconfinement assume nothing specific to weak coupling.

In this paper, we considered only one fixed value of the temperature.
It is important to extend the analysis to various different values and study the temperature dependence at the maximum of the free energy, which describe the states realized in the microcanonical ensemble.

We saw some qualitative similarities between the Gaussian matrix model and the Yang-Mills matrix model, 
namely the same functional form of the Polyakov line phase distribution $\rho^{\rm (P)}(\theta)$ and very weak $\frac{M}{N}$-dependence.
We expect that such similarities are specific to those models.
Indeed, there are other examples which have more complicated forms of  
$\rho^{\rm (P)}(\theta)$ and more involved dependences on $\frac{M}{N}$. See e.g.~Refs.~\cite{Hanada:2019czd,Hanada:2019kue}. 

A natural future direction is to investigate QCD.
Suppose, as usual, $N=3$ is not too far from $N=\infty$.
Because the QCD phase transition is not of first order~\cite{Aoki:2006we}, the partially deconfined phase is thermodynamically stable~\cite{Hanada:2018zxn}.
In the large-$N$ limit, the size of the deconfined sector can be read off from the distribution of the Polyakov loop phases~\cite{Hanada:2020uvt}.
If we (perhaps too naively) adopt the relation of weakly-coupled Yang-Mills theory on S$^3$, $0<P<\frac{1}{2}$, to roughly identify the partially deconfined phase, partial deconfinement would persist up to several hundred MeV.
(See e.g.~Ref.~\cite{Bazavov:2013yv} regarding the numerical estimate of the renormalized Polyakov loop.)
In finite-temperature QCD, partial deconfinement may have consequences for flavor symmetry.
Ref.~\cite{Hanada:2019czd} discussed such a possibility.
A more recent proposal \cite{Glozman:2020ujx} considers another symmetry enhancement mechanism related to partial deconfinement based on numerical data from lattice QCD simulation.
Yet another interesting observation which might be related to partial deconfinement is the existence of a deconfined and CP-broken phase in SU(2) Yang-Mills theory at $\theta=\pi$ \cite{Chen:2020syd}. 
A natural possibility would be that the deconfinement and CP-breaking temperatures considered in Ref.~\cite{Chen:2020syd} correspond to the Hagedorn and GWW temperature. 

A reasonable starting point to understand the implication of partial deconfinement for QCD would be to study the response of probe fermions to partial deconfinement in simple models such as the Yang-Mills matrix model.
Another reasonable starting point would be an applied-holography-like approach.
Although known examples which have a controllable gravity dual exhibit first order transitions unlike actual QCD, there may be some universal features which can be addressed by the analysis on the gravity side.
For example, Refs.~\cite{Dias:2016eto,Jokela:2015sza} contain useful comments about the gravity side, while first order transitions in holography are studied by Refs.~\cite{Faedo:2017fbv,Elander:2020rgv} and
exhibit similar features as partial deconfinement.
It would be interesting to find out whether these holographic descriptions can be understood in the framework we studied in this paper.

\begin{figure}[htbp]
  \centering
  \includegraphics[width=0.5\textwidth]{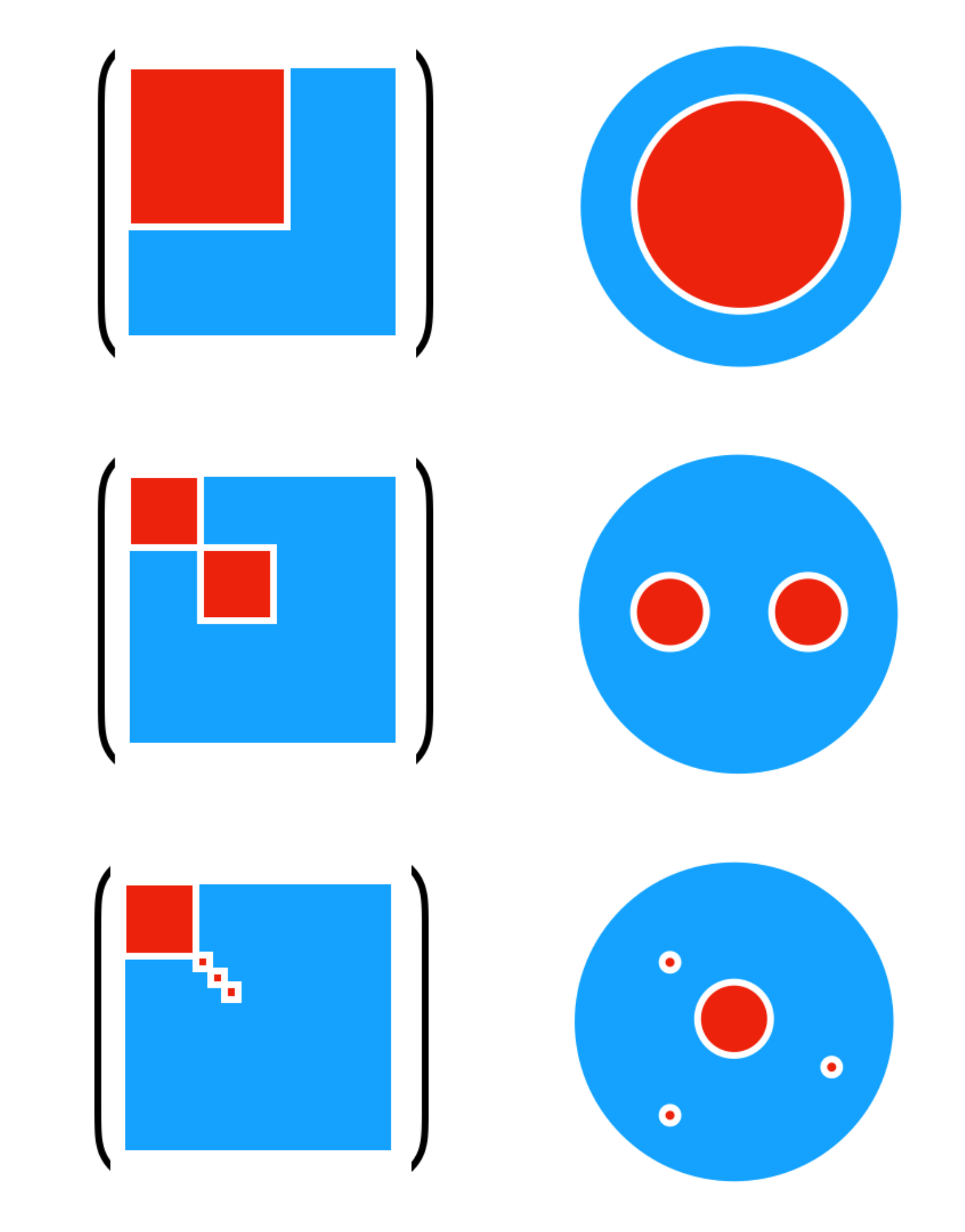}
  \caption{
[Top] Partially deconfined phase should correspond to the small black hole in the bulk.
[Middle] A state with two partially-deconfined sectors would describe two small black holes.
[Bottom] Radiation from black hole would be described by small blocks.
}\label{fig:Matrix-vs-Geometry}
\end{figure}

The master fields should be related to classical geometry, when the theory admits a weakly-curved gravity dual.
Previously, the classical dynamics of the Yang-Mills matrix model has been studied~\cite{Asplund:2011qj,Asplund:2012tg,Aoki:2015uha,Gur-Ari:2015rcq,Aprile:2016mis} with the expectation that typical configurations in the classical theory capture the aspects of the gravitational geometry.
Therefore, the specific properties of the master field discussed in this paper should have some geometric interpretation.
Clarifying partial deconfinement within the framework of AdS/CFT is another promising direction of research.
At present, we have not yet succeeded to do so, but just to illustrate what the outcome might be, we end this outlook with some speculations.
The natural counterpart of the partially-deconfined phase on the gravity side is the small black hole phase~\cite{Aharony:1999ti,Aharony:2003sx,Dias:2016eto} (the top row of Fig.~\ref{fig:Matrix-vs-Geometry}).
Actually, the original motivation to introduce the partially-deconfined phase~\cite{Hanada:2016pwv} was to find the dual of the small black hole phase.
Up to $1/N$ corrections, all the entropy comes from the deconfined sector (black hole), and it appears to be consistent~\cite{Hanada:2016pwv} with the Bekenstein-Hawking entropy~\cite{Bekenstein:1973ur,Hawking:1974sw}.
Hence, it is natural to interpret the deconfined and confined phases as the black hole and its exterior.
Note that, in general, the confined and deconfined sectors are interacting with each other.\footnote{
Many examples studied previously were weakly-coupled theories and hence the interaction was not important; see e.g.~Ref.~\cite{Hanada:2019czd}.
}
Such interaction, which was discussed in Sec.~\ref{sec:constrained_simulation} when we considered the energy, could explain the change of the geometry of the exterior compared to the vacuum, due to the existence of the black hole.
According to the BFSS proposal~\cite{Banks:1996vh}, block-diagonal configurations, which are partially Higgsed, describe multi-body state.
The same interpretation would make sense for multiple partially-deconfined sectors (the middle and bottom rows of Fig.~\ref{fig:Matrix-vs-Geometry}).
Hawking radiation would be described by ripples on the confined sector, or tiny deconfined blocks (strings which are not too long).
Local operators can excite tiny deconfined blocks, which propagate in the bulk (confined sector).
It would give a natural generalization of the philosophy of BFSS --- everything is embedded in matrices --- to gauge/gravity duality \`{a} la Maldacena.
Note also that the color degrees of freedom in the confined sector can naturally be entangled, and hence, the scenario that the entanglement is responsible for the emergence of the bulk geometry in holography \cite{Maldacena:2001kr,VanRaamsdonk:2010pw} would naturally fit this point of view~\cite{Hanada:2019czd,Alet:2020ehp}.
It would be fun to imagine that the tensor network representing the bulk geometry~\cite{Swingle:2009bg} is hidden in the space of colors.
The ideas proposed in Refs.~\cite{Das:2020jhy,Mazenc:2019ety} may be related to this speculation.
Partial deconfinement may also be related to other mechanisms of emergent geometry such as the ones in the Eguchi-Kawai model ~\cite{Eguchi:1982nm,Kovtun:2007py} or the IKKT matrix model~\cite{Ishibashi:1996xs,Aoki:1998vn,Nishimura:2001sx,Kim:2012mw}, because the eigenvalue distribution plays an important role there as well.

Numerically tractable targets useful for quantum gravity are the BMN matrix model~\cite{Berenstein:2002jq}, and, perhaps, the BFSS matrix model \cite{Banks:1996vh,deWit:1988wri,Itzhaki:1998dd}.
In the BFSS matrix model, only the large black hole phase (type IIA black zero-brane) is expected in the 't Hooft large-$N$ limit (i.e.~$T$ and $\lambda=g_{\rm YM}^2N$ are fixed)~\cite{Itzhaki:1998dd}.
In the very strongly-coupled region ($T\lesssim\lambda^{1/3}N^{-5/9}$), M-theory becomes a better description, and the eleven-dimensional Schwarzschild black hole should describe thermodynamics~\cite{Itzhaki:1998dd}.
It would be natural to expect that this parameter region, which has negative specific heat, is partially deconfined.
Practically, it is not easy to study such a strongly-coupled region numerically.
The situation is better in the BMN model, which is a deformation of the BFSS matrix model with the flux parameter $\mu$.
For finite $\mu$, we expect a first order transition at $T\sim\mu$ \cite{Costa:2014wya}.
At larger $\mu$, the transition temperature is higher, the lattice size needed to study the transition region is smaller, and hence, simulations near the transition temperature are numerically less demanding (see Refs.~\cite{Catterall:2010gf,Asano:2018nol,Schaich:2020ubh} for previous attempts), and it might be possible to study the details of the phase transition in the near future.

It would also be interesting if the partially deconfined phase could be seen in classical gauge theories, which serve as starting points for learning about real-time dynamics.
Different but analogous situations can be seen in the classical real-time dynamics of two-dimensional Yang-Mills theory with adjoint scalar fields \cite{Hanada:2018qpf}.
This theory exhibits the `non-uniform black string phase' \cite{Aharony:2004ig} which shares a few essential features with the partially-deconfined phase, including the separation in the space of color degrees of freedom \cite{Hanada:2019kue}.

\acknowledgments
M.~H. thanks Pavel Buividovich, Leoniz Glozman, Raghav Jha, David Schaich, Hidehiko Shimada, Yuya Tanizaki and Masaki Tezuka for discussions and comments.
G.~B. acknowledges support from the Deutsche Forschungs-gemeinschaft (DFG) Grant No.~BE 5942/2-1.
N.~B. was supported by an International Junior Research Group grant of the Elite Network of Bavaria.
The work of M.~H. was partially supported by the STFC Ernest Rutherford Grant ST/R003599/1 and JSPS  KAKENHI  Grants17K1428.
P.~V. was supported by DOE LLNL Contract No. {DE-AC52-07NA27344}.
The numerical simulations were performed on ATHENE, the HPC cluster of the Regensburg University Compute Centre; 
the HPC cluster ARA of the University of Jena; and `pochi' at  the University of Tsukuba.
Computing support for this work came also from the Lawrence Livermore National Laboratory (LLNL) 
Institutional Computing Grand Challenge program.

\bibliographystyle{JHEP}
\bibliography{partial-deconfinement-test}

\appendix
\section{Details of lattice simulation}\label{sec:lattice_setup}

\subsection{Yang-Mills matrix model}
\hspace{0.51cm}

The action is the same as the one used in Ref.~\cite{Bergner:2019rca}, except for $\Delta S$ added for the constrained simulation.
This is the bosonic version of the tree-level improved action used for the study of the D0-brane matrix model~\cite{Berkowitz:2016jlq,Berkowitz:2018qhn}.
This lattice regularization utilizes the static diagonal gauge, similarly to the study of the D0-brane matrix model in Refs.~\cite{Hanada:2007ti,Anagnostopoulos:2007fw,Hanada:2013rga}.
For more details, see Sec.~2.~2 of Ref.~\cite{Berkowitz:2018qhn}.
We used the Hybrid Monte Carlo algorithm~\cite{Duane:1987de}.

\subsubsection{A technical remark regarding the constrained simulation}
\hspace{0.51cm}

The Faddeev-Popov term associated with the gauge fixing is $S_{\rm FP}=-\sum_{i<j}\log\left|\sin^2\left(\frac{\theta_i-\theta_j}{2}\right)\right|$.
This term becomes infinitely large when neighboring $\theta$'s coincide.
In the HMC simulation, this infinity leads to an infinitely strong repulsive force, which prevents the ordering of $\theta$'s from changing.
This is not a problem in the original model without the constraint term $\Delta S$, because of the S$_N$ permutation symmetry.
However this is a problem when we add $\Delta S$; for example, if the initial condition is taken such that $\theta_1<\theta_2<\cdots<\theta_N$, the target distribution --- $\frac{1+\cos\theta}{2\pi}$ for $\theta_1,\cdots,\theta_M$ and $\frac{1}{2\pi}$ for $\theta_{M+1},\cdots,\theta_N$ --- cannot be realized.

To avoid this problem, we randomly choose $1\le p\le M$ and $M+1\le q\le N$, exchange $p$-th and $q$-th row/column and perform a Metropolis test.
Between each HMC step, 100 random exchanges are performed.
Note that only $\Delta S$ matters in this Metropolis test.
It is easy to see that this procedure does not violate any condition in the Markov Chain Monte Carlo.
Therefore, the correct distribution is obtained.
That this procedure works shows the non-uniqueness of `gauge fixing', as in the Gaussian matrix model.
When the coefficient of the constraint term $\gamma$ is very large, the permutation takes place only when $\theta_p\simeq\theta_q$, which corresponds to the residual permutation symmetry.

\subsection{Gaussian matrix model}
\hspace{0.51cm}
We have just replaced the potential term of the Yang-Mills matrix model with
\begin{eqnarray}
S_{\rm mass,\ lattice}
=
\frac{aN}{2}\sum_t {\rm Tr}X_I(t)^2.
\end{eqnarray}

\section{More on $\rho^{\rm(X)}$ in Yang-Mills matrix model}\label{sec:comments_rho(x)}
\hspace{0.51cm}
In this Appendix, we give a few observations regarding the distribution $\rho^{\rm(X)}$ in the Yang-Mills matrix model, which appear to be very different from the case of the Gaussian matrix model.

We compare the distributions $\rho^{\rm(X)}_{\rm con}(x)$ and $\rho^{\rm(X)}_{\rm dec}(x)$ with Gaussian distributions whose variances are given by Eq.~\eqref{eq:sigma^2_con} and Eq.~\eqref{eq:sigma^2_dec}.
\begin{figure}[htbp]
  \centering
  	\begin{subfigure}[b]{0.5\textwidth}
      \includegraphics[width=\textwidth]{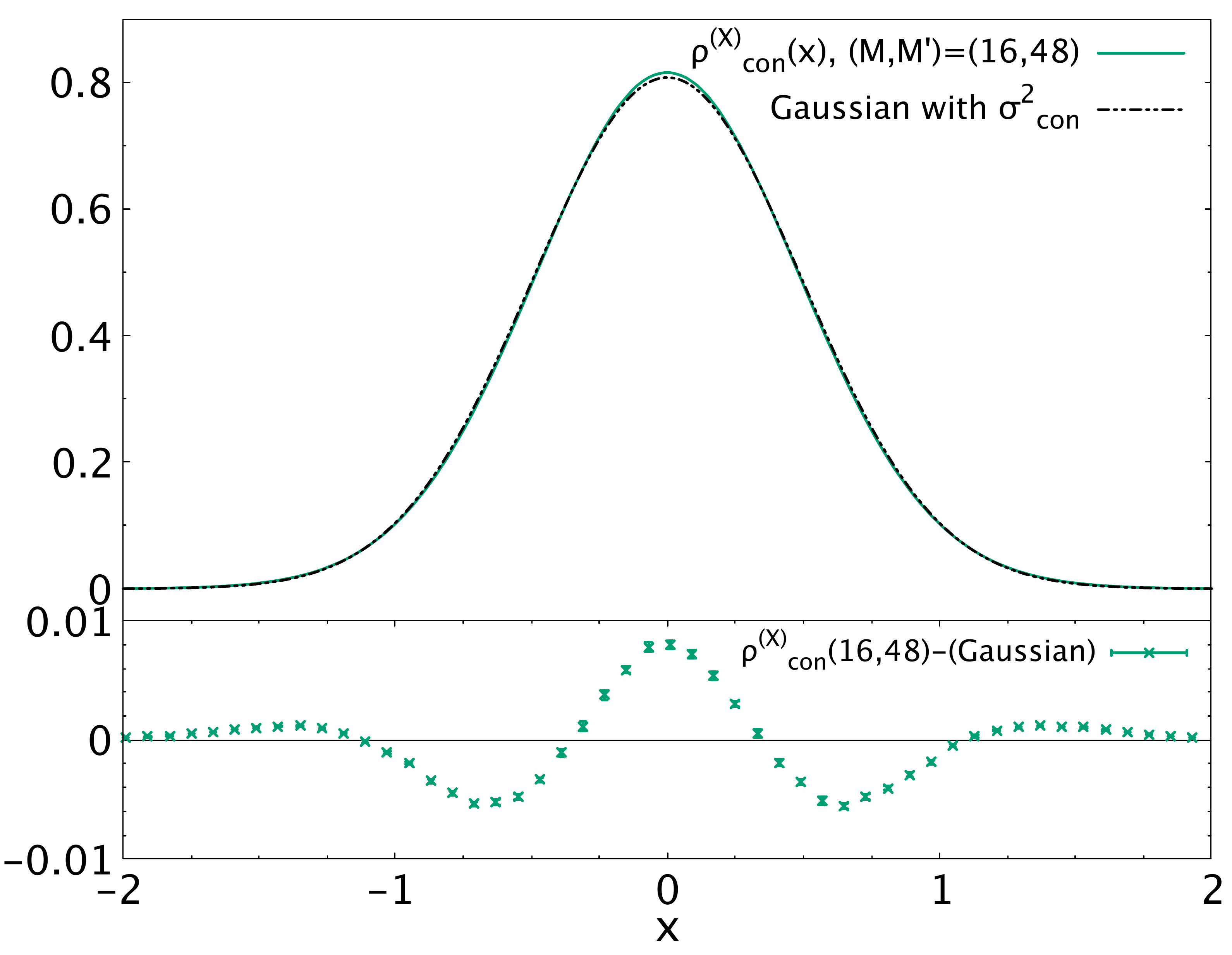}
      \caption{$\rho^{\rm(X)}_{\rm con}(x)$}
  	\end{subfigure}
  	\begin{subfigure}[b]{0.5\textwidth}
      \includegraphics[width=\textwidth]{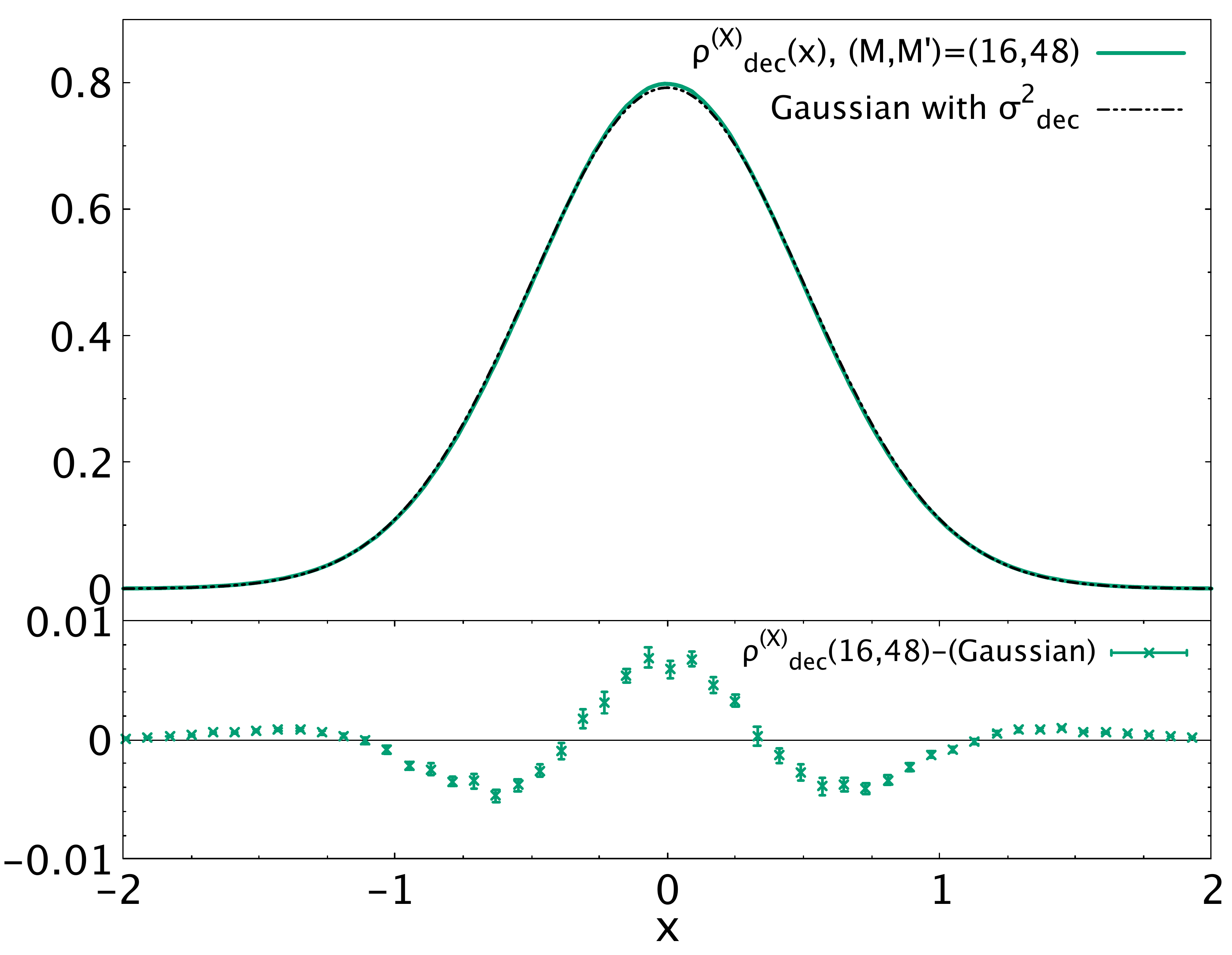}
      \caption{$\rho^{\rm(X)}_{\rm dec}(x)$}
  	\end{subfigure}
\caption{Comparison between $\rho^{\rm(X)}_{\rm con}(x)$ and $\rho^{\rm(X)}_{\rm dec}(x)$ from the Yang-Mills matrix model and Gaussian distributions.
The histograms of $\rho^{\rm(X)}_{\rm con}(x)$ and $\rho^{\rm(X)}_{\rm dec}(x)$ are identical to the ones in Fig.~\ref{fig:P_constrained-BFSS-rho-scalar}, namely $N=64$, $L=24$, and $T=0.885$.
The variances of the Gaussian are chosen as $\sigma^2_{\rm con} = 0.244$ and $\sigma^2_{\rm dec} = 0.254$.
}
\label{fig:comparison_with_Gaussian_Pconst}
\end{figure}
The results are shown in Fig.~\ref{fig:comparison_with_Gaussian_Pconst}.
We show only the distributions $\rho^{\rm(X)}_{\rm con}(x)$ and $\rho^{\rm(X)}_{\rm dec}(x)$ computed by a pair of $(M,M')=(16, 48)$ because the $M$-dependence is small.
We can see that $\rho^{\rm(X)}_{\rm con}(x)$ and $\rho^{\rm(X)}_{\rm dec}(x)$ are close to Gaussian distributions, while we can see a small but non-vanishing deviation which may be a finite-$N$ or finite-lattice-spacing effect.
Note that, in the Gaussian matrix model, $\rho^{\rm(X)}_{\rm con}(x)$ and $\rho^{\rm(X)}_{\rm dec}(x)$ at $T=T_c$ are far from being Gaussian distributions.

\begin{figure}[htbp]
  \centering
  	\begin{subfigure}[b]{0.5\textwidth}
      \includegraphics[width=\textwidth]{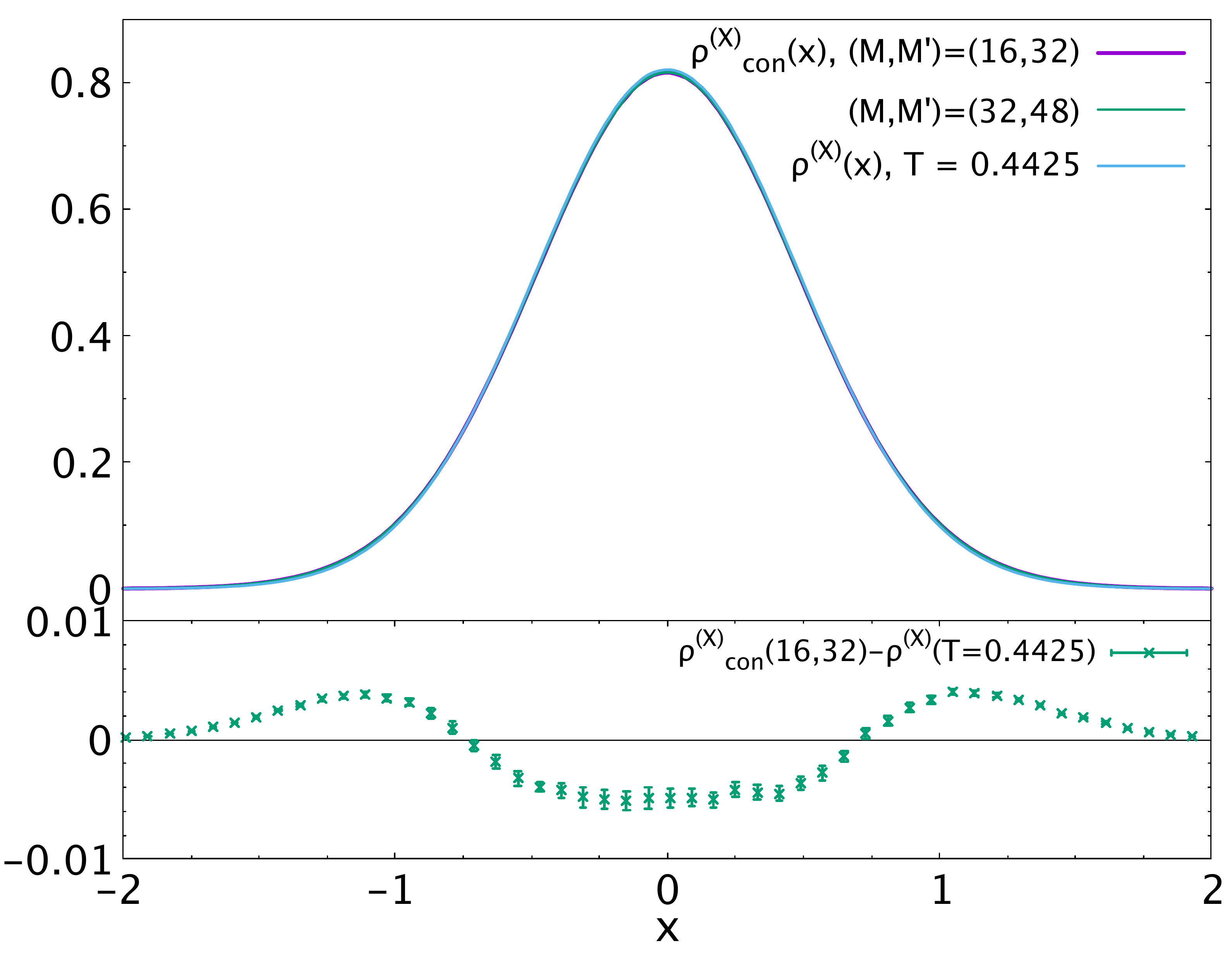}
      \caption{$\rho^{\rm(X)}_{\rm con}(x)$}
  	\end{subfigure}
  	\begin{subfigure}[b]{0.5\textwidth}
      \includegraphics[width=\textwidth]{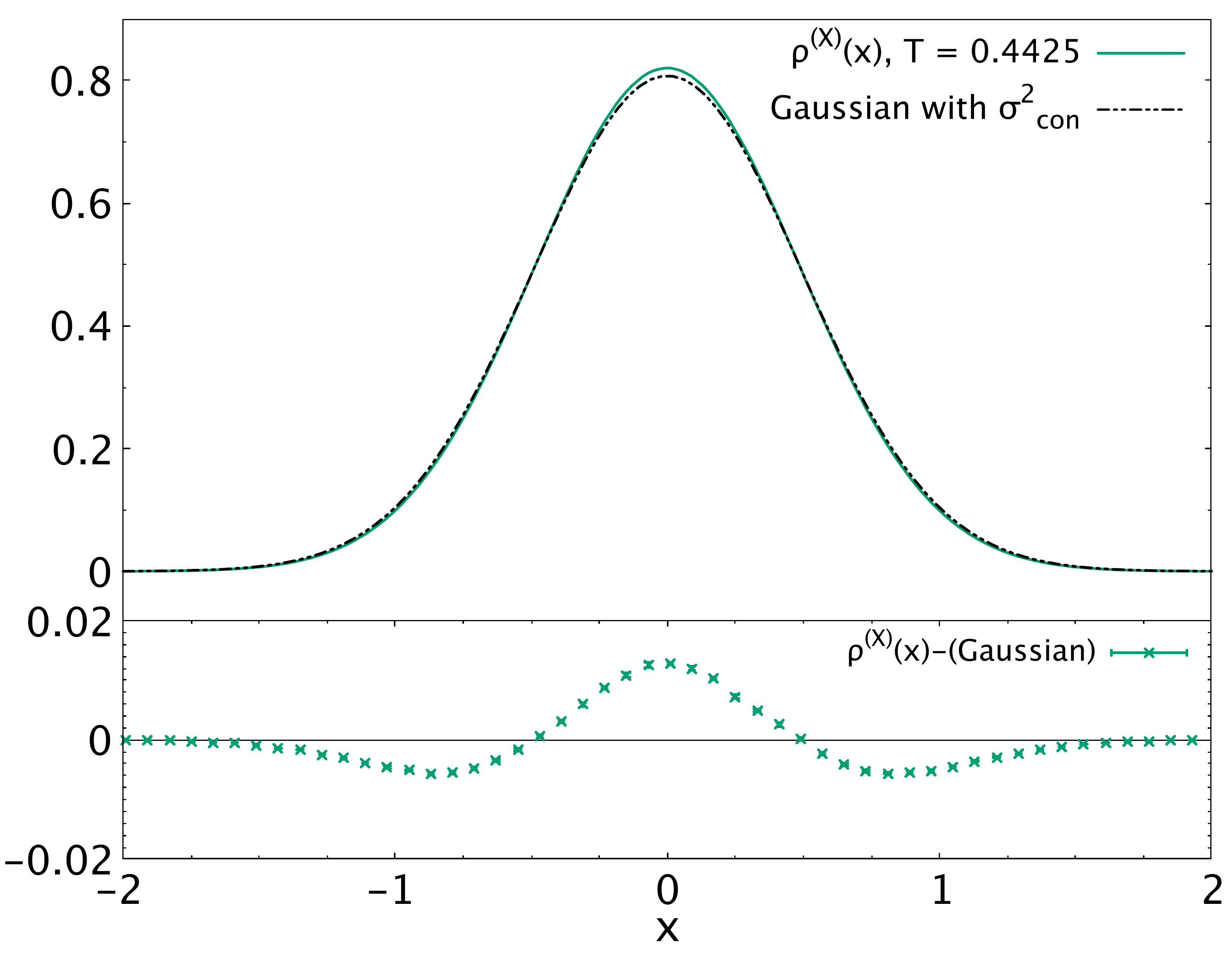}
      \caption{$\rho^{\rm(X)}_{\rm dec}(x)$}
  	\end{subfigure}
\caption{
(a) Comparison between $\rho^{\rm(X)}(x)$ at low temperature ($T=0.4425 \approx 0.5\cdot T_c$) and the distribution $\rho^{\rm(X)}_{\rm con}(x)$ at $T=0.885$.
The histogram $\rho^{\rm(X)}(x)$ is made using 500 configurations, while $\rho^{\rm(X)}_{\rm con}(x)$ is identical to the one in Fig.~\ref{fig:P_constrained-BFSS-rho-scalar}.
(b) Comparison $\rho^{\rm(X)}(x)$ at $T=0.4425$ with the Gaussian distribution.
The variance is chosen as $\sigma^2_{\rm con} = 0.244$.
The error bars in each figure are obtained by jackknife analysis.
}
\label{fig:comparison_lowtemp_with_Gaussian_Pconst}
\end{figure}

Next, we compare $\rho^{\rm(X)}_{\rm con}$ in the transition region with $\rho^{\rm(X)}(x)$ obtained at low temperature, which is in the completely confined phase.
In the left panel of Fig.~\ref{fig:comparison_lowtemp_with_Gaussian_Pconst}, we show the results of this comparison.
We can see that the temperature dependence is small.
Again, this is different from the Gaussian matrix model; see Fig.~\ref{fig:rho_con_low_T} regarding a large temperature-dependence in the Gaussian matrix model.

\end{document}